     \documentclass[oneside]{book}
     \pdfoutput=1
     \usepackage{graphicx}
     \usepackage{epsfig}
     \usepackage{amssymb}
     \usepackage{amsmath}
     \usepackage{amsfonts}
     \usepackage{amsthm}

     \usepackage{latexsym}
\usepackage{graphicx}
\usepackage{verbatim}

\usepackage{xy}
\xyoption{all}

\usepackage{rotating}

\theoremstyle{definition}
\theoremstyle{definition}
\theoremstyle{definition}
\theoremstyle{definition}
\theoremstyle{definition}
\theoremstyle{definition}
\theoremstyle{definition}
\theoremstyle{definition}
\theoremstyle{definition}
\theoremstyle{definition}

\renewcommand{\cup}{\raisebox{4.5pt}{\ \ \ \mbox{\begin{rotate}{180}$\curvearrowleft$\end{rotate}}}}
\renewcommand{\cap}{\curvearrowright}

\newcommand{\bit}{\begin{itemize}}
\newcommand{\eit}{\end{itemize}\par\noindent}
\newcommand{\ben}{\begin{enumerate}}
\newcommand{\een}{\end{enumerate}\par\noindent}
\newcommand{\beq}{\begin{equation}}
\newcommand{\eeq}{\end{equation}\par\noindent}
\newcommand{\beqa}{\begin{eqnarray*}}
\newcommand{\eeqa}{\end{eqnarray*}\par\noindent}
\newcommand{\beqn}{\begin{eqnarray}}
\newcommand{\eeqn}{\end{eqnarray}\par\noindent}

\def\II{{\rm I}}  

\usepackage{color}  

\def\bR{\begin{color}{red}}
\def\bB{\begin{color}{blue}}
\def\bM{\begin{color}{magenta}} 
\def\bC{\begin{color}{cyan}}
\def\bW{\begin{color}{white}}
\def\bBl{\begin{color}{black}}
\def\bG{\begin{color}{green}} 
\def\bY{\begin{color}{yellow}}
\def\e{\end{color}}

\begin{document}
     \chapter[
     %
     %
              Deep Beauty---Coecke (rev. yyyy Mmm dd)
     %
     %
     ]{\huge{
     A universe of 
     processes\\  
     and some of its guises 
     }\bigskip
     \\ \large{
                             Bob Coecke
     }\\
     \bigskip
     \small{
                             Oxford University Computing Laboratory
     \\
                             OX1 3QD Oxford, UK
   \\
                             coecke@comlab.ox.ac.uk
     }}
%


\section{Introduction}

Our starting point is a particular `canvas' aimed to `draw' theories of physics, which has \em symmetric monoidal categories \em as its mathematical  backbone.  In this paper we consider the \em conceptual foundations \em for this canvas, and how these can then  be converted into mathematical structure.   

With very little structural effort (i.e.~in very abstract terms) and in a very short time  span the \em categorical quantum mechanics \em (CQM) research program, initiated by Abramsky and the author 
in \cite{AC}, has reproduced a surprisingly large fragment of quantum theory  \cite{deLL, Selinger, Vicary, CPav, CPaq, CPV, CD, AbrClone, CPer}.  It also provides new insights both in \em quantum foundations \em and in \em quantum information\em, for example in \cite{CPaqPav, CPaqPer, CE, CES, RossSimon, CWWWZ, CK, DP2}, and has even resulted in automated reasoning software called {\tt quantomatic} \cite{DixonDuncan, quanto, quantodemo} which exploits the deductive power of CQM, which is indeed a  \em categorical  quantum logic \em \cite{Duncan}.  

In this paper we complement the available material by not requiring prior knowledge of category theory, and by pointing at connections to previous and current developments in the foundations of physics.

This research program is also in close synergy with developments elsewhere, for example in representation theory \cite{DoplicherRoberts1}, quantum algebra \cite{StreetBook}, knot theory \cite{YetterBook}, topological quantum field theory \cite{Kock} and several other areas.

Philosophically  speaking, this framework achieves the following:
\bit
\item
It shifts the conceptual focus 
from `material carriers'  such as particles, fields, or other `material stuff', to `logical flows of information', by mainly encoding how things stand in \em relation \em to each other.  
\item
Consequently,
it privileges \em processes \em over states. 
The chief structural ingredient of the framework is the \em interaction structure \em on processes. 
\item 
In contrast to other ongoing \em operational \em approaches  (\cite[and references therein]{CMW}, \cite{HardyTempleton}, \cite{DArianoaxioms} etc.), we do not  take probabilities, nor properties,  nor experiments as a priori, nor as generators of structure,
but everything is encoded within the interaction of processes. 
\item 
In contrast to other ongoing \em structural \em approaches  (\cite[and references therein]{CMW},\cite{ConvexApproach1}, \cite{CorbettI,CorbettII,IshamI,IshamII,Landsman} etc.), we do not start from a notion of system, 
systems now being `plugs' within a web of interacting processes.     
Hence systems are organized within a structure for which compoundness is a player and not the structure of the system itself: a system is implicitly defined in terms of its relation(ship)/interaction with other systems.
\eit

So for us,  \em composition of relation(ship)s \em is the carrier of all structure, that is,  how several relations make up one whole.  For example, if $x_1, x_2, x_3, a$ are in relation(ship) $R_1$ and 
$y_1, y_2, y_3, a$ are in relation(ship) $R_2$ then this induces a relation(ship) between 
$x_1, x_2, x_3,y_1, y_2, y_3$.
\[
\epsfig{figure=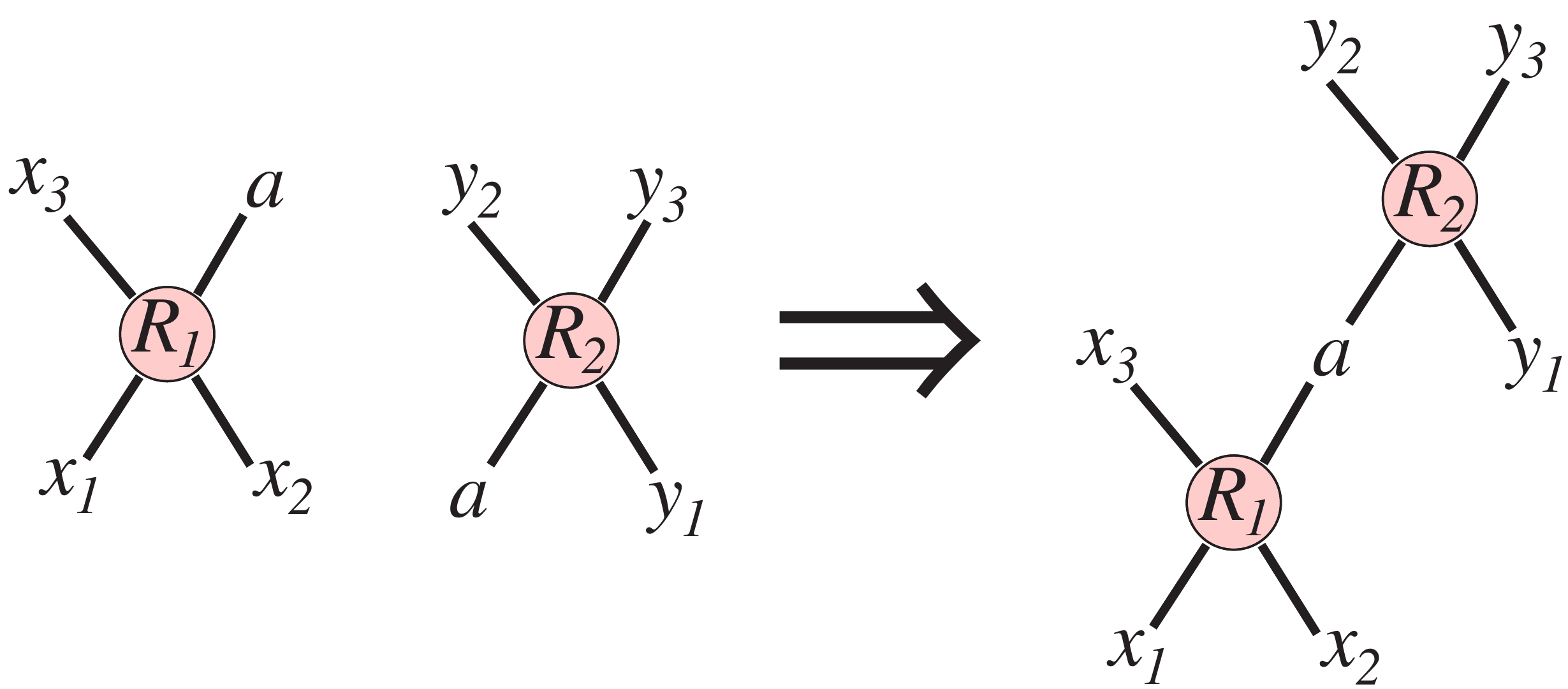,width=210pt}  
\]
These relation(ship)s are much more general than the usual mathematical notion of a relation. A mathematical relation  only tells us whether or not a thing relates to another thing, while for us also `the manner in which' a thing relates to another thing matters.

\em Processes \em are   special kinds of relations, which make up the actual `happenings'.  \em Classicality \em is an attribute of certain processes, and \em measurements \em are special kinds of processes, defined in terms of their capabilities to correlate other processes to these classical attributes. 

So rather than \em quantization\em, what we do is \em classicization  \em within a universe of processes. 
For a certain theory, classicality and measurements may or may not exist, since they are not a priori.
For example, in analogy to `non-quantized field theories', one could consider \em non-classicized \em theories within our setting.



Our attempt to spell out conceptual foundations is particularly timely given that other work in quantum foundations and ours are converging, most notably Hardy's recent work \cite{Hardy2,Hardy3, HardyTempleton} and Chiribella, D'Ariano and Perinotti's even more recent work \cite{DArianoTalk,DArianoPaper}.  Also proponents of the   `convex set approach' \cite[and references therein]{ConvexApproach1,ConvexApproach2} as well as those  of the more traditional `Birkhoff-von Neumann style quantum logic' \cite{BvN, Mackey2, Jauch, Piron} have meanwhile adopted  an essential component of our framework \cite{BarnumWilceCats, Barnum, HardingCats, HeunenJacobs, Heunen, HinesBraunstein}. 

The mathematical  flexibility of our framework   allows one to craft hypothetical non-physical universes, a practice which turns out to provide important insights in the theories of physics that we know, and which recently gained popularity, e.g.~\cite{CBH, Spekkens, Barrett, ConvexApproach1}. Such approaches provide an arena to explore how many physical phenomena arise within a theory from very few assumptions. Our approach has been particularly successful in this context, not only by producing many phenomena from little assumptions, but also by casting theories that initially were defined within distinct mathematical frameworks within a single one. 
For example,  it unifies quantum theory as well as Spekkens' toy theory \cite{Spekkens} within a single framework \cite{CE}, which enabled to identify the key differences \cite{CES}, and also substantially simplified the presentation of the latter.



This chapter is structured as follows.  Section \ref{sec:history} briefly sketches some earlier developments, be it because they  provided ideas, or because they exposed certain sources of failure.   
Section \ref{sec:systrelations} introduces the primitives of our framework: systems, relations (and processes) and their composition. We show how these can be used to encode identical systems, symmetries and dynamics,  variable causal structure, and an environment.  
%
Section \ref{sec:mathmodels}  shows that in mathematical terms these concepts give rise to symmetric monoidal categories.
Next, in \ref{sec:class}, we define classicality and  measurement processes.

As the author is not a professional philosopher but a hell of a barfly, the philosophical remarks  throughout this chapter, of which there are plenty, should be taken with a grain of salt.  

We will purposely be somewhat vague on many fronts, in order to leave several options available for the future; the reader is invited to fill in the blanks.   

\section{Some (idiosyncratic) lessons from the past}\label{sec:history}

We will  in particular focus on the role of \em operationalism \em in quantum theory reconstructions, the \em formal definition \em of a physical property as proposed by the Geneva School (e.g.~\cite{JauchPiron, Moore}), the role of \em processes \em therein and, forefront role of processes in  quantum information, the manner in which algebraic quantum field theory \cite{HaagKastler, Haag} retains the notion of a \em system\em, the modern logical view on the different guises of the \em connective \em `and' for systems,  ideas of \em relationalism \em in physics \cite{Barbour, Rovelli}, the options of \em discreteness \em and \em pointlessness \em in quantum gravity, and the status of \em foundations of mathematics \em in all of this.

While we will make some reference to mathematical concepts in category theory, order theory, C*-algebra, quantum logic, linear logic and quantum information, neither of these  are prerequisites for the remainder of this paper.  

\subsubsection{To measure or not to measure}

While nature hasn't been created by us, the theories which describe it have been, 
and hence, unavoidably these will have to rely on concepts that make reference to our senses, or some easy to grasp generalizations thereof. For example,  as humans we experience a three-dimensional space around us, hence the important role of geometry in physics.  Similarly, the symmetries which we observe around us have led to the importance of group theory in physics.  

A fairly  radical stance in this light is that of the typical \em operationalist\em.    His/her take on quantum theory (and physics in general) is that measurement apparatuses constitute our window on nature.  Different `schools' of operationalists each isolate an aspect of the measurement processes which they think causes the apparent  non-classicality of quantum theory, be it the structure of the space of probabilities, or the structure of the verifiable propositions, etc.  

This practice traces back to the early days of Hilbert space quantum mechanics.    In \em Mathematische Grundlagen der Quantenmechanik \em \cite{vN} von Neumann stressed  that it are the projectors which make up self-adjoint operators that should be the fundamental ingredient of whatever formalism that describes the quantum world.  Indeed, while he himself crafted Hilbert space quantum mechanics, he was also the first to denounce it in a letter  to Birkhoff \cite{Birk,Redei}:
\begin{quote}
``I would like to make a confession which may seem immoral: I do not believe
absolutely in Hilbert space any more.'' 
\end{quote}
This focus on projectors, led to a sharp contrast with happenings in logic \cite{BvN}:  
\begin{quote}
``... whereas for logicians the orthocomplementation properties of negation were the ones least able to withstand a critical analysis, the study of mechanics points to the distributive identities as the weakest link in the algebra of logic.'' ,
\end{quote}
and ultimately resulted in Birkhoff-von Neumann quantum logic \cite{BvN}.  

Via Mackey \cite{Mackey1, Mackey2} several structural paradigms emerged:  the \em Geneva School \em\cite{JauchPiron, Piron} inherited its lattice theoretic paradigm directly from Birkhoff and von Neumann, the \em Ludwig School \em \cite{Ludwig1, Ludwig2} associated the convex structure of state spaces attributed to experimental situations, and the \em Foulis-Randall School \em \cite{FoulisRandall1, FoulisRandall2, Wilce} considered the intersection structure of outcome spaces.

But this key role of the measurement process is rejected by many \em realists \em for whom physical properties of a system exist independent  of any form of observation. E.g.~a star still obeys quantum theory even when not (directly) observed, and, a red pencil does not stop being  red when we are not observing it.  More boldly put: Who  measures the (entirely quantum) universe?\footnote{This utterance is regularly heard 
as a motivation for various histories interpretations \cite{Griffiths, Gell-Mann, IshamConHist}, which, in turns, motivated  the so-called topos approach  to quantum theory \cite{IshamI,IshamII,Landsman} -- we briefly discuss this approach at the end of this section.} 

The realist and operationalist views are typically seen as somewhat conflicting.  But attributing properties to systems which are not being observed, while still subscribing to a clear operational meaning of basic concepts, was already explicitly realized within the Geneva School Mark II \cite{Aerts, Moore}.  While its formal guise was still quite similar to Birkhoff-von Neumann quantum logic, the lattice structure is \em derived \em  from an in-operational-terms precisely stated conception of `what it means for a system to possess a property'.

The following example is due to Aerts \cite{Aerts}, and its pure classicality makes it intriguing in its own right.   Consider a block of wood and the properties `floating' and `burning'.  If, with certainty, we want to know whether the block of wood possesses either of these properties, then we need to, respectively, throw it in the water and observe whether it floats, or, set it on fire and observe whether it burns.  But obviously, if we observed either, we altered the block of wood in such a manner that we won't be able anymore to observe the other.  Still, it makes perfect sense for a block of wood to both be burnable and floatable.

In the Geneva School, one considers a system  $A$ and the `yes/no'-experiments $\{\alpha_i\}_i$ one can perform thereon.  These experiments are related to each other in terms of a  preordering: for experiments  $\alpha$ and $\beta$ we have that $\alpha\preceq\beta$ if and only if,  when we would perform $\alpha$ and obtain a `yes'-answer with certainty, then we would also have obtained a `yes'-answer with certainty when performing $\beta$.
A \em property \em is then defined as an equivalence class for this preordering. 
The lattice structure on the induced partial ordering follows from  the existence of certain product experiments.\footnote{The meet of a collection of properties arises from the experiment consisting of choosing among experiments which correspond to these properties \cite{Aerts, Moore}.  Since these are arbitrary meets, it also follows that the lattice has arbitrary joins (see e.g.~\cite{CM}).}  Such a property is called \em actual \em if the physical system possesses it, and \em potential \em otherwise.

\subsubsection{Measurement among other processes}

So in the Geneva School Mark II properties are a secondary notion emerging from considering experimental procedures on a given system $A$.   The Geneva School Mark III emphasized the role of processes 
 \cite{Daniel, FMP, CInt, CMS, CSmets}. 
Faure, Moore and Piron were able to derive unitarity of quantum evolution by cleverly exploiting the definition of a physical property.\footnote{Roughly, this argument goes as follows: if $\alpha_2$ is an experiment at time $t_2$ and $U$ is the unitary operation which describes how the system evolves from time $t_1$ to time $t_2$, then we can consider the experiment $\alpha_1$ at time $t_1$ which consists of first evolving the system according to $U$ and then performing $\alpha_2$.  More generally, $U$ induces a mapping from experiments at time $t_2$ to experiments at time $t_1$, and one can show that from the definition of a property it follows that this map must preserve all infima.  Using the theory of Galois adjoints it then follows that the map which describes how properties propagate during $U$ must preserve all suprema.  The final purely technical step then involves using Wigner's theorem \cite{Wigner} and a modern category-theoretic account on projective geometry \cite{FaureFrolicher, StubbeVanSteirteghem}.}  Also the  (in)famous orthomodular law  of quantum logic is about how properties propagate in measurement processes.\footnote{Explicitly, for $L$ the lattice of closed subspaces of a Hilbert space ${\cal H}$ and $P_a$ the projector on the subspace $a$ lifted to an operation on $L$, we have 
\[
[P_a:L\to L::b\mapsto a\wedge(a^\perp\vee b):] \dashv [(a\rightarrow_{\mbox{\tiny Sasaki}}-):L\to L::b\mapsto a^\perp\vee(a\wedge b)]\,, 
\]
with $(-\rightarrow_{\mbox{\tiny Sasaki}}-)$ the (in)famous Sasaki hook  \cite{CSmets}.  In the light of the above argument, $P_a$ now plays the role of how properties propagate in quantum measurements, while $(a\rightarrow_{\mbox{\tiny Sasaki}}-)$ is that map which assigns to each property after the measurement one before the measurement.} These results were a key motivation to organize physical processes within certain  \em categories\em, which lift the operationally motivated lattice structure from systems to processes \cite{CMS}.  

The crucial mathematical concept in the above is \em Galois adjunctions\em,\footnote{A survey in the light of the Geneva School approach is in \cite{CM}}   the order-theoretic counterpart to \em adjoint functors \em between categories \cite{Kan}. These are by many category-theoreticains considered as the most important concept provided by category theory, in that almost all known mathematical constructions can be formulated in a very succinct manner in terms of these. Galois adjunctions were already implicitly present in the work by Pool in the late 1960s \cite{Pool}, which arguably was the first attempt to replace the quantum formalism by a formalism  in which \em processes \em are the key players.\footnote{More details on this are in \cite{MooreValckenborgh}.}

 From a more conceptual perspective, the idea that the structure of processes might help us to get a better understanding of nature  was already present in the work of Whitehead in the 1950s \cite{Whitehead} and the work of Bohr in the early 1960's \cite{Bohr}.  It  became more prominent in the work of Bohm in the 1980s and later also in Hiley's \cite{Bohm, BohmHiley1, BohmHiley2}, who is still pursuing this line of research  \cite{Hiley}.


So why did Pool's work nor that by the Geneva School Mark III had ever any real impact?  
As discussed in great detail in \cite{Moore}, the entire Geneva School program only makes sense when considering `isolated systems' on which we can perform the experiments.  This immediately makes it inappropriate to describe a system in interaction with another one.  This is a notorious flaw of most quantum logic programs,  which all drastically failed in providing a convincing abstract counterpart   
to the Hilbert space tensor product.  
In the approach outlined in this paper, we will consider an abstract counterpart   to the Hilbert space tensor product as primitive. It encodes 
how systems interact with other systems, so rather than explicitly given,  its character is implicitly encoded in the structure on processes.  

Today, the \em measurement-based quantum computational model \em (MBQC) poses a clear challenge for a theory of processes.  MBQC is one of the most fascinating  quantum computational architectures, which relies on the dynamics of the measurement process for transforming the quantum state.\footnote{Recently, Rau realized a reconstruction of Hilbert space based on a set of axioms which takes the fact that the one-way measurement-based quantum computational model can realize arbitrary evolutions as its key axiom \cite{Rau}, and proposes this dynamics-from-measurement-processes as a new paradigm for quantum foundations.} By modeling  quantum process interaction in a  \em dagger compact closed category\em, in  \cite{AC, Kindergarten} Abramsky and the author trivialized computations within the Gottesman-Chuang \em logic-gate teleportation \em MBQC model  \cite{Gottesman}.   The more sophisticated Raussendorf-Briegel \em one-way \em MBQC model \cite{Briegel, RBB} was accounted for within a more refined categorical setting, by Duncan, Perdrix and the author \cite{CD, RossSimon, CPer, DP2}. 

\subsubsection{Systems from processes}

Less structurally adventurous than the Ludwig School, the Foulis-Randall School, and the Geneva School, are the 
C*-algebra disciples, who prefer to stick somewhat closer to good old Hilbert space quantum mechanics.  This path was again initiated by von Neumann  after denouncing Birkhoff-von Neumann style quantum logic.\footnote{For a discussion of the what and the why of this we refer the reader to \cite{Redei}.} 
A highlight of the C*-algebraic approach is  \em algebraic quantum field theory \em (AQFT) \cite{HaagKastler, Haag, Halvorson}, mainly due to Haag and Kastler.  In contrast with most other  presentations of quantum field theory, not only is AQFT mathematically solid, but it also has a clear conceptual foundation.  

This approach takes as its starting point  that every experiment takes place within some region of space-time.  Hence to each space time region $R$\footnote{Which is typically restricted to open diamonds in Minkowski space-time.} we assign the C*-algebra $A(R)$ of all observables associated to the experiments that potentially could take place in that region.\footnote{It is a natural requirement that inclusion of regions $R\subseteq R'$ carries over to C*-algebra embeddings $A(R)\hookrightarrow A(R')$, since any experiment that can be performed within a certain region of space-time can also be performed within a larger region of space-time.  The key axiom of    algebraic quantum field theory  is that space-like separated regions correspond to commuting C*-algebras.  All these C*-algebras are then combined in a certain manner to form a giant C*-algebra ${\cal A}$.  The connection with space-time is retained by a  mapping which sends each space-time region on the corresponding sub-C*-algebra of ${\cal A}$, and the embeddings of C*-algebras now become themselves  inclusions.}       While,  quantum field theory does not support the quantum mechanical notion of system due to the creation and annihilation of particles, AQFT re-introduces by means of regions of space-time and associated algebras of observables a meaningful notion of system $(R, A(R))$.\footnote{Compact closed categories play a key role within AQFT \cite{DoplicherRoberts1, DoplicherRoberts2, Halvorson}, but their  role in AQFT is conceptually totally different from this role in our framework.  The natural manner to recast AQFT as a monoidal category, somewhat more in the spirit of the  developments of this paper, would be to replace the C*-algebra ${\cal A}$ by a monoidal category with the sub-C*-algebras of ${\cal A}$ as the objects, and completely positive maps as morphisms, subject to some technical issues to do with the non-uniqueness of the tensor product of C*-algebras.  
}

In \cite{CBH} C*-algebras also provided an arena for Clifton, Bub and Halvorson's to address  Fuchs' and Brassard's 
challenge  to reconstruct the quantum mechanical formalism in terms of information-theoretic constraints \cite{Fuchs1, Brassard, Fuchs2}.
Meanwhile it has been recognized by at least one of the authors that most of the work in this argument is done by the C*-algebra structure rather than by axioms  \cite{HavorsonP}, hence a more abstract mathematical arena is required.


 
 
\subsubsection{The logic of interacting processes}

What does it mean to have two or more systems?  I.e.~what is ``$A$ \emph{and}  $B$'':
\bit
\item[1.] I have a choice between $A$ \emph{and}  $B$.
\item[2.] I have both $A$ \emph{and}  $B$.
\item[3.] I have an unlimited availability of both $A$ \emph{and}  $B$.
\eit  
Developments in logic have started to take account of these sorts of issues.  In particular, Girard's \em linear logic \em \cite{Girard, Troelstra, AbrLin, LNPAT} (which originated in the late 1980s) makes the difference between either having the availability of one out of two alternatives, or having both alternatives available.\footnote{Since its birth, linear logic did not only radically change the area of logic, but has immediately  played a very important role in computer science, and still does \cite{LinCS}.  The first occurrence of linear logic in the scientific literature was in Lambek's  mathematical model for the grammar of natural languages \cite{Lambek} in the 1950s.}
The first of the two conjunctions in linear logic, the \em non-linear \em conjunction, is denoted by $\&$, while the second one, the \em linear \em conjunction, is denoted by $\otimes$.  The difference is:
\[
A \vdash A \& A   \qquad  A\& B\vdash A \qquad \mbox{while} 
\qquad A \not\vdash A \otimes A   \qquad  A\otimes B \not\vdash A\,.
\]
 That is, in words, from the fact that $A$ (resp.~$A\& B$) holds we can derive that also $A \& A$ (resp.~$A$) holds, but from the fact that $A$ (resp.~$A\otimes B$) holds we cannot derive that also $A \otimes A$ (resp.~$A$) holds.  Hence, the linear conjunction treats its arguments as genuine \em resources\em, that is,  they cannot freely be copied nor discarded. It is a \em resource sensitive \em connective.

From a naive \em truth-based \em view on logic,  where $A$ merely  stands for the fact that this particular proposition holds, the failure of the last two entailments might look weird.  However, a more modern view on logic is obtained  in terms of \em proof theory\em.  In this perspective $A$ stands for the fact that one possesses a proof of $A$, $A\otimes B$ stands for the fact that one possesses a proof of $A$ and a proof of $B$, and $A\otimes A$ stands for the fact that one possesses two proofs of $A$.

In proof theory propositions mainly play a supporting role.
What is of particular interest in proof theory is the \em dynamics of proofs\em: how to turn a long proof into a short one, how to eliminate lemmas etc.   In other words, the \em derivation process \em (i.e.~proof) is the key player, and it is all about how proofs \em compose \em to make up another proof. 
The mathematical arena where all of this takes place is that of \em closed symmetric monoidal categories \em e.g.~\cite{Seely}.

  One  indeed can take the view that `states' stand to `systems' in physics as  `proofs' stand to `propositions' in logic.  `Physical processes' which turn a system into another in physics then correspond to `derivation processes' of one proposition into another in logic.   In this view, systems mainly serve as things along which physical processes can be composed,  a view that we shall adopt here. 

 \subsubsection{Processes as relations} 
 
 Once one considers processes and their interactions as  more fundamental than systems themselves one enters the realm of \em relationalism\em. 
 
 One well-known recent example of relationalism is Barbour and Bertotti's take on relativity theory in terms of Mach's principle \cite{Barbour}, which states that inertia of a material system is only meaningful in relation to its interaction  with other material systems \cite{Mach}.  Rovelli's relational interpretation of quantum theory \cite{Rovelli} considers all systems as equivalent, hence not subscribing to a classical-quantum divide, and all information carried by systems as relative to other systems.   Here we  will also adopt this relational view on physics.
 
  One thing that  relationalism provides is an alternative to the dominant ``matter in space-time''-view on physical reality, by taking space-time to be a secondary construct.  What it also does, is that it relaxes the constrains imposed by no-go theorems on accounts of the measurement problem \cite{GNoGo, JPNoGo, KSNoGo}.\footnote{It is a common misconception that the Kochen-Specker theorem \cite[(1967)]{KSNoGo} would be in any way the first result of its kind.   It is in fact a straightforward corollary of Gleason's theorem \cite[(1957)]{GNoGo}, and a crisp direct no-go theorem was already provided by Jauch and Piron \cite[(1963)]{JPNoGo}.  A discussion of this is in Belinfante's book \cite{BelinfanteNoGo}.}  For example, if a systems' character is defined by its relation to other systems, \em contextuality\em , rather than being something weird, becomes not just perfectly normal, but a \em fundamental requirement for a theory not to be trivial\em.\footnote{Obviously this paragraph may be for many the most controversial, challenging or interesting one in this paper. They would have probably liked to see more on it. We expect to do this in future writings once we have obtained some more formal support for our claims.}
 
 The main problem with relationalism seems  to be that, while it is intuitively appealing, there is no clear formal conception.   This is where category theory \cite{EilenbergMacLane} provides a natural arena, in that it abstracts over the internal structure of \em objects \em (cf.~the properties of a single physical system), and instead considers the structure of \em morphisms \em between systems (cf.~how systems relate to each other).   Monoidal categories \cite{Benabou, MacLane0} moreover come with an intrinsic  notion of compound system. In their diagrammatic incarnation, these categories translate `being related' into the topological notion of `connectedness'.  The `non-free' part of the structure then provides the modes in which things can be related.   It seems to us that the \em dagger compact \em symmetric monoidal  structure \cite{AC2, Selinger} in particular provides a formal counterpart to the relational intuition.  A more detailed and formal discussion of this issue is in  Section \ref{sec:Axmeetmodel}.





%


\subsubsection{Mathematical rigor}

One of the favorite activities of operationalists is to reconstruct quantum theory by imposing reasonable axioms on families of experimental situations. Some recent examples of such reconstructions are \cite{Hardyaxioms, DArianoaxioms, Rau}. 

This tradition was initiated by Mackey \cite{Mackey1} around 1957, with Piron's 1964 theorem as the first success \cite{PironThm}. The different attempts vary substantially in terms of their mathematical guise, in that some reconstructions start from the very foundations of mathematics, e.g.~\cite{PironThm, Piron,Soler}, while others will take things like the real continuum as God-given in order to state the axioms in a very simple language, e.g.~\cite{Hardyaxioms}.   Quoting Lucien Hardy on this \cite{Hardyaxioms}:
\begin{quote}
``Various authors have set up axiomatic formulations of quantum theory, [...] The advantage of the present work is that there are a small number of simple axioms, [...] and the mathematical methods required to obtain quantum theory from these axioms are very straightforward (essentially just linear algebra).'' 
\end{quote}
Quoting Tom Yorke,  singer of the Oxford based band Radiohead \cite{Radiohead}:
\begin{quote}
``Karma Police, arrest this man. He talks in Maths.'' 
\end{quote}

It is an undeniable fact that mathematical rigor is one of the key cornerstones of science.  But on the other hand, very important science has been developed long before there existed anything like a foundation of mathematics.
Even in recent history scientific progress was only possible by not subscribing to mathematical rigor, of which the problem of renormalization in quantum field theory is the most prominent witness, even leading to a Nobel Prize.

Ultimately this boils down to the respect one gives to mathematics.  Roughly put, is mathematics an a priori given thing which we can use to formulate our theories of physics, or, is it something secondary that intends to organize our experiences, be it when reasoning, exploring nature, or whatever, and that should be adjusted to cope with our evolving spectrum of experiences?   Simpler put, do we serve mathematics or does mathematics serves us?

Our approach will be to assume a physical reality, with the things `out there' truly happening.  We will consider certain physical primitives, namely relations and composition thereof.  These primitives come with a notion of `sameness' which will play the role of equality, i.e.~it will tell us when compositions of relations are equal.\footnote{Let us mention that currently,  even within the foundations of mathematics we don't really know what the sign `$=$' stands for. In universal algebra it is a binary predicate, but once one goes beyond classical logic this breaks down. In first order logic equality is a distinguished binary relation. In higher-order logic it is given by Leibniz identity which identifies things with the same properties \cite{LambekScott}.  In Martin-L\"of type theory  \cite{MartinLof} and Bishop-style constructive mathematics \cite{Bishop} one uses yet again other notions of equality.  In categorical logic \cite{Jacobs} several options are still being explored. Credits for this concise summary go to Phil J.~Scott.}
As a second step, we will try to match these physical primitives with a mathematical structure, namely particular kinds of categories. This,  despite the great flexibility of category theory,  will come at a certain cost. 

In our view, Hilbert's proposal to \em axiomatize physics\em\,\footnote{Cf.~Hilbert's 6th problem  \cite{Hilbert}.} is a very different ball game than \em axiomatizing mathematics\em,\footnote{Cf.~Hilbert's 2nd problem on the consistency of axiomatic arithmetic \cite{Hilbert}. G\"odel later showed  that this issue cannot be settled within arithmetic itself \cite{Godel}.}
something which also proved to be a far more delicate business than one imagined at first.

Our goal is also quite different from the reconstructionists.  Rather than reproducing quantum theory with a set of reasonable axioms, our goal is rather to reproduce as much as possible physical phenomena with as little as possible `structural effort' or `axiomatic compromise', hence providing a very flexible setting that may be better adjusted to the theories of the future.


\subsubsection{The continuous or the discrete?}

In the light of future theories of quantum gravity it has been argued that we may have to abandon our reliance on the continuum, be it either with respect to the structure of the space of states, spectra of observables, space-time, or even probability valuation. Quoting Isham and Butterfield \cite{IB00}:
\begin{quote}
``... the succes of [the edifice of
physics] only shows the `instrumentalist utility' of the continuum --- and not that physical quantities
have real-number values ... there is no good {\it a priori}\, reason why space should be a continuum;
similarly, {\it mutatis mutandis} for time.'' 
\end{quote}
\begin{quote}
``...  limiting relative frequency interpretation seems problematic 
in the quantum gravity regime ... for the other
main interpretations of probability --- subjective, logical or propensity --- there seems
to us to be no compelling {\it a priori}\, reason why it should be real numbers.''
\end{quote}
Once one abandons the continuum as a mathematical default we need a paradigm and/or mechanism to either reproduce it or replace it by.   

One option are `spaces without points',  which both have a  topological and geometric incarnation, respectively called locales and  frames  \cite{Johnstone1}.\footnote{Locales and  frames are a beautiful example of how the nature of a mathematical structure can change merely by changing the nature of its relation to other structures of the same kind,  rather than by changing the structure itself: in category theoretic terms, locales and frames are exactly the same objects, but live in a different category; one obtains the category of locales simply by reversing the direction of the arrows in the category of frames.}  These spaces have been used both to model spectra as well as truth-values in the so-called topos approach to physical theories, which rose to prominence some ten years ago with the still ongoing work of Isham, collaborators and followers.\footnote{The first work in this area seems to be by Adelman and Corbett in the 1993's \cite{CorbettI, CorbettII}.} Both the locales/frames as well as topos theory also provide a mathematical foundation for  intuitionistic logic \cite{Vickers, Johnstone2}. In all of their guises they have been particularly popular among computer scientists.   

Also popular among computer scientists are discrete combinatorial spaces. In fact,  computer scientists proposed various discrete space-time structures \cite{Lamport, Petri} well before physicists did so (e.g.~Sorkin et al.~\cite{Sorkin}). 

Our setting is flexible enough to accommodate both perspectives.  For example, a topos gives rise to a so-called  alegory of generalized relations \cite{Aligator}, and similar, categories arise when organizing combinatorial species \cite{JoyalSpecies1, JoyalSpecies2, FioreSpecies}.  

In fact, even at a much more basic level categories 
abstract over concrete well-pointed spaces, by abstracting over the actual structure of objects.  They obviously 
also immediately provide a rich variety of combinatorial structures, in that they themselves always form a graph.

\section{Systems $\leftarrow$ relations $\leftarrow$  composition}\label{sec:systrelations}

We mentioned that operational approaches appeal to our everyday experiences, or some easy to grasp generalization thereof. 
Also here 
we will make some reference to our perceptions, but at a much more abstract level than in all of the above mentioned examples.   Not measurement devices, nor probabilities, nor propositions, nor classical mechanics concepts such as 3D space, concrete observables such as position, nor the real continuum will play any role. 

We assume as primitive a  flexible notion of \em system\em, a very general notion of \em relation \em between these, and two modes of \em composition \em of the latter, one which typically imposes  dependencies between the processes that one composes and one which excludes dependencies.  In graphical terms these will correspond with the primal topological distinction between `connected' and `disconnected', cf.:
\[
connected\ \sim\ \raisebox{-0.91cm}{\epsfig{figure=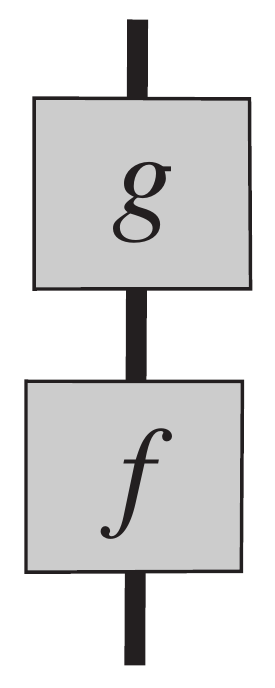,width=22pt}}
\qquad\qquad
disconnected\ \sim\ \raisebox{-0.50cm}{\epsfig{figure=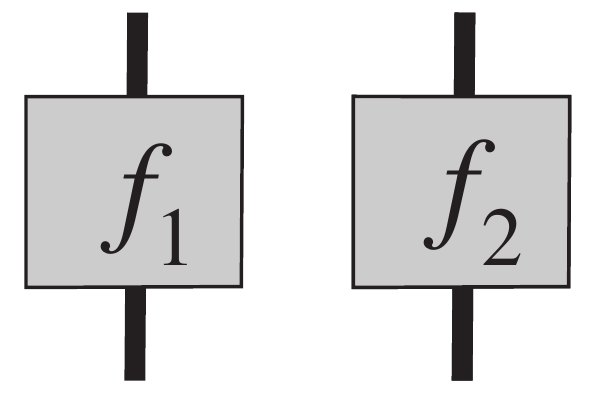,width=50pt}}
\]
Within our approach, which models how relations compose to make up other relations,  systems play the role of the `plugs' by means of which we can create dependencies between relations in one of the two modes of composition. 

So while it is in `bottom-up'  order in which we introduce the basic concepts:
\begin{center}
systems $\rightarrow$ relations $\rightarrow$  composition
\end{center}
in order to appeal to the reader's intuition, the most important concept is  composition. Relations are then those things that we can compose, and systems the things along which we can compose these relations in a dependent manner.   

This top-down view may seem to go in the opposite direction of a physicists'  reductionist intuition.  Nonetheless  it is something the physicist is well acquainted with.  For symmetry groups, it is not the elements of the group that are essential, but the way in which they multiply ($\sim$ compose), since the same set of elements may in fact carry many different group structures.    In a similar manner that group structure conveys the shape of a space, the composition structure on relations will convey the `shape' of the `universe of processes'.

In support of the reader's intuition we refer to `properties' of a system when discussing concepts, but this has no defining status whatsoever.  For this discussion, we will inherit the `actual' versus `potential' terminology from the Geneva School, 
 the first saying something about  the state of the system, while the second says something about  the system itself.

\subsection{Systems}

So the prime purpose of a notion of \em system \em is to support the notion of  a relation, 
systems being those things along which relations can be composed.  

More intuitively, by a system we mean something identifiable about which we can pose questions, and hence about which it makes sense to speak about `properties'.  It is the latter which are usually stated relative to our world of experiences. This however does not mean that a system is completely determined by that what we consider to be its actual properties, nor that necessarily there exists an experiment by means of which we can verify these.\footnote{There are many things we can speak about without being able to set up an experiment, for example, simply because the technology is not (yet) available.  One could consider speaking in terms of hypothetical or idealized experiments, but we don't know the technologies of the future yet.  These will be based on theories of the future, and since crafting these theories of the future is exactly the purpose of this framework, guessing would lead to a circularity.} An example of a system that is not completely  determined by its actual  properties is one which is part of a larger system, that is, when considering  `parts of a larger whole'.

We denote systems by $A$, $B$, $C$, ...  

\paragraph{Example: quantum systems.}  Quantum systems are the entities that we describe in Hilbert space  quantum theory e.g.~position, momentum or spin.  Here, systems that are not completely determined by their actual properties are those described by density operators, which  arise due to a lack of knowledge as well as by tracing out part of a compound system.  The need for a concept of system which is not characterized in terms of its actual properties becomes even more important in the case of quantum field theory, where we want to be able to consider what is relevant about the field for a certain region of space-time.


\paragraph{Example: AQFT and beyond.} In AQFT the systems are the C*-algebras associated to a region of space-time \cite{Haag, Halvorson}.  So a system is a pair $(R, A(R))$ where $R$ is a region of space-time and $A(R)$ represents the observables attributed to that region.  This idea of a pair consisting of a space-time region and another mathematical object 
which encodes observables can be generalized to other manners of encoding observables, for example, in terms of \em observable structures \em i.e.~special commutative dagger Frobenius algebras  (see below) on an object in a dagger symmetric monoidal category (see below) as is done in categorical quantum mechanics \cite{CPav, CPV, CPaqPav}. These two perspectives are not that far apart, given that Vicary has shown in \cite{VicaryCstar} that finite dimensional C*-algebras are precisely the non-commutative generalizations of observables structures in the dagger symmetric monoidal category ${\bf   FHilb}$ (see below).
\bigskip


By $\II$ we denote the system which represents  everything that we do not explicitly consider within our theory.  One may refer to this as the \em environment\em, i.e.~what is not part of our domain of consideration. Intuitively put, it is the system which represents everything to which we do not attribute any properties whatsoever. 
Formally it will play the role of the `trivial system' in that composing it with any other system $A$ will yield that system $A$ itself. Obviously, what will be considered as $\II$ may in part be a cognitive decision, or a technological constraint, or maybe a fundamental physical principle.\footnote{For example, the disciples of the so-called `Church of the larger Hilbert space' seem to believe that system $\II$ could always be eliminated from any situation in quantum theory.}  
Here these interpretational issues won't  matter.  What does matter is that there is a  \em domain of consideration\em, and that everything else  falls under the umbrella of $\II$.

\paragraph{Example: open systems.}   In  open systems theory (e.g.~\cite{Davies,Krausbook}) $\II$ stands for the environment. In quantum theory it is $\II$ which is responsible for decoherence.  Sections \ref{sec:environment} and \ref{sec:class} elaborate on this issue in great detail.

\bigskip
 
Our account on systems as `a bag of things' may sound naive and it indeed is.  A more realistic account which involves the notion of subsystem is discussed in Section \ref{sec:subsystems}.  This will require that we first introduce some other concepts.

 We denote ``system $A$ \emph{and} system $B$'' by $A\otimes B$.    The precise meaning of $A\otimes B$ will become clear below from what we mean by composition of processes.  In particular, we will see that $A$ and $B$ in $A\otimes B$ will always be independent and hence distinct i.e.~we cannot conjoin a systems with itself. 
 
 The notation $A_1\otimes A_2$ (wrongly) indicates that $A_1$ and $A_2$ are ordered.  This is an unavoidable artifact of the 1 dimensional linear notation which is employed in most natural languages as well as in the majority of mathematical notation.  Hence $A_1\otimes A_2$ is to be conceived as `a set of two systems' rather than  as `an ordered pair of two systems'.  
 
 \paragraph{Example: AQFT and beyond.} AQFT considers an inclusion order on diamond-shaped regions which carries over on inclusion for C*-algebras.  Intuitively, the joint system would consist of the union of the two regions and the corresponding union of C*-algebras, at least in the case that the regions are space-like separated.  But two regions do not make up a diamond anymore, so this naive notion of system $A$ and system $B$ would already take one beyond the AQFT framework.  A paper on this subject is in preparation with Samson Abramsky, Rick Blute, Marc Comeau, Timothy Porter and Jamie Vicary \cite{ABC}.  



\subsection{Processes and their composition}

\em Processes \em are relations that carry the `genuine physical substance' of a theory. It are  those entities we think of as  actually `happening' or `taking place' (as opposed to the \em symmetry relations \em discussed below in Sec.~\ref{sec:logicalrelations}). They arise by `orienting' a relation, that is, by assigning input/output-roles to the systems it relates, i.e.~it is a relation that `happens' within a by us perceived \em causal structure \em -- cf.~a partial ordering, or more generally, a directed graph.  

Intuitively, a process embodies how properties of system $A$ are transformed into those of system $B$.   The environment may play an important role in this. 

The \em type \em of a process is the specification of the input system $A$ and the output system $B$, and is denoted as $A\to B$. We call $A$ the \em input \em and $B$ the \em output \em of the process.  
Processes themselves are denoted as $f:A\to B$.

\paragraph{Example: operations.} Processes can be the result of performing an operation on a system $A$   in order to produce system $B$, e.g.~measuring, imposing evolution, or any other kind of experimental setup. Our whole framework could be given a more radical operational connotation, by restricting to processes arising from operations. It would then match Hardy's recent proposal \cite{Hardy3}. 

\paragraph{Example: quantum processes.} These include state preparations, evolutions, demolition and non-demolition measurement processes etc.  

\paragraph{Example: de-instrumentalizing Geneva School Mark II.} One can modify the Geneva School Mark II approach by replacing the experimental projects with any process $f$ that may cause a particular other process $f_{yes}$ to happen thereafter.  Roughly put: a property of a system would then be an equivalence class of those processes which cause $f_{yes}$ to happen with certainty.  
\par\bigskip

The trivial process from system $A$ to itself is denoted by $1_A: A\to A$. It `happens' in the sense that it asserts the existence of system $A$, and it trivially obeys causal order. These trivial processes  are useful in that they provide a bridge between systems and processes, by associating to each system a process. 

For all other non-trivial processes the input and the output  are taken to be non-equal i.e.~if the type of a process is $A\to A$ then it is (equal to) $1_A$. 


By a \em state \em we mean a process of type $\II \to A$, and by an \em effect \em we mean a process of type $A\to\II$.  What is important for a state is indeed what it is, and not its origin, which can consequently be comprehended within $\II$. 

By a \em weight \em we mean a process of type $\II\to \II$.   

\paragraph{\em Sequential composition.} The \em sequential \em or \em causal \em or \em dependent composition \em of processes  $f:A\to B$ and  $g:B\to C$ is the process which relates input system $A$  to output system $C$. 
We denote it by:
\[
g\circ f:A\to C\,.
\]
We will also refer to $g\circ f$ as ``$g$ \em after \em $f$'' or as ``\em first \em $f$ \em and then \em $g$''.

\paragraph{Example: operations.} For processes resulting from operations, $g\circ f$ is the result of \em first \em performing  operation $f$ \em and then \em performing  operation $g$.  The operations corresponding to states are preparation procedures. 

\paragraph{Example: weights as probabilities.} 
When we compose a state $\psi:\II\to A$ and an effect $\pi:A\to\II$ then the resulting weight $\pi\circ\psi:\II\to\II$ can be interpreted as the probability of the sequence ``$\pi$ after $\psi$'' to happen.  That a projective measurement effect  in quantum theory may be impossible for certain states and certain for others boils down to   $\langle\phi| \circ |\psi\rangle=0$ while $\langle\phi| \circ |\phi\rangle\not=0$ for $|\psi\rangle\perp|\phi\rangle$. More generally, these weights can articulate likeliness of processes. 

 \paragraph{\em Separate composition.} The \em separate \em or \em acausal \em or \em independent composition \em of processes
 $f_1:A_1\to B_1$ and $f_2:A_2\to B_2$ is the process which relates input system  $A_1\otimes A_2$  to output system $ B_1\otimes B_2$. We denote it by:
 \[
 f_1\otimes f_2: A_1\otimes A_2\to B_1\otimes B_2\,.
 \]
 
The key distinction between sequential and separate composition in terms of `dependencies' between processes is imposed by the following constraint. 

\paragraph{\em Independence constraint on separate composition.}  \em A process is independent from any process to which it is not `connected via sequential composition', and the same holds for the systems that make up the types of these processes, with the exception of the environment $\II$. 
In particular, within the compound process $f\otimes g$ the processes $f$ and $g$ are independent. \em

\bigskip
We will precisely define what we mean by  `connected via sequential composition' in Section \ref{sec:graphrepresentation}, by relying on the topological notion of `connectedness'.
The spirit of this constraint is that causal connections can only be established via dependent composition, not by means of separate composition.  

 \paragraph{Example: operations.} 
When considering processes resulting from operations,  $f_1$ and $f_2$ in $ f_1\otimes f_2$ are realized by two independent operations.  This means that the setup in which one realizes one operation should be sufficiently isolated from the one that realizes the other operation.  
\par \bigskip
 
The independence of $f_1$ and $f_2$ in $f_1\otimes f_2$  imposes  independence of $A_1$ and $A_2$ in $A_1\otimes A_2$ and  of $B_1$ and $B_2$ in $B_1\otimes B_2$,
which indeed forces  $A_1$ and $A_2$ in  $A_1\otimes A_2$ and $B_1$ and $B_2$ in  $B_1\otimes B_2$ to always be distinct.

 \paragraph{Example: AQFT.} For intersecting regions $R$ and $R'$ the systems $(R, A(R))$ and $(R', A(R'))$ are obviously not independent.  Neither are they for regions $R$ and $R'$ that are causally related.
 \par \bigskip
 
 While in $ f_1\otimes f_2$ the two processes have to be independent, this does not exclude that via causal composition with other processes dependencies can emerge:

\paragraph{Example: quantum entanglement.} Although two quantum processes $f_1:A_1\to B_1$ and $f_2:A_2\to B_2$ are independent in $f_1\otimes f_2$, it may of course be the case that due to common causes in the past measurements on their respective output systems $B_1$ and $B_2$ may expose correlations.
In that case we are in fact considering $(f_1\otimes f_2)\circ |\Psi\rangle$ where $ |\Psi\rangle:\II\to B_1\otimes B_2$.  In this case, as we shall see below, $B_1$ and $B_2$ \em are \em connected via sequential composition.
 \par \bigskip

 Note that the exception of $\II$ in the independence constraint allows for:
\[
A\otimes \II =A\,,
\]
which affirms that  the environment may always play a certain role in a process.

On a more philosophical note, within our setting the independence constraint replaces the usual conception of \em sufficient isolation \em within the \em scientific method \em \cite{Poincare, Moore}: we do not assume that  systems or processes are sufficiently isolated, but our formal vehicle which represents when we compose them 
 implicitly requires that they are independent, the environment excluded.
 


Now consider the four processes: 
\beq\label{post:intere}
f_1:A_1\to B_1\quad f_2:A_2\to B_2\quad g_1:B_1\to C_1\quad g_2:B_2\to C_2
\eeq
 Note that causal composition of the processes $f_1\otimes f_2$ and $g_1\otimes g_2$ resulting from separate composition,  which implies  matching intermediate types, is  well-defined since $B_1$ and $B_2$ will  always be taken to be distinct in  \mbox{$B_1\otimes B_2$}.  But there is another manner in which we can compose these processes to make up a whole  of type $A_1\otimes A_2\to C_1\otimes  C_2$, namely by    separate   composition of  $g_1\circ f_1$ and  $g_2\circ f_2$. While symbolically these two compounds are represented differently, \em physically \em they represent \em the same \em overall relation, and hence:

\paragraph{\em Interaction rule for compositions.}  For  
processes (\ref{post:intere}) we have:
\[
( g_1\circ f_1)\otimes( g_2\circ f_2)=( g_1\otimes g_2)\circ( f_1\otimes f_2)\,.
\]
Similarly, for  systems $A_1$ and $A_2$ we also have:
\[
1_{A_1}\otimes 1_{A_2}=1_{A_1\otimes A_2}\,.
\]



\subsubsection{Non-isolated systems and probabilistic weights of states}

The natural way to assert inclusion of \em non-isolated \em (or \em open\em) systems within a theory of processes  is in terms of  a particular kind of process:
\[
\top_A:A\to \II
\]
that `feeds' a system into the environment $\II$, and hence explicitly realizes such a non-isolated system. Feeding a system $A$ into the environment can be achieved by taking a process $f:A\to B$ and by then `deciding' to consider $B$ as part of the environment $\II$.  So as was the case for $\II$, these feeding-into-the-environment processes may involve a cognitive component. 

What characterizes such a  \em feed-into-environment process\em?  
Firstly, it is easily seen that we can always set:
\[
\top_{A\otimes B}:=\top_A\otimes\top_B\qquad\qquad \mbox{and}\qquad\qquad\top_\II:=1_\II\,.
\]
Secondly, 
it should be allowed to happen with certainty, independently on the state of the system, as opposed to, for example, the \em projective measurement effects \em  in quantum theory discussed above. 
Denoting weights by $\mathbb{W}$, 
given a  measure that assigns weights to each process, and in particular to states $\mathbb{S}$:
\[
|-|:\mathbb{S}\to\mathbb{W}\,,
\]
a feed-into-environment process $\top_A$ should be such that applying it leaves the weight of the state it is applied to invariant, i.e.~concretely:
\beq\label{eq:post:non-isol}
\top_A\circ \psi=|\top_A\circ \psi|= |\psi|
\eeq
for all states $\psi:\II\to A$, where the first equality merely says the weight of a weight is itself. 
But (\ref{eq:post:non-isol}) can now also be dually interpreted: 
feed-into-environment processes are characterized in that they provide the measure for assigning weights to states, by post-composing states with them.  

\paragraph{Example: traces and probabilities in quantum theory.}
In  quantum theory,   the completely positive maps that  trace out spaces play the role of  the  feed-into-environment processes:
\[
tr_{\cal H}::\rho_{\cal H}\mapsto\sum_i\langle i|\rho_{\cal H}|i\rangle\,.
\]
These indeed stand for ignoring part of a system as well as for measuring  the overall probability of a non-normalized density matrix.  Note in particular that these are the only completely positive maps which satisfy (\ref{eq:post:non-isol})  for every possible state.  In the language of \cite{DArianoPaper}, $tr_{\cal H}$ is the \em unique deterministic effect \em on ${\cal H}$.

\par \bigskip


\subsection{Graphical representation of processes}\label{sec:graphrepresentation}

The data specified above can be given a diagrammatic representation:
\[
f\equiv\ \raisebox{-0.91cm}{\epsfig{figure=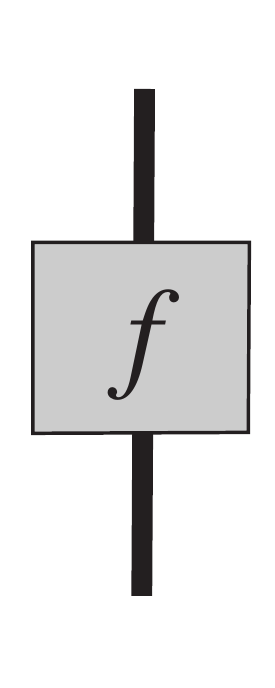,width=22pt}}
\qquad\quad
g\circ f\equiv\ \raisebox{-0.91cm}{\epsfig{figure=Ober1.pdf,width=22pt}}
\qquad\quad
f_1\otimes f_2\equiv\ \raisebox{-0.50cm}{\epsfig{figure=Ober2.pdf,width=50pt}}
\]
That is:
\bit
\item a process is represented by a box with inputs and outputs\,;
\item $-\circ-$ is represented by connecting outputs to inputs\,;
\item $-\otimes-$ is represented by not connecting boxes.
\eit
The object $\II$ will be represented by `no wire', and:
\[
\psi \equiv\ \raisebox{-0.41cm}{\epsfig{figure=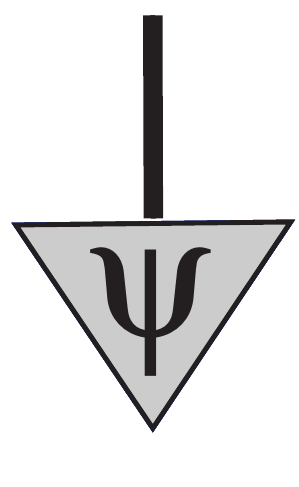,width=18pt}}\ : \II\to A
\qquad\qquad
\pi \equiv\ \raisebox{-0.41cm}{\epsfig{figure=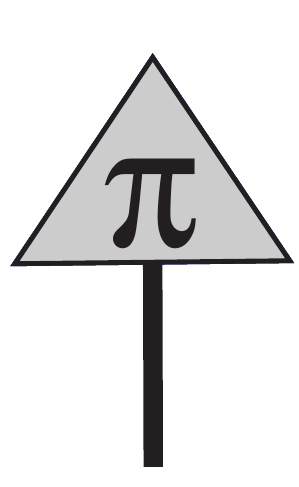,width=18pt}}\ : A\to \II
\qquad\qquad
\omega \equiv\ \raisebox{-0.41cm}{\epsfig{figure=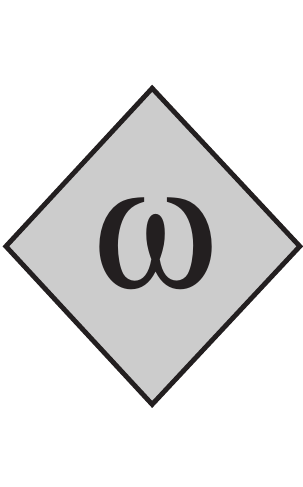,width=18pt}}\ : \II\to\II
\]

What is particularly nice in this graphical representation is that the interaction rule 
automatically holds, since translating both its left-hand-side and its 
right-hand-side into the graphical calculus both result in the same: 
\[
( g_1\circ f_1)\otimes( g_2\circ f_2)
=( g_1\otimes g_2)\circ( f_1\otimes f_2)
\quad\Leftrightarrow \quad
\raisebox{-0.91cm}{\epsfig{figure=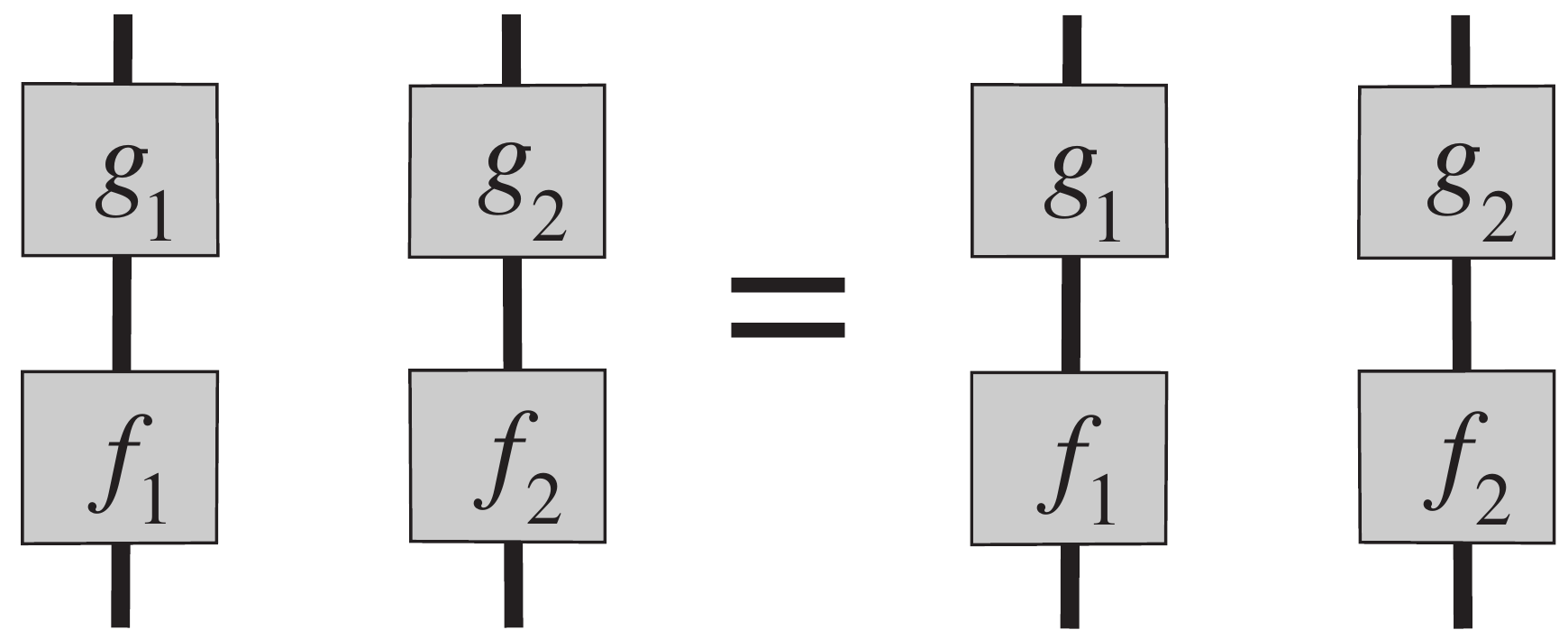,width=140pt}}
\]
This shows that also the interaction rule is in fact nothing more than an artifact of one-dimensional symbolic notation!

Here is the definition we promised earlier:

\paragraph{Definition.} Two processes are \em connected via sequential composition \em if in the graphical representation they are topologically connected.  
\bigskip

There also  is a direct translation of this graphical representation of processes to directed graphs and vice versa.  The rules to do this are:
\bit
\item processes (i.e.~boxes, triangles, diamonds etc.) become the nodes\,; 
\item systems (i.e.~wires between boxes) become directed edges, with the direction pointing from what used to be an output to an input.
\eit
Such a directed graph makes the underlying causal structure on processes explicit.
Here's an example of this:
\[
\epsfig{figure=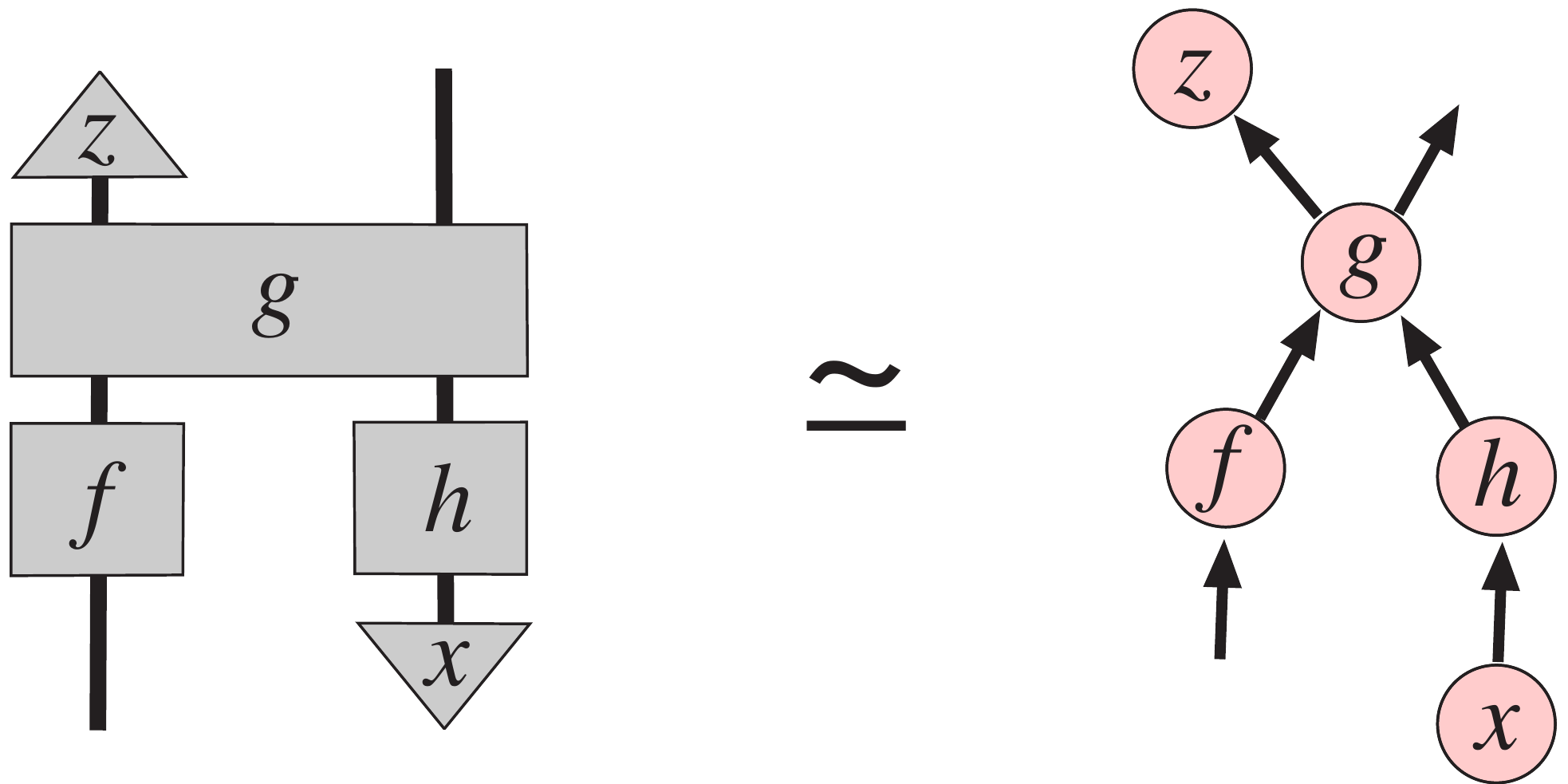,width=160pt} 
\]
In this example all processes are connected via sequential composition as the picture is as a whole connected. 

Special processes can be give special notations, for example, a   feed-into-environment process  $\top_C:C\to\II$ could be denoted as:
\[
\top_C\equiv\ \raisebox{-3pt}{\epsfig{figure=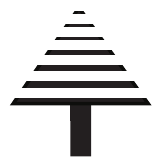,width=15pt}} \qquad \mbox{so} \qquad ( 1_B \otimes \top_C)\circ f \equiv \raisebox{-16pt}{\epsfig{figure=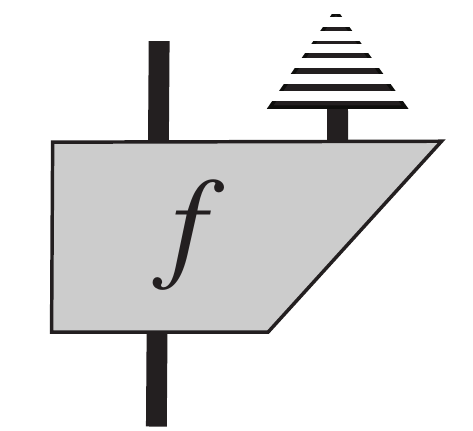,width=45pt}}
\]
for $f: A\to B\otimes C$.

In two-dimensional  graphical language, as it is the case for symbolic notation, systems  appear in a certain order (cf.~from-left-to-right) which has no direct ontological counterpart.  However, in the graphical notation this order can be exploited
to identify distinct systems in terms of their position within the order, hence in part omitting the necessity to label the wires.  More on this will follow below.
One can of course also think of these pictures as living in 3 dimensions rather than in 2 dimensions, or some even more abstract variation thereof. One calls graphs which exploit a third dimension \em non-planar \em \cite{SelingerSurvey}. Planar graphs are subject to  Kuratowski's characterization theorem \cite{Kuratowski}.

\subsection{Physical scenarios,  snapshots and subsystems} \label{sec:subsystems}

A \em physical scenario \em is a collection of processes together with the   \em composition structure \em in which they happen.  By the \em resolution \em of a scenario we mean the resulting overall process. A scenario comprises more information than its resolution, in that it also comprehends the manner in which the overall process is decomposed in subprocesses. Obviously, the selection of a particular scenario which has a given process as its resolution has no actual physical content, but is merely a subjective choice of what to consider as the parts of a whole.  

Given such a physical scenario, one can consider a 
subset of the systems appearing within it, neither of which are causally related, e.g.~those connected by the hand-drawn line in the following picture:  
\[
\epsfig{figure=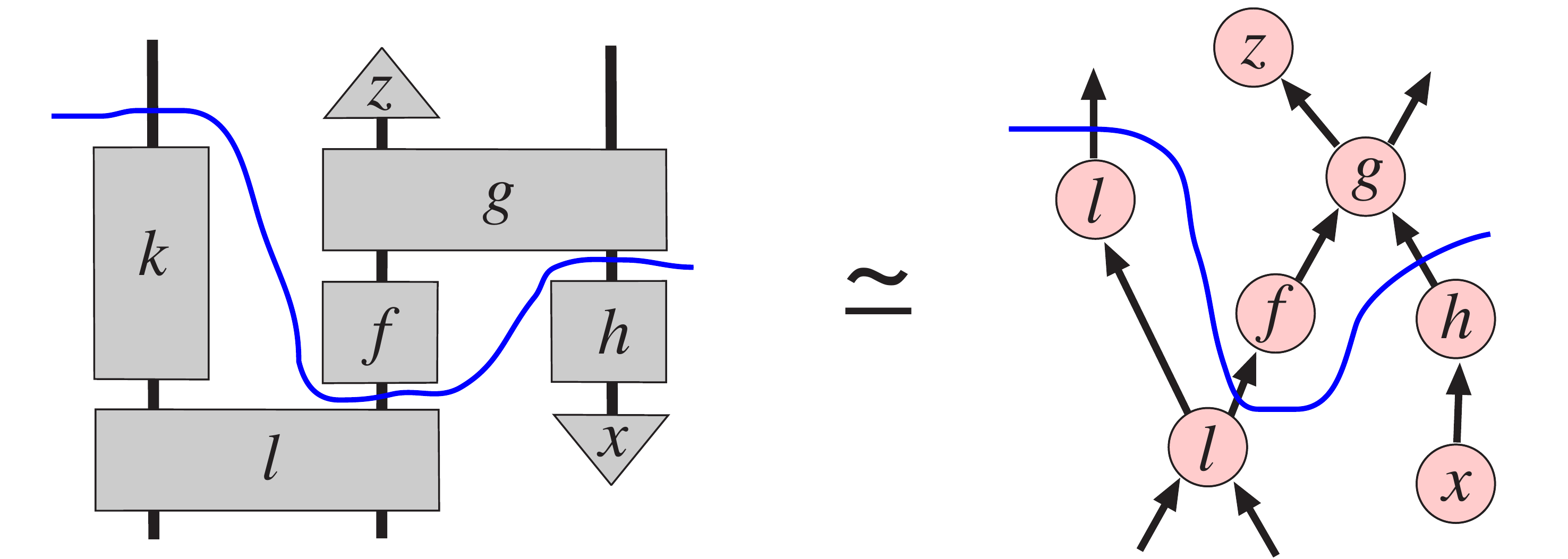,width=240pt}  
\]
We call such a collection of systems appearing in a scenario a \em snapshot\em.  

Note that it is not excluded that snapshots resulting from distinct scenarios are the same, for example, the hand-drawn line:
\[
\epsfig{figure=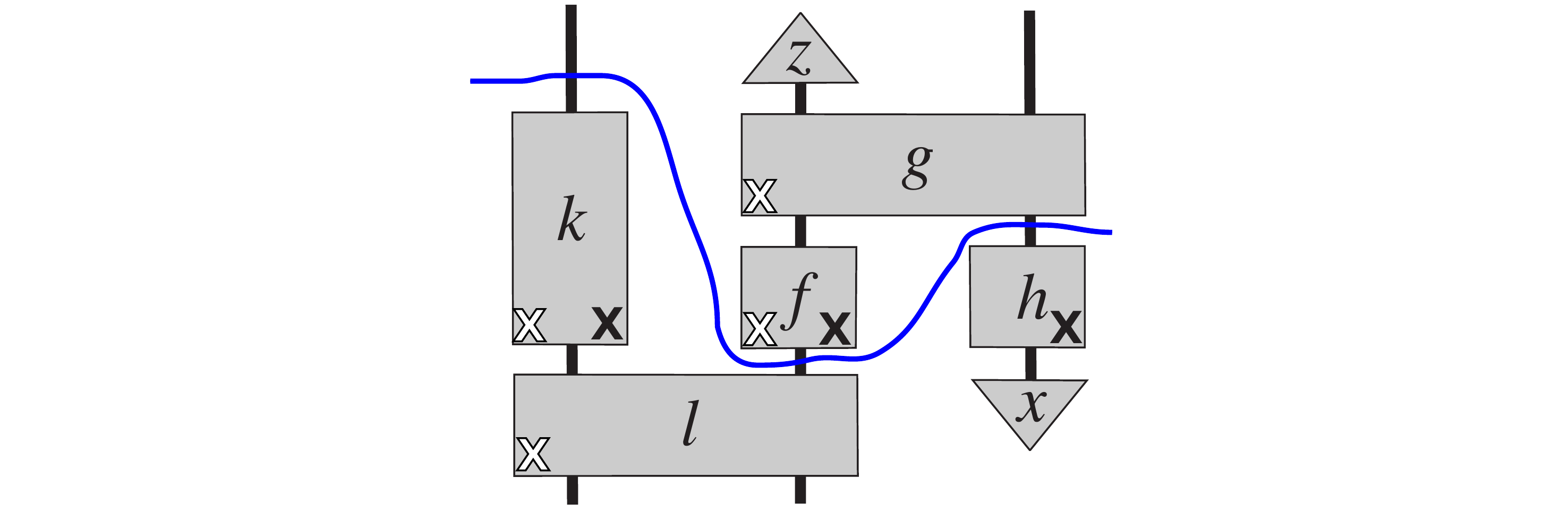,width=240pt}  
\]
represents a snapshot both for the restriction of the boxes to those with a white cross as well as for the restriction  to those with a black cross.

\paragraph{Example: relativistic causal histories.} These snapshots are Hardy's `systems' within his instrumental framework \cite{Hardy3, HardyTempleton}.  In turns, these generalize Blute, Panangaden and Ivanov's `locative slices' within their framework which endows standard quantum mechanical operations with a causal ordering \cite{BIP, Prakash}. A dual point of view was earlier put forward by Markopoulou  \cite{Markopoulou}.
\bigskip

These snapshots are indeed systems as much as any other system.  But as mentioned at the very beginning of this chapter, they are not anymore the primal physical concept, but things along we `decide' to decompose processes.  It is the resolution of the snapshot which is physically the only primal concept. 

Note that there exists  a partial  order on systems in terms of inclusion of snapshots.  For example, 
\[
\epsfig{figure=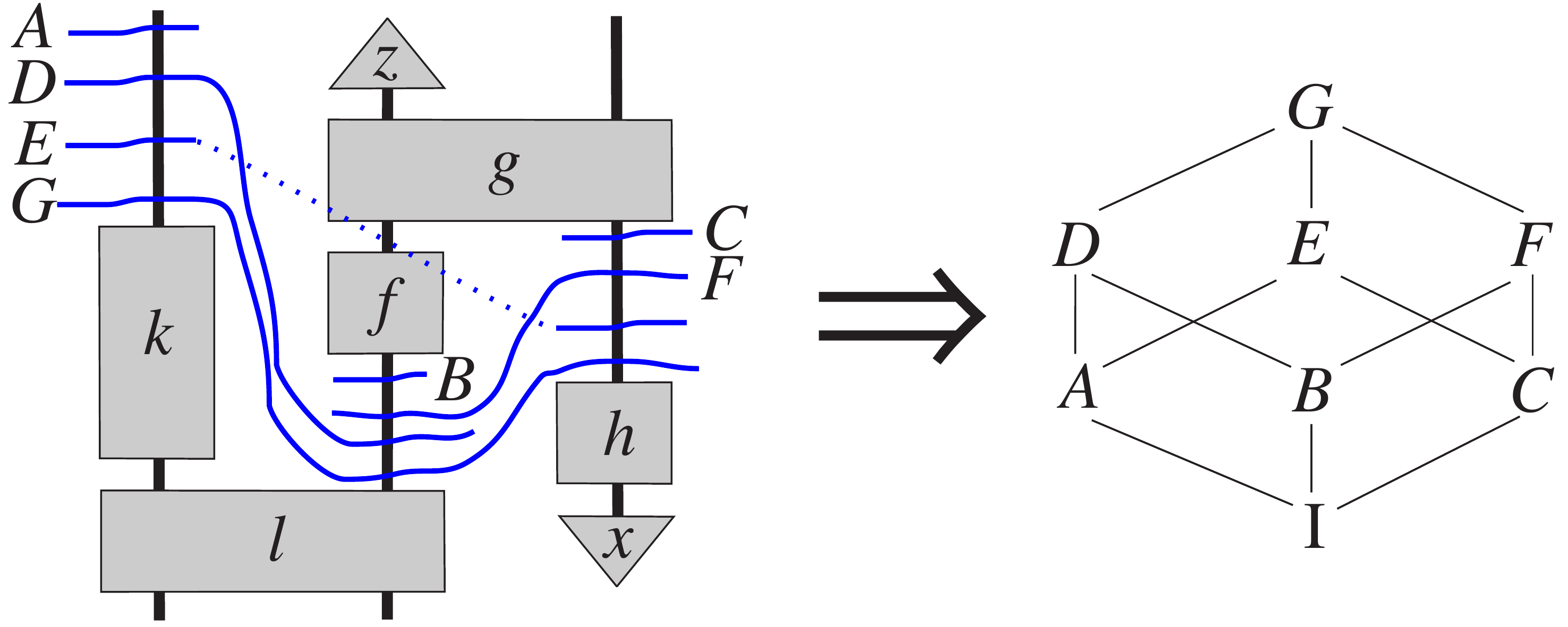,width=240pt}  
\]
This in particular implies that `being non-equal' for systems, as for example in $A\otimes B$,  does not capture independence.  In mathematical terms, it should be replaced by disjointness within the Boolean algebra structure arising from these snapshots.  A formal account on this is being developed with Ray Lal in \cite{CauCat}.


\subsection{Symmetry relations}\label{sec:logicalrelations}

Unlike processes, \em symmetry relations \em do not represent actual `happenings', nor do they have to respect any by us perceived causal structure.  Rather, they will enable one to express structural properties (i.e.~symmetries) of the processes which make up the physical universe,  in a manner similar to how the Galileo/Lorentz group conveys the shape of space-time.  
 But rather than being a structure on processes, in our approach they will interact with processes in the same manner as processes interact with each other, in terms of $\circ$ and $\otimes$, so that we can treat them as `virtual processes' within an `extended universe' which does not only consist of processes but also of  symmetry relations, as well as the relations arising when composing these.  The interaction of processes and symmetry relations would for example embody how usually dynamics is derived in terms of representations of the Galileo/Lorentz group.  Here, such a virtual process could for example be a Lorentz boost along a space-like curve.

 Intuitively, symmetry relations relate properties of one system to those of another system, and since the content carried by a process is how properties of system $A$ are transformed in those of system $B$, it indeed makes perfect sense to treat processes and symmetry relations on the same footing.   Consequently, we  can extend dependent composition to symmetry relations, but it obviously loses its causal connotation.  Also separate composition can evidently be extended to symmetry relations, separation now merely referring to some formal independence.    Consequently we can also still speak of scenarios and snapshots. 
 
  For some the distinction between process and symmetry relation might seem somewhat artificial.  But this would in fact even more advocate our framework. 
  
\paragraph{Example: active and passive rotations in classical mechanics.}   In classical mechanics rotations of a rigid body are `processes'  modeled in $SO(3)$, while the $SO(3)$ fragment of the Gallilei group consists of `symmetry relations' which assert the rotational symmetry of three-dimensional Euclidean space.

\paragraph{Example: inverses to processes.}  We define an \em inverse \em to a process $f: A\to B$ as the symmetry relation $f^{-1}:B\to A$ which satisfies:
\beq\label{eq:inverse}
f^{-1}\circ f=1_A     \qquad\mbox{and}\qquad   f\circ f^{-1}=1_B\,.
\eeq
It immediately follows that such an inverse, if it exists, is unique. 

\paragraph{Example: identical systems.}  How can we describe  distinct but  `identical' systems?  A pair of systems $A_1$ and $A_2$ is \em identical \em  if it comes with a pair of mutually inverse  relations $1_{A_1,A_2}:A_1\to A_2$ and $1_{A_2,A_1}:A_2\to A_1$. Explicitly: 
\[
1_{A_2,A_1}\circ 1_{A_1,A_2}=1_{A_1}   \quad   \qquad\mbox{and}\quad \quad 1_{A_1,A_2}\circ 1_{A_2,A_1}=1_{A_2}\,.
\] 
Let  $f: C\to A_1$ and $g: A_1\to C$ be any processes.  We set:
\beq\label{eq:identical}
F_{A_1,A_2}f:=1_{A_1,A_2}\circ f    \quad   \qquad\mbox{and}\quad \quad G_{A_1,A_2}g:= g\circ 1_{A_2,A_1}\,
\eeq
These can also be represented  in a \em commutative diagram \em \cite{MacLane1, MacLane2}:
\[
\xymatrix@=0.34in{
& D &\\
A_1 \ar@/_0.4em/[rr]_{1_{A_1,A_2}} \ar[ru]^{g} &  &A_2 \ar@/_0.4em/[ll]_{1_{A_2,A_1}} \ar[lu]_{G_{A_1,A_2}g}\\
& \ar[lu]^{f} C \ar[ru]_{F_{A_1,A_2}f} &
}
\]
i.e.~a diagram in which any two paths 
which take you from one system to another are equal.  It also follows that:
\[
F_{A_2,A_1}F_{A_1,A_2} f = f \quad   \qquad\mbox{and}\quad \quad G_{A_2,A_1}G_{A_1,A_2} g= g\,.
\]
Intuitively, the relations $1_{A_1,A_2}$ and $1_{A_2,A_1}$ identify the  potential properties  of systems $A_1$ and $A_2$, and  do this in a mutually inverse manner due to $1_{A_1,A_2}\circ 1_{A_2,A_1}=1_{A_2}$ and $1_{A_2,A_1}\circ 1_{A_1,A_2}=1_{A_1}$.
The assignments $F_{A_1,A_2}(-)$ (respectively $G_{A_1,A_2}(-)$) and $F_{A_2,A_1}(-)$ (respectively $G_{A_2,A_1}(-)$) identify processes involving $A_1$ and $A_2$  as output (resp.~input) in a similar manner. 

\paragraph{Example: identical processes.}  We leave it to the reader to combine the notion of inverse and that of identical systems into \em identical processes\em.   

\paragraph{Example: bosonic states.} The \em symmetric states \em
\[
\Psi:\II\to A_1\otimes \ldots \otimes A_n
\]
which describe non-isolated bosons can now be defined. For any permutation 
\[
\sigma:\{1, \ldots, n\}\to\{1, \ldots, n\}
\]
we have:
\beq\label{eq:boson}
(1_{A_1,A_{\sigma(1)}}\otimes \ldots \otimes 1_{A_n,A_{\sigma(n)}})\circ\Psi=\Psi\,,
\eeq
i.e.~when permuting the roles of the (identical) systems that make up the joint system, then that state should remain invariant.  Graphically, we represent the symmetry relation $1_{A_1,A_{\sigma(1)}}\otimes \ldots \otimes 1_{A_n,A_{\sigma(n)}}$ induced by the permutation $\sigma$ as `re-wiring' according to $\sigma$, e.g.:
\[
1_{A_1,A_3}\otimes 1_{A_2,A_1}\otimes 1_{A_3,A_4}\otimes 1_{A_4,A_2}
\equiv\ \raisebox{-0.34cm}{\epsfig{figure=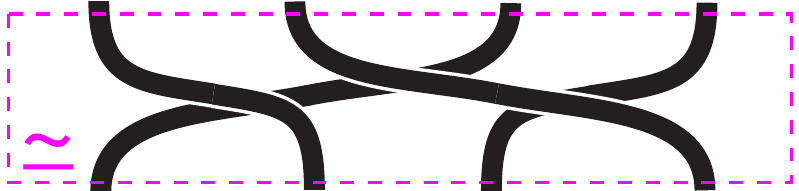,width=100pt}}
\]
The dotted box and the `$\simeq$'-sign refer to the fact that the wires are different from those we have seen so far, which represented systems.  Here they encode a relation which changes systems. Equation (\ref{eq:boson})  now becomes:
\[
\epsfig{figure=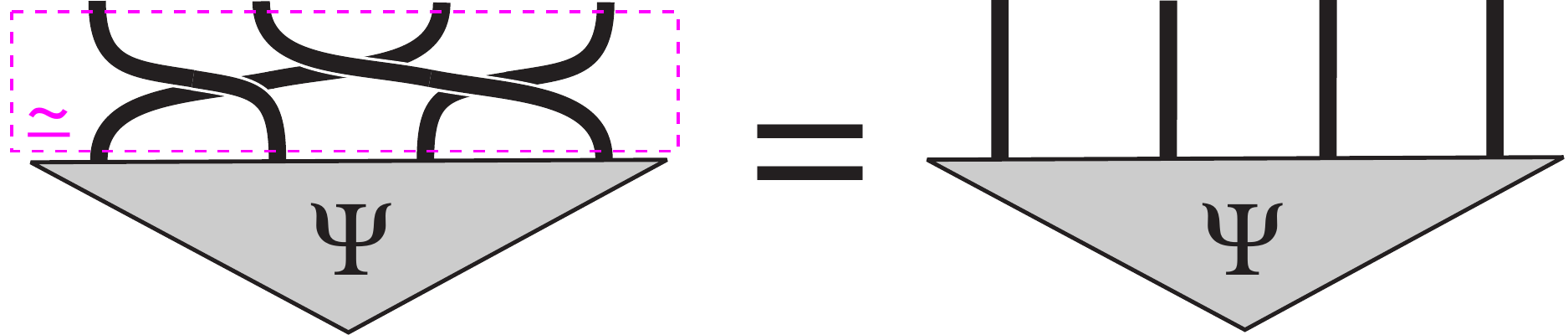,width=220pt}
\]
We indeed now truly exploit the fact that in the graphical language systems  appear in a certain order, which can be used to  identify systems.  Symmetry relations which identify distinct identical systems now identify different positions within the order.  
More on this ordering and identity of systems is in \S \ref{sec:mathmodels}.
\bigskip

We now combine symmetry relations representing  identical systems with the notion of a  process, to derive the crucial notion of an evolution:

\paragraph{Example: evolutions.}  By an \em evolution \em we mean a scenario only  involving causal composition and for which all maximal snapshots are identical  in the above sense.  Consider such a scenario with the process $f:A_0\to B$ as its resolution, and let  $A_\eta$ be a maximal snapshot distinct from $A_0$.   Now consider the scenario  that one obtains by restricting to those processes that happen before $A_\eta$, including $A_\eta$ itself, let the process $f_\eta:A_0\to A_\eta$ be its resolution, and now consider the symmetry relation: 
\[
e_\eta:= 1_{A_\eta, A} \circ f_\eta:A_0\to A_0\,.
\]
If the collection of all labels $\eta$ carries the structure of the real continuum we obtain a generalization of the standard notion of an evolution in terms of a one-parameter family of `things', here symmetry relations, which, intuitively,  relate  potential properties  of a system  at time $\eta$, here $A_0$, to those at time $0$.

\paragraph{Example: symmetry groups.}  The maps $e_\eta:A_0\to A_0$ in the previous example are special in that they relate a system to itself, while typically not being identities. One could associate  to each system a collection of such symmetry \em endo\em-relations which are closed under $\circ$ and each of which comes with an inverse, i.e.~for $f:A\to A$ there is $f^{-1}:A\to A$ such that $f\circ f^{-1}=f^{-1}\circ f=1_A$\,.  Such a collection plays the role of the symmetry groups in existing theories. It follows that a symmetry group of a system carries over to a symmetry group of an identical, and that  evolutions respect symmetries.

\paragraph{Example: variable causal structure.}  This example addresses a particular challenge posed by Lucien Hardy at a lecture in Barbados, spring 2008 \cite{Hardy1}.

Thus far processes were required to respect some perceived causal structure. However, several authors argue that a framework that stands a chance to be of any use for describing quantum gravity should allow for variable causal structure e.g.~\cite{Hardycausaloid,  Hardy2}.  Once we `solved' Einstein's equations in general relativity then the causal structure is of course fixed, so varying causal structure doesn't boil down to merely dropping it, but rather to allow for a variety of causal structures.  

This is what we will establish here, namely to introduce processes which have the potential to adopt many different causal incarnations, while still maintaining the key role of composition within the theory.  In other words,  a certain causal incarnation becomes something like potential property. 

For each system we introduce two symmetry relations:
\[
\cup_A: \II\to A^*\otimes A\qquad\mbox{and}\qquad\cap_A\ : A\otimes A^*\to \II\,,
\] 
to which we respectively refer to as \em input-output reversal \em and  \em output-input reversal\em.
These are subject to the following equations:
\beq\label{eq:compact}
(\cap_A\otimes 1_A)\circ(1_A\otimes \cup_A)=1_A \quad\mbox{and}\quad
(1_{A^*}\otimes \cap_A)\circ(\cup_A\otimes 1_{A^*})=1_{A^*}
\eeq
which state that reversing twice yields no reversal.  For a process \mbox{$f:C\otimes A\to B$} (resp.~$g:A\to B\otimes C$) we can use reversal to produce a variation on it where the input $C$ (resp.~output $C$) has become an output  (resp.~input) $C^*$\,:
\[
\tilde{f}=(1_{C^*}\otimes f)\circ(\cup_C\otimes 1_A):A\to C^*\otimes B
\]
\[
\tilde{g}=(1_B\otimes \cap_C)\circ(g\otimes 1_{C^*}):A\otimes C^*\to B
\]
 Here, the `$*$' tells us that while $A$ was an input  (resp.~output) for process  $f$ (resp.~$g$) that it is now converted into an output  (resp.~input).  This is crucial when composing  $f$ (resp.~$g$) with other processes. 

Putting this in pictures we set:
\[
\cup_A
\equiv\ \raisebox{-0.34cm}{\epsfig{figure=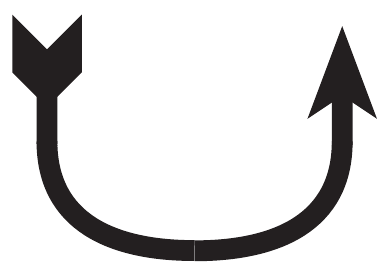,width=30pt}}
\qquad\qquad\mbox{and}\qquad\qquad
\cap_A\ \equiv\ \raisebox{-0.34cm}{\epsfig{figure=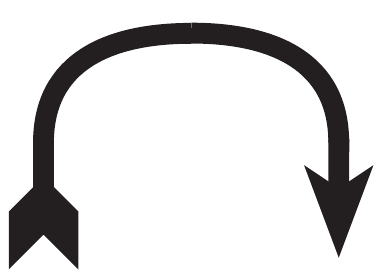,width=30pt}}
\]
where the directions of the arrows represent the $*$'s.  The equations then become:
\[
\raisebox{-0.34cm}{\epsfig{figure=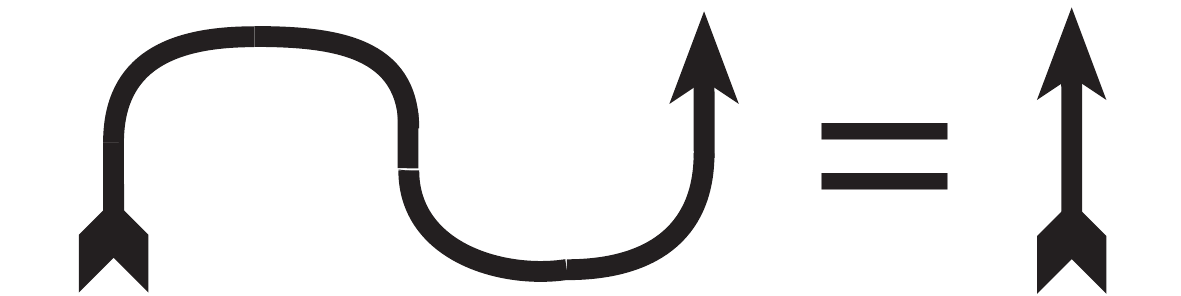,width=100pt}}
\qquad\mbox{and}\qquad
\raisebox{-0.34cm}{\epsfig{figure=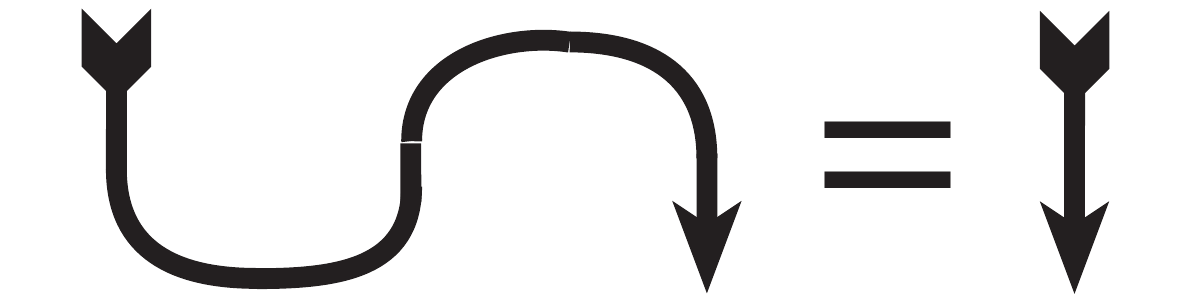,width=100pt}}
\]
and the converted processes depict as:
\[
\tilde{f}
\equiv\ \raisebox{-0.54cm}{\epsfig{figure=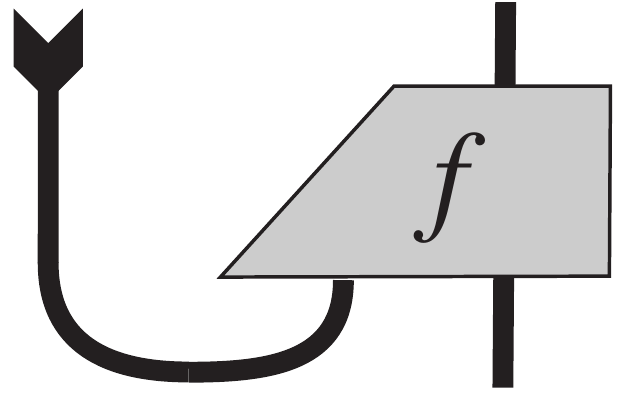,width=52pt}}
\qquad\qquad\mbox{and}\qquad\qquad
\tilde{g}\ \equiv\ \raisebox{-0.46cm}{\epsfig{figure=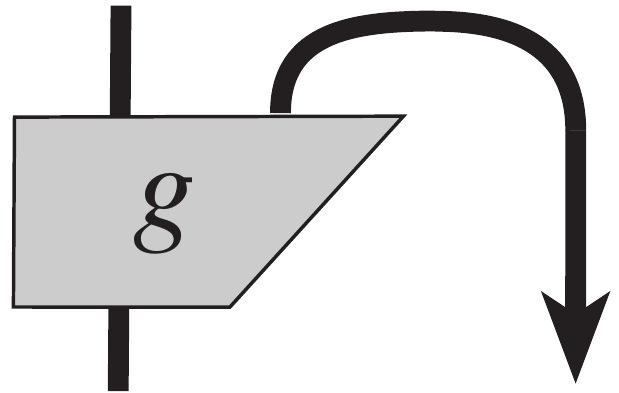,width=52pt}}
\]

By a \em pre-causal process \em we mean a collection of potential  processes which is closed under reversal of all inputs and outputs.  For example, graphically:
\[
\left[ \raisebox{-0.8cm}{\epsfig{figure=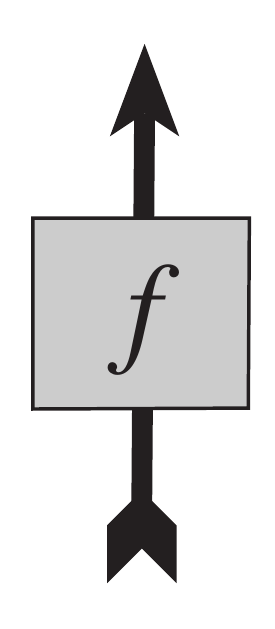,width=22pt}}\right]:=
\left\{  \raisebox{-0.8cm}{\epsfig{figure=precausal1.pdf,width=22pt}}\ \,, 
 \raisebox{-0.8cm}{\epsfig{figure=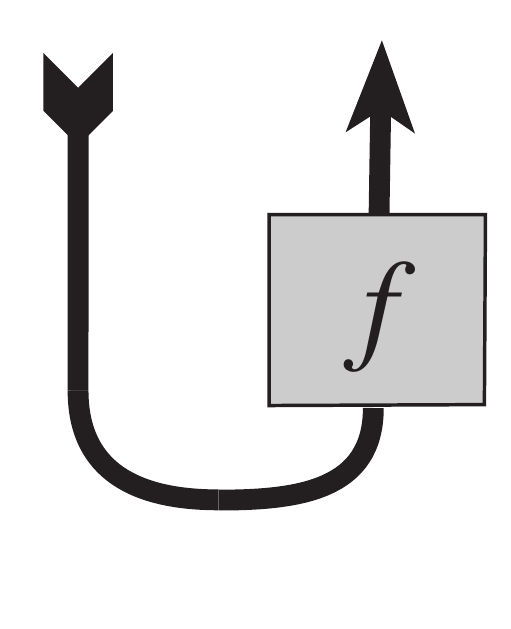,width=44pt}}\ \, , \, 
  \raisebox{-0.8cm}{\epsfig{figure=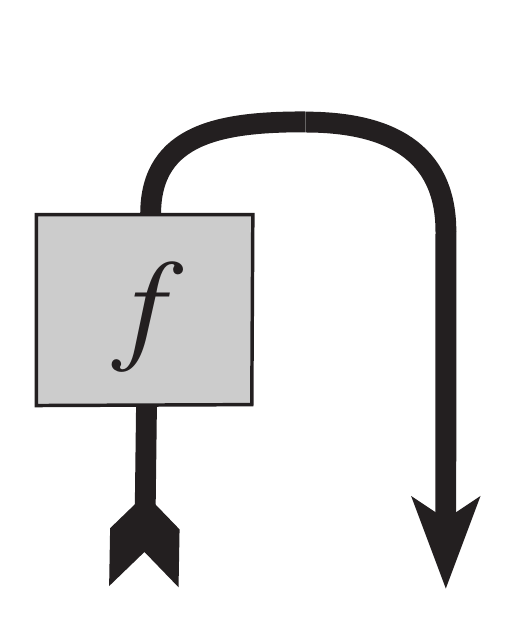,width=44pt}}\,, 
   \raisebox{-0.8cm}{\epsfig{figure=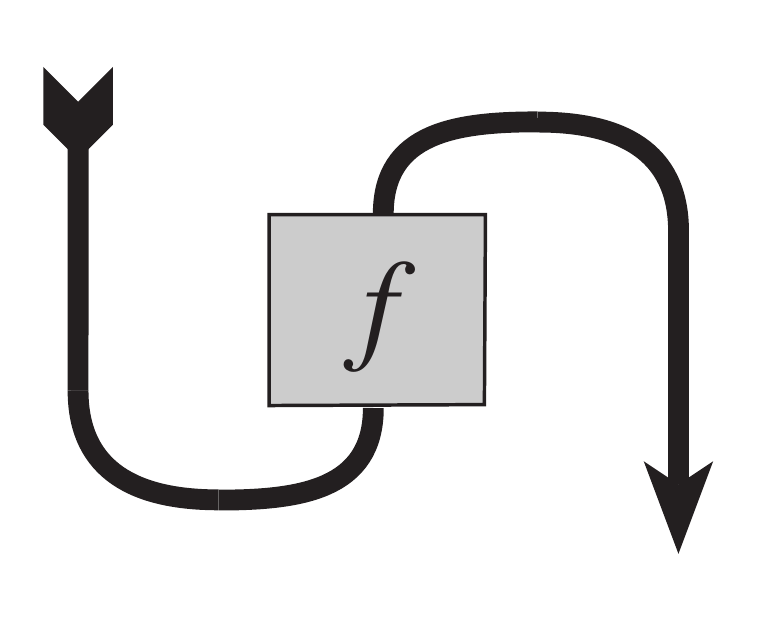,width=66pt}}\!\!\right\}
\]
is such a pre-causal process.  Similarly to how composition of processes could be represented by directed graphs, one can show that composition of pre-causal processes can be represented by \em un\em directed graphs:
\[
\epsfig{figure=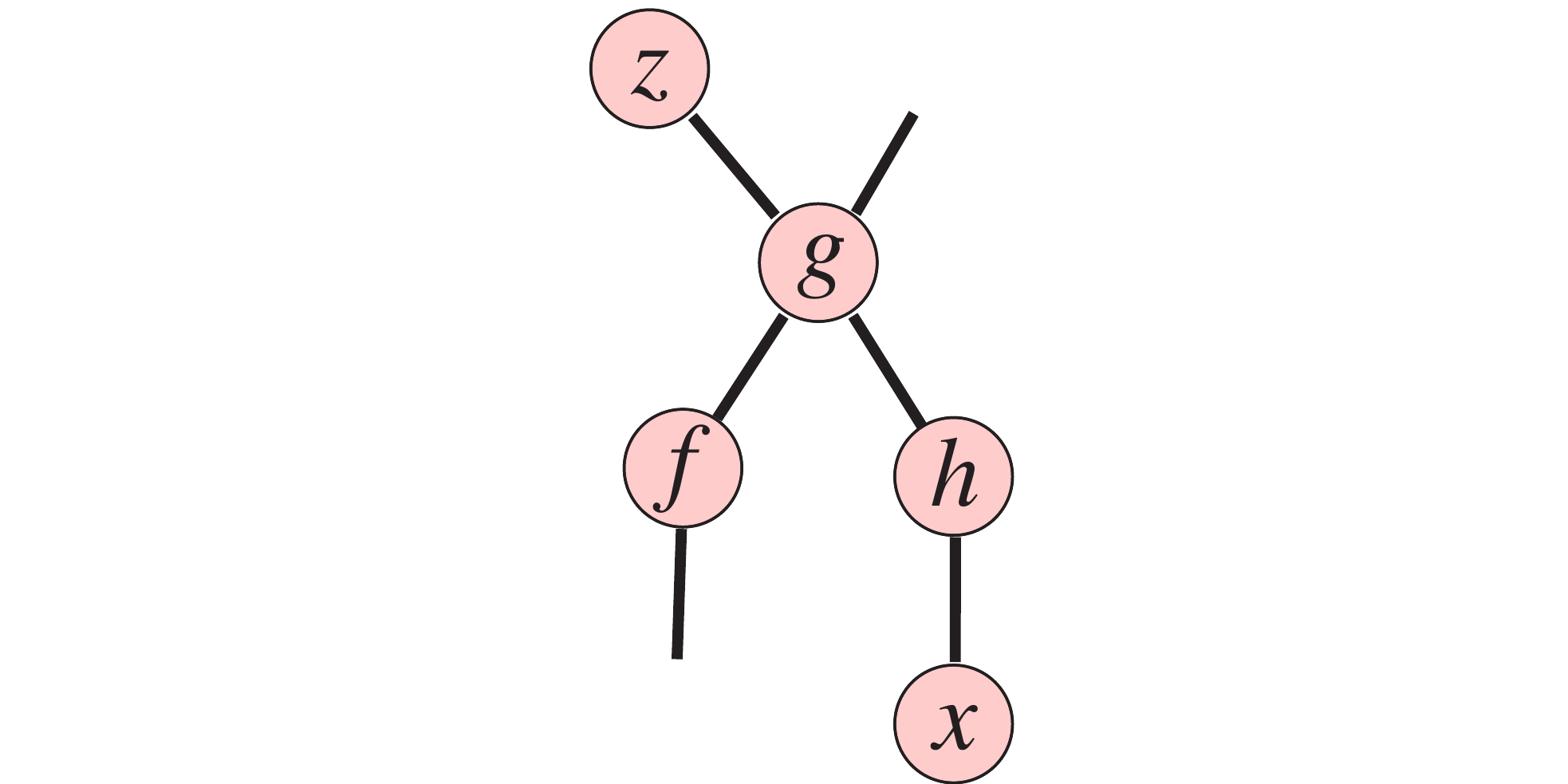,width=160pt}
\]
Indeed, by considering a node as representing a pre-causal process, that is, all of its potential causal incarnations, we obtain:
\[
\!\!\epsfig{figure=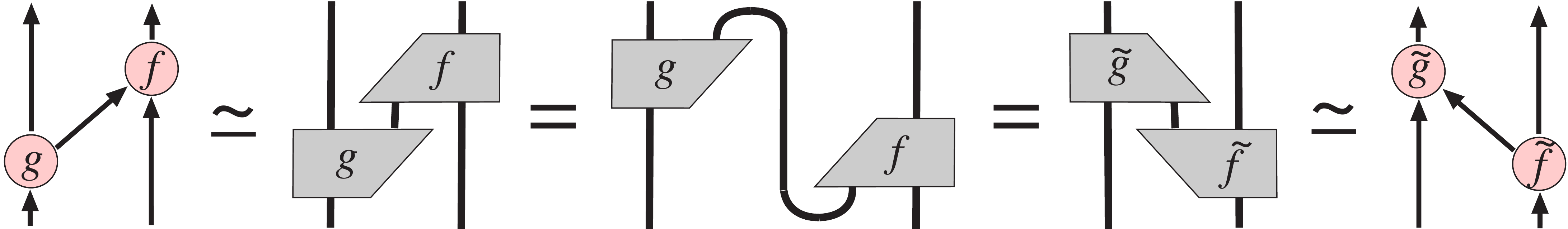,width=350pt}
\]
Consequently, the directions on arrows carry no content.

While the presentation of scenarios as nodes of undirected graphs is of course more concise than as collections of causal incarnations, the latter has the conceptual advantage that causal structure is attributed to processes.  Since in our setup these are the things that really `happen', whereas systems only play a supporting role, it is the processes which should carry the causal structure.


\subsection{Vacuous relations: correcting denotational artifacts}

Both processes and symmetry relations carry structural content of the theory under consideration.  We mention one more kind of relation of which the sole purpose is to correct artifacts due to a particular choice of denotation.  Above we already pointed at the fact that when we denote separate composition either symbolically or diagrammatically this unavoidably comes with some ordering due to the fact that the points of a line are totally ordered.  

To undo this, we need to state that all orderings are equivalent.  Therefore we introduce for each pair of systems $A_1$ and $A_2$ an invertible relation $\sigma_{A_1, A_2}$ which \em exchanges \em the order:
\[
\sigma_{A_1, A_2}:A_1\otimes  A_2\to A_2\otimes  A_1 \qquad\mbox{with}\quad \sigma_{A_2,A_1}\circ \sigma_{A_1,A_2}=1_{A_1\otimes A_2}\,. 
\]
These relations  then generate arbitrary permutations e.g.:\footnote{Note the difference with the example of bosonic  states earlier in this paper in that now the wires relate a system with itself, just like identities do.  They just shift the order. Hence:
\[
\epsfig{figure=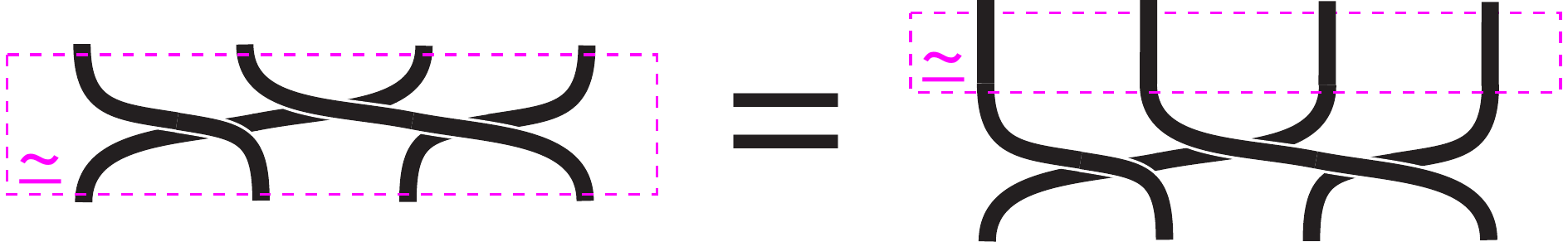,width=220pt}
\]
where the straight wires in the dotted box stand for change of system.}
\[
\raisebox{-0.34cm}{\epsfig{figure=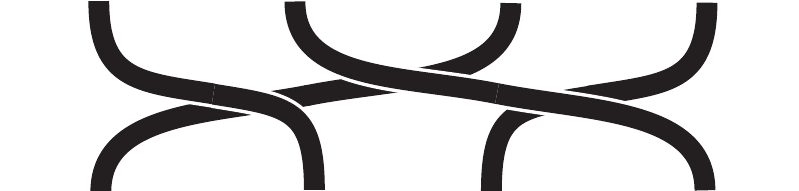,width=80pt}}:A_1\otimes  A_2\otimes  A_3\otimes  A_4\to A_2\otimes  A_4\otimes  A_1\otimes  A_3\,.
\]
To state that these exchanges of order are indeed vacuous we have to assert that they do not affect the structural content of the theory, that is, the two compositions.  Firstly, separate compositions should be preserved:
\[
\sigma_{B_1,B_2}\circ (f_1\otimes f_2) = (f_2\otimes f_1)\circ \sigma_{A_1,A_2}\,.
\]
For example, for the above permutation of four systems we have:
\[
\epsfig{figure=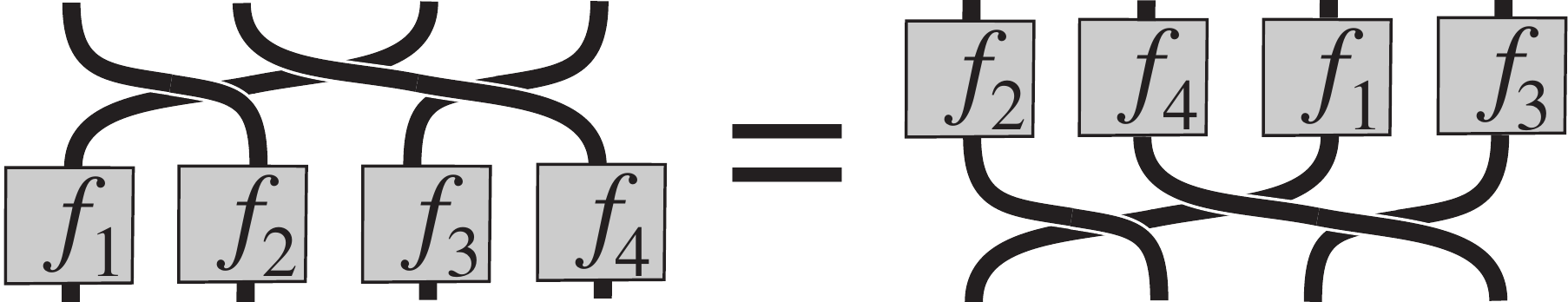,width=184pt}
\]
That causal composition is also preserved then trivially follows:
\[
\sigma_{C_1,C_2}\circ (g_1\otimes g_2)\circ (f_1\otimes f_2) = 
(g_1\otimes g_2)\circ\sigma_{B_1,B_2} \circ (f_2\otimes f_1)= 
(g_1\otimes g_2)\circ (f_2\otimes f_1)\circ \sigma_{A_1,A_2}.
\]

\subsection{Summary of this section}

Within the proposed framework a \em physical theory \em has the following ingredients:
\bit
\item a collection of \em relations \em with two \em compositions \em $\circ$ and $\otimes$ thereon, subject to an \em independence constraint\em, as well as additional \em equations \em that specify for which scenarios the corresponding resolutions are equal\,;
\item certain relations called \em potential \em (or \em candidate\em) \em processes \em which will act as the `actual physical substance' of the theory, according to: 
\[
{\mbox{actual\ process}\over\mbox{actual\ property}}
\, \ {\raisebox{-4pt}{$\,\cdot\,$}\over\raisebox{4pt}{$\,\cdot\,$}} \ \,
{\mbox{might\ happen}\over\mbox{might\ be}}\,;
\]
Examples of processes are:
\bit
\item states, effects and weights\,,
\item processes resulting from performing an operation\,,
\item feed-in-environment processes witnessing non-isolation, etc.
\eit
\item certain relations called  \em symmetry relations \em which carry additional structure of the theory\,;  examples of symmetry relations are:
\bit
\item those that identify symmetries\,,
\item those that identify identical systems\,,
\item those that vary the causal structure, etc.
\eit
\item certain relations called \em vacuous relations \em that carry no physical content whatsoever but undo artifacts that are merely due to denotation.
\eit 
Further below we will identify some more ingredients but first we will see how we can cast these ones within standard mathematical structures.

\section{The mathematical guise of physical theories}\label{sec:mathmodels}

In set theory \cite{Bourbaki, Devlin} a \em class \em is a collection of which the members are defined by a predicate which they all obey. For example, the class of groups is defined as sets that come equipped with a binary and a unary operation which obey the usual axioms of groups.  By Russell's paradox, which can restated as the fact that the collection of all sets itself does not form a set, it immediately follows that the collection of all groups together do not form a set, but a \em proper class\em.

\paragraph{Modeling concession 1.} The collection of all systems together forms a class and the collection of all relations of the same type forms a set. 
\bigskip

This concession reflects standard mathematical practice,\footnote{There exist proposals to generalize this, e.g.~the \em universes \em as in  \cite{Borceux1}\S1.1.} and hence is essential when trying to provide the `informal' ideas in the previous section with a more standard formal backbone, either in terms of axiomatics or in terms of more concrete models obeying this axiomatics.\footnote{We refer the reader to \cite{CatsII} for a more detailed discussion of the sense in which we use `axiomatics' and `concrete models', where rather than `axiomatics' we used the term `abstract'.}

\subsection{Axiomatics}\label{sec:axiomatics}

The physical framework outlined above, when subjected to the stated modelling concession 1, can be represented as a so-called \em strict symmetric monoidal category\em.  For a more detailed discussion we refer the reader again to \cite{CatsII}.

\paragraph{Definition.}  A \em strict symmetric monoidal category \em ${\bf C}$ consists of a class of \em objects \em $|{\bf C}|$, for each pair of objects  $A, B\in|{\bf C}|$ a set of \em morphisms \em ${\bf C}(A,B)$,\footnote{Such a set ${\bf C}(A,B)$ of morphisms is usually referred to as a \em homset\em.} a privileged \em unit \em object $\II\in|{\bf C}|$, for each object $A\in|{\bf C}|$ a privileged \em identity \em morphism $1_A\in{\bf C}(A,A)$, and the following operations and axioms:
\bit 
\item an associative binary operation $\otimes$ on $|{\bf C}|$ with unit $\II$\,;
\item an associative binary operation $\otimes$ on $\bigcup_{A,B}{\bf C}(A,B)$ with unit $1_\II$, and with $f_1\otimes f_2\in{\bf C}(A_1\otimes A_2, B_1\otimes B_2)$ for $f_1\in{\bf C}(A_1,B_1)$ and  $f_2\in{\bf C}(A_2,B_2)$\,;
\item a partial associative binary operation $\circ$ on $\bigcup_{A,B}{\bf C}(A,B)$ restricted to pairs in 
${\bf C}(B,C)\times {\bf C}(A,B)$ where  $A, B, C\in|{\bf C}|$ are arbitrary, and for all $A, B\in|{\bf C}|$, all $f\in {\bf C}(A,B)$ have  \em right identity \em $1_A$  and \em  left identity \em $1_B$\,.
\eit
Moreover, for all $A, B, A_1, B_1, C_1, A_2, B_2, C_2\in|{\bf C}|$, $f_1\in{\bf C}(A_1,B_1)$,  $g_1\in{\bf C}(B_1,C_1)$, $f_2\in{\bf C}(A_2,B_2)$ and  $g_2\in{\bf C}(B_2,C_2)$ we have:
\[
( g_1\circ f_1)\otimes( g_2\circ f_2)=( g_1\otimes g_2)\circ( f_1\otimes f_2)\qquad\mbox{and}\qquad
1_{A_1}\otimes 1_{A_2}=1_{A_1\otimes A_2}\,.
\]
Finally, for all $A_1, A_2\in|{\bf C}|$ there is a privileged morphism 
\[
\sigma_{A_1,A_2}\in{\bf C}(A_1\otimes A_2 , A_2\otimes A_1)
\qquad\mbox{with}\qquad
\sigma_{A_2,A_1}\circ \sigma_{A_1,A_2}=1_{A_1\otimes A_2}
\]
such that for all $A_1, A_2, B_1, B_2\in|{\bf C}|, f_1\in{\bf C}(A_1,B_1), f_2\in{\bf C}(A_2,B_2)$ we have:
\beq\label{eq:symnat}
\sigma_{B_1,B_2}\circ (f_1\otimes f_2) = (f_2\otimes f_1)\circ \sigma_{A_1,A_2}\,.
\eeq

\bigskip
This is quite a mouthful but there are very short more elegant ways to say this which rely on higher-level category theory.\footnote{For example, a (not-necessarily strict) symmetric monoidal category, which takes a lot more space to explicitly define than a strict one (see \cite{MacLane1, LNPAT, LNPBS} for the usual definition and \cite{CatsII} for a discussion), is  an \em internal commutative monoid in the category of all categories\em.}  It is also a well-known fact that these strict monoidal categories are in exact correspondence with the kind of graphical calculi that we introduced to describe relations \cite{JS}. While the use of this calculi traces back to Penrose's work in the early seventies  \cite{Penrose}, they only became a genuine formal discipline within the context of monoidal categories with the work of Joyal and Street  \cite{JS} in the nineties. However, the first comprehensive detailed account on them was only produced this year  by Selinger  \cite{SelingerSurvey}, which provides an even nicer presentation.  We say something more about these graphical presentations below in Section \ref{sec:Axmeetmodel}.

We now show how the above discussed framework, subject to the modeling concession, can be interpreted in the language of strict symmetric monoidal categories. Recall here that an \em isomorphism \em in a category is a morphism $f:A\to B$ which has an inverse, precisely in the sense of (\ref{eq:inverse}).

Systems are represented by objects of the symmetric monoidal category, relations by morphisms, and the compositions have been given matching notations.   We discuss the role of some of the privileged morphisms:
\bit
\item The \em symmetry natural isomorphism\em\footnote{The significance of the word `natural' here precisely boils down to validity of (\ref{eq:symnat}).} 
\[
\{\sigma_{A_1,A_2}:A_1\otimes A_2 \to A_2\otimes A_1\mid A_1, A_2\in|{\bf C}| \}\,.
\]
plays the role of the symmetry relation that  undoes the unavoidable a priori ordering on systems when composing them with $\otimes$.
\item There may be several occurrences of the same object within a string of tensored objects e.g.~$A\otimes A$. To align  this with the fact that all systems occurring in such an expression must we independent, 
we either:
\bit
\item[c1] not assign any meaning to all objects and morphisms of the symmetric monoidal category, 
but rather consider a subcategory of it with a partial tensor, an approach which is currently developed in \cite{CauCat};
\item[c2] represent distinct identical systems by the same object, which allows for the two $A$'s in $A\otimes A$ to be interpreted as independent.    
\eit
\eit
Above, c1 and c2 can also be seen as modeling concessions.

\paragraph{Example: compactness models variable causal structure.} A \em compact (closed) category \em \cite{Kelly, KellyLaplaza} is a symmetric monoidal category in which every object $A$ has a \em dual \em $A^*$, that is, there are morphisms $\cup_A: \II\to A^*\otimes A$ and $\cap_A\ : A\otimes A^*\to \II$ satisfying equations (\ref{eq:compact}).  Equivalence classes of morphism then enable to model variable causal structure as indicated above.

\paragraph{Example: symmetry and compactness in communication protocols.} Above, the `symmetry' and `compact' structure represented relations which respectively undo the order on systems within scenarios or a causal structure.  These morphisms can also play a more constructive role as special kinds of communication processes.  It was this which initially motivated the use of compact closed categories to model quantum protocols in \cite{AC}.
To this end we will treat the ordering of objects relative to the tensor as genuine locations in space-time, represented by two agents, respectively named Ali and Bob. Then the morphism:
$\sigma_{A_1,A_2}:A_1\otimes A_2 \to A_2\otimes A_1$ means that the agents \em exchange \em their physical systems. We can represent the agents by regions in the plane which extend vertically i.e.~in the direction of causal composition:
\[
\, (\raisebox{-3mm}{$\overbrace{\,\hspace{14mm}\,}$})\,  \otimes\, (\raisebox{-3mm}{$\overbrace{\,\hspace{14mm}\,}$})
\]
\[
\epsfig{figure=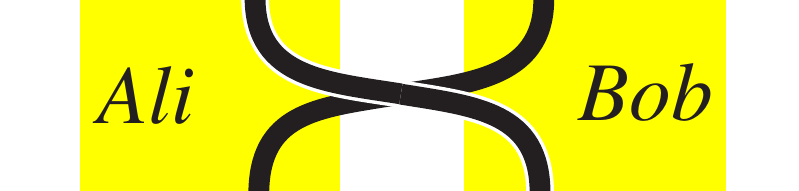,width=140pt}
\]
If the category is moreover compact closed then by the axioms of compact closure we have the following equation between scenarios: 
\[
\epsfig{figure=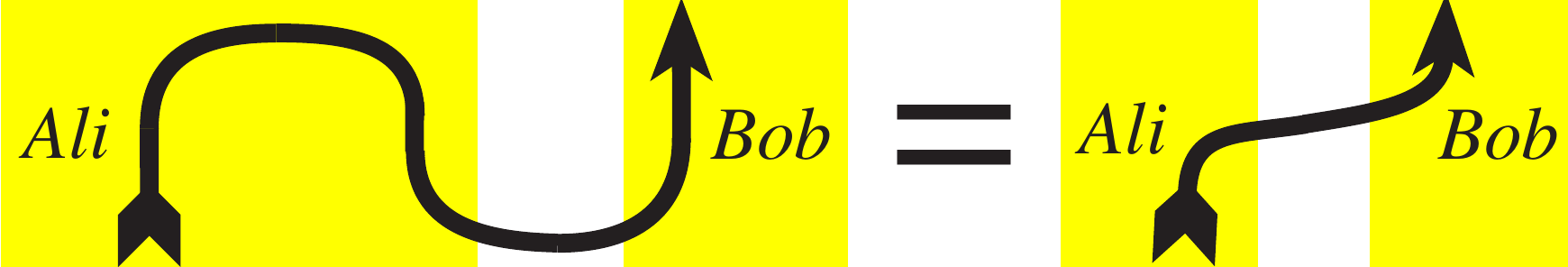,width=240pt}
\]
which can be interpreted as a correctness proof of \em post-selected quantum teleportation\em.  Here, $\cup_A: \II\to A^*\otimes A$ represents a \em Bell-state \em and $\cap_A\ : A\otimes A^*\to \II$ represents a \em post-selected Bell-effect\em.  A more detailed analysis as well as more sophisticated variations on the same theme which involve varying the entangled state and allowing for non-determinism of the effects are in \cite{AC, Kindergarten, ContPhys, CPaqPer, CPaqPav, CWWWZ}.

\paragraph{Example: explicit agents.} While the previous example gives a very concise presentations of protocols, it is not completely consistent with our earlier interpretation of symmetry and compactness as symmetry relations.  One possible manner to accommodate  the use of these morphisms both as symmetry relations as well as processes is by explicitly introducing agents.  To model agents, respectively named Ali and Bob, we take objects to be pairs consisting of an entry which represents the physical system together with an entry which represents the agent that possesses that system for that snapshot.  Morphisms will be  pairs consisting of the manner in which physical systems are processed, as well as specification of which agents possess it at the beginning and the end of the processes.  We provide a rough idea of how naively this can formally be established, skimming over certain technical details. Take the \em product category\,\em:
\[
{\bf C}\times F_{\sf SM}{\bf Agents} 
\]
of the symmetric monoidal category ${\bf C}$ in which we model physical systems, and the \em free symmetric monoidal category over a category \em 
${\bf Agents}$,\footnote{An overview of free constructions for the categories which we consider here is in \cite{AbrFree}.  The objects of the free symmetric monoidal category $F_{\sf SM}{\bf D}$ over a category ${\bf D}$ are finite lists of objects of ${\bf D}$ and the morphisms are finite lists of morphisms of ${\bf D}$ together with a permutation of objects.  Concretely we can write these as 
$\sigma_\pi\circ(f:A_1\to B_1, \ldots, A_n\to B_n): (A_1,\ldots, A_n)\to (B_{\pi(1)},\ldots, B_{\pi(n)})$ where 
$\pi:\{1, \ldots, n\}\to\{1, \ldots, n\}$ is a permutation. The permutation component alone provides the symmetry natural isomorphism.} 
which has two objects $Ali$ and $Bob$ and only identities as morphisms:
\[
\ {\bf Agents}(Ali,Ali)=\{1_{Ali}\}\qquad\quad\
{\bf Agents}(Ali, Bob)=\emptyset
\]
\[
{\bf Agents}(Bob, Bob)=\{1_{Bob}\} \qquad\quad {\bf Agents}(Bob,Ali)=\emptyset
\]  
The category ${\bf C}\times F_{\sf SM}{\bf Agents}$ inherits symmetric monoidal structure component-wise from 
${\bf C}$ and $F_{\sf SM}{\bf Agents}$, with a symmetry morphism now of the form:
\[
\left(\sigma_{A_1, A_2}\,,\, \sigma_\pi\right)
:(A_1, Ali)\otimes (A_2, Bob)\to(A_2, Bob)\otimes(A_1, Ali)\,,
\]
where $\pi$ is the (only) non-trivial permutation of two elements. This now represents the symmetry relation that undoes the ordering on objects.  On the other hand,  the exchange process can now be differently represented, namely by:
\[
\left(\sigma_{A_1, A_2}\,,\, 1_{Ali\otimes Bob}\right)
:(A_1, Ali)\otimes (A_2, Bob)\to(A_2, Ali)\otimes(A_1, Bob)\,,
\]
So we have  distinct morphisms representing both symmetry relations and  processes, and the same can be done for   
post-selected quantum teleportation.
\bigskip

\subsection{Concrete models}

Thus far we treated categories as a structure in their own right, and consequently, also the diagrammatic calculi.  
However, to realize existing theories such as quantum theory, we need to consider \em concrete models \em of these.  That is, the objects constitute some kind of mathematical strucure, for example Hilbert spaces, while the morphisms constitute mappings between these, for example linear maps.  The monoidal tensor is then a binary construction on these.

But what we obtain in this manner are not strict symmetric monoidal categories, but\ \, strict\hspace{-0.94cm}--------- \,\em symmetric monoidal categories\em.  In particular,  we lose (strict) associativity and (strict) unitality of the tensor:
\[
A\otimes (B\otimes C)\not =(A\otimes B)\otimes C \qquad \II\otimes A\not= A \qquad
A\otimes\II\not= A
\]
\[
\,\ \ \ f\otimes (g\otimes h)\not =(f\otimes g)\otimes h\ \  \qquad 1_\II\otimes f\not= f \qquad
f\otimes 1_\II\not= f\,.
\]
This is a consequence of the fact that in set-theory: 
\[
\ \ \ \ \ (x ,(y,z))\not =((x,y),z)\ \  \qquad\ \ \ \ (*,x)\not=x   \qquad\   (x,*)\not= x\,.
\]
For a detailed discussion of this issue we refer to \cite{CatsII}.  We mention here that the main consequence of this is the fact that in any standard textbook the definition of a symmetric monoidal category may stretch many pages.  The reason is that in one way or another we need to articulate that $A\otimes (B\otimes C)$ and $(A\otimes B)\otimes C$ are in a very special way related, similarly to how $A\otimes B$ and $B\otimes A$ relate was captured above by the symmetry `natural isomorphisms'.   

Five examples of models of symmetric monoidal categories are:
\bit
\item
${\bf{\rm(}F{\rm)}Set}$:=
\bit
\item Objects:= (finite) sets
\item Morphisms:= functions between these
\item Tensor:= the Cartesian product of sets
\eit
\item
${\bf{\rm(}F{\rm)}Rel}$:=
\bit
\item Objects:= (finite) sets
\item Morphisms:= (ordinary mathematical) relations between these
\item Tensor:= the Cartesian product of sets
\eit
\item 
${\bf{\rm(}F{\rm)}Hilb}$:=
\bit
\item Objects:= (finite dimensional) Hilbert spaces
\item Morphisms:= linear maps between these
\item Tensor:= the Hilbert space tensor product
\eit
\item 
${\sf WP}{\bf{\rm(}F{\rm)}Hilb}$:=
\bit
\item Objects:= (finite dimensional) Hilbert spaces
\item Morphisms:= linear maps between these up to a global phase
\item Tensor:= the Hilbert space tensor product
\eit
\item${\sf CP}{\bf{\rm(}F{\rm)}Hilb}$:=
\bit
\item Objects:= (finite dimensional) Hilbert spaces
\item Morphisms:= completely positive maps between these\footnote{These can for example be defined in terms of the Kraus representation $f::\rho\mapsto \sum_ iA_i^\dagger \rho A_i$ on the space of density matrices, where each $A_i$ is an $n\times n$-matrix \cite{Davies,Krausbook}.}
\item Tensor:= the Hilbert space tensor product
\eit
\eit
 The reason for restricting to finite sets/dimensions is explained in Section \ref{sec:Axmeetmodel}.

Mappings from one of these models, which take each object $A\in|{\bf C}|$ to an object $FA\in|{\bf C}|$ and which take each morphism $f\in{\bf C}(A,B)$ to a morphism $Ff\in{\bf C}(FA,FB)$, and which preserve  the full symmetric monoidal structure, are called \em strict monoidal functors\em.  If a strict monoidal functor is injective on homsets it is called \em faithful\em.  These strict monoidal functors allow one to relate different models to each other.  For example, there are the identity-on-objects faithful strict monoidal functors:\footnote{To see this  for ${\sf WP}{\bf   FHilb}\hookrightarrow {\sf CP}{\bf   FHilb}$, note that ${\sf WP}{\bf   FHilb}$ can be presented as a category with the same objects as ${\bf   FHilb}$ but with maps of the form $f\otimes \bar{f}: {\cal H}\otimes {\cal H}\to {\cal H}'\otimes {\cal H}'$ (where $f: {\cal H}\to {\cal H}'$ is any linear map) as the morphisms in ${\sf WP}{\bf   FHilb}({\cal H}, {\cal H}')$
 \cite{deLL}.  Similarly, also ${\sf CP}{\bf   FHilb}$ can be presented as a category with the homset ${\sf CP}{\bf   FHilb}({\cal H}, {\cal H}')$ containing maps of type ${\cal H}\otimes {\cal H}\to {\cal H}'\otimes {\cal H}'$, but now more general ones \cite{Selinger}.}
\beq\label{eq:Setembeddngs}
F_{func}:{\bf Set}\hookrightarrow {\bf Rel}\qquad \qquad \qquad F_{pure}:{\sf WP}{\bf   FHilb}\hookrightarrow {\sf CP}{\bf   FHilb}
\eeq
as well as object-squaring (i.e.~${\cal H}\mapsto{\cal H}\otimes{\cal H}$)   faithful strict monoidal functors:\footnote{This again relies on the presentations mentioned in the previous footnote.}
\beq\label{eq:Hilbembeddngs}
F_{cp}:{\sf CP}{\bf   FHilb}\hookrightarrow {\bf   FHilb} \qquad \qquad
F_{cp}\circ F_{pure}:{\sf WP}{\bf   FHilb}\hookrightarrow {\bf   FHilb} \,.
\eeq

But in our view the physical theories should primarily be  formulated axiomatically rather than in terms of these models, since it is at the axiomatic level that the conceptually meaningful entities live, and hence it is on those that structures should be imposed, rather than providing concrete presentations of them which typically would carry more information than necessary/meaningful.  Ultimately, one would like to equip a strict symmetric monoidal category with enough structure so that we can  derive all observable physical phenomena, without the necessity to provide a concrete model.  

Then, the choice of a particular model such as ${\sf WP}{\bf   FHilb}$ can be seen as a choice of \em coordinate system \em which might enable one to solve a certain problem better than other coordinate systems. Hence for us the non-strictness of the mathematical models is an unfortunate artifact, while the strictness which we took for granted when setting up the formalism, which is also implicitly present in the diagrammatic calculi, reflect the true state of affairs.

The category ${\bf   FHilb}$ is the one that we typically have in mind in relation to quantum mechanics.  
But other models may provide the same features. These other models, in particular those of a more combinatorial nature, might give some useful guidance towards, say, a theory of quantum gravity.  Also, discrete models are also extremely useful for computer simulations.

\subsection{Where axioms and models meet: a theorem}\label{sec:Axmeetmodel}

Consider the following four devices:
\bit
\item[{\bf(1)}] Axiomatically described strict symmetric monoidal categories, possibly equipped with additional structure. 
\item[{\bf(2)}] Axiomatically described\ \, strict\hspace{-0.94cm}--------- \,symmetric monoidal categories (for which we refer to the many available textbooks and survey papers \cite{MacLane1, LNPAT, LNPBS}) possibly equipped with additional structure.
\item[{\bf(3)}] The diagrammatical calculus of strict symmetric monoidal categories (of which precise descriptions can be found in \cite{JS, Selinger, SelingerSurvey}) possibly equipped with additional graphical elements.
\item[{\bf(4)}] The concrete category ${\bf   FHilb}$.
\eit
If we establish an equation in one of these, what do we know about the validity of equations in one of the others?

\paragraph{Theorem.} \em An equation between two scenarios in the language of symmetric monoidal categories follows from the axioms of symmetric monoidal categories,  {\bf if and only if}, the corresponding equation between two scenarios in the language of strict symmetric monoidal categories follows from the axioms of strict symmetric monoidal categories,  {\bf if and only if},  the corresponding equation in the graphical language follows from  isomorphisms of diagrams.\em
\bigskip

The reader who wants to understand the nitty-gritty of this statement can consult \cite{SelingerSurvey}.  The main point that we wish to make here is that for all practical purposes, strict symmetric monoidal categories, general  symmetric monoidal categories, and the corresponding graphical language, are essentially one-and-the-same thing!  The first `if and only if' is either referred to as \em MacLane's strictification theorem \em or \em coherence for symmetric monoidal categories \em \cite{MacLane1, CatsII}.  

So what about the concrete category ${\bf   FHilb}$?  As it is an example of a symmetric monoidal category, by the above result, whatever we prove about a strict one, or within the diagrammatic language, will automatically also hold for ${\bf   FHilb}$.  Obviously, one would expect the converse not to hold since we are only considering about a very particular symmetric monoidal category.

However, recently Selinger \cite{SelingerCompleteness} elaborated on an existing result due to Hasegawa, Hofmann and Plotkin \cite{Plotkin} to show that there is in fact a converse statement, provided one adds some extra structure.  

\paragraph{Definition.}  A \em dagger compact (closed) category\em\footnote{Strict dagger
compact (closed) categories appeared in the work of Baez and Dolan \cite{BaezDolan}
as a  special case for n = 1 and k = 3 of  k-tuply monoidal n-categories
with duals.} \cite{AC, AC2, Selinger} is a compact (closed) category ${\bf C}$ together with a \em dagger functor\em, that is, for all $A, B\in|{\bf C}|$ a mapping
\[
\dagger_{A,B}:{\bf C}(A,B)\to {\bf C}(B,A)\,,
\]
which is such that $\dagger_{A,B}$ and $\dagger_{B,A}$ are mutually inverse, and which moreover preserves the composition and the tensor structure, including units and identities.

Graphically, the dagger functor merely flips things upside-down \cite{Selinger, ContPhys}:
\[
\epsfig{figure=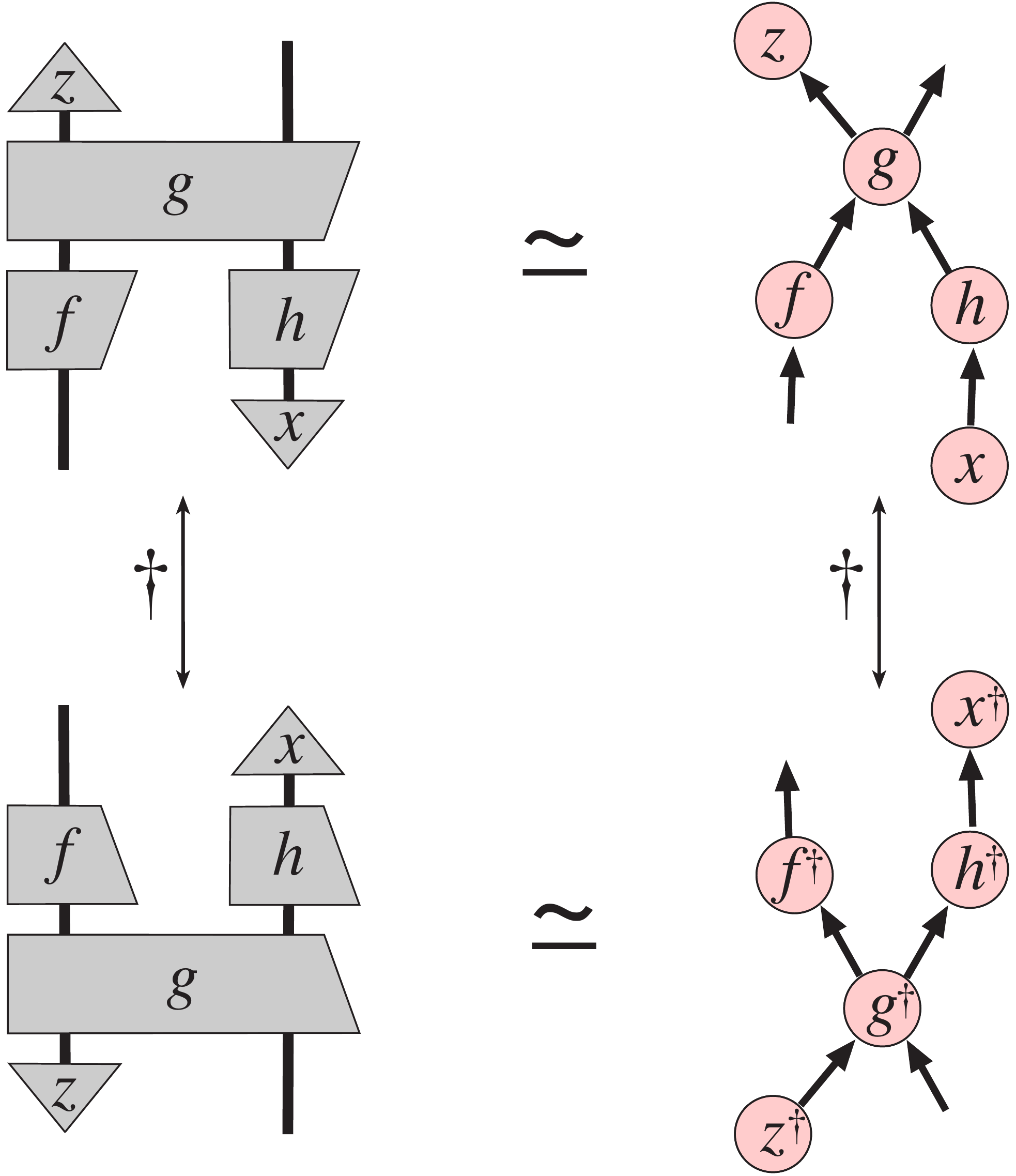,width=160pt} 
\]
In the representation on the left we made the boxes asymmetric to distinguish between a morphism and its dagger.  In the one on the right, since nodes have no a priori orientation in the plane, we used an explicit involution on the symbols.

\paragraph{Theorem.} \em An equation between two scenarios in the language of dagger compact  categories follows from the axioms of dagger compact categories,  {\bf if and only if}, the corresponding equation in the graphical language follows from  isomorphisms of diagrams,  {\bf if and only if}, 
an equation between two scenarios in the language of dagger compact (closed) categories holds in ${\bf   FHilb}$\em.
\bigskip 

Since there are faithful strict monoidal functors which embed ${\sf WP}{\bf   FHilb}$ as well as ${\sf CP}{\bf   FHilb}$ within ${\bf   FHilb}$ the correspondence with the diagrammatic language also carries over to these models.

The question whether we can carry this through for richer languages than that of dagger compact categories remains open.  Still, the language of dagger compact  categories already captures many important concepts: trace, transpose, conjugate, adjoint, inner-product, unitarity, (complete) positivity \cite{ContPhys}.

While admittedly, the conceptual significance of the dagger is still being discussed,\footnote{Also mathematically, there are some issues with the dagger to which some refer as `Evil'.  The main problem is that the structure of the dagger, in particular its strict action on objects, is not preserved under so-called \em categorical equivalences\em.  This has been the subject of a recent long discussion on the categories mailing list involving all the big shots of the area.}  besides \em compactness\em,  the dagger is what truly gives a theory it's \em relational character\em.  In particular, it is a key property of the category ${\bf FRel}$ which the category ${\bf FSet}$ fails to admit: each relation has a converse relation.  This is also the reason why ${\bf FRel}$ and ${\bf FHilb}$ are so alike in terms of their categorical structure, while ${\bf FRel}$ and ${\bf FSet}$ are very different in terms of categorical structure, despite the fact that ${\bf FRel}$ and ${\bf FSet}$ have the same objects, the same compositions, and that the morphisms of ${\bf FSet}$ are a subset of those of ${\bf FRel}$.\footnote{A detailed analysis of the similarities between ${\bf FRel}$ and ${\bf FHilb}$ and the differences between ${\bf FRel}$ and ${\bf FSet}$ is in \cite{CatsII}. To mention one difference: in ${\bf FSet}$ the Cartesian product behaves like a non-linear conjunction while in ${\bf FRel}$ it behaves like a linear conjunction.}   Intuitively, the reason for this is that linear maps (when conceived as matrices) can be seen as some kind of generalized relations, in that they do not just encode whether two things relate, but also in which manner that they relate, by means of a complex number.  In contrast to the many who conceive quantum theory as a generalized probability theory, for us it is rather a theory of generalized relations, the latter now to be taken in its mathematical sense. 

\paragraph{Example: Spekkens' toy qubit theory.} In \cite{SpekkensBis} Spekkens suggested dagger duality as an axiom for a class of theories which would generalize his toy qubit theory \cite{Spekkens}.  The concrete presentation of Spekkens' qubit theory as a dagger compact category ${\bf Spek}$ is in \cite{CE, CES, Edwards}. This presentation enabled a clear comparison with a dagger compact  category ${\bf Stab}$ which encodes stabilizer qubit theory, from which it emerged that the only difference between the toy qubit theory and stabilizer qubit theory is the different group structure of the \em phase groups \em \cite{CES},\footnote{These correspond to the two available four-element Abelian groups, the four-element cyclic group for stabilizer qubit theory and the Klein four group for the toy qubit theory.} a concept introduced by Duncan and the author in \cite{CD}. 

\bigskip
 We are also in a position explain why we restricted to finite sets/dimensions.  While ${\bf Rel}$ is compact closed and has a dagger structure, ${\bf Hilb}$ is neither compact closed nor has a dagger.\footnote{We do obtain a dagger when restricting to bounded linear maps and there are also category-theoretic  technical tricks to have something very similar to compact structure \cite{ABP}.} While some may take this as an objection to the dagger compact structure, we think that the fact that ${\bf Hilb}$ fails to be dagger compact may be an artifact of the Hilbert space structure, rather than a feature of nature. Having said this, we do agree that dagger compactness surely isn't the end of the story.  In particular, we would like to conceive also the dagger as some kind of relation, rather than as an operation on a category as a whole.\footnote{There are manners to do this but we won't go into them here.}

\subsection{Non-isolation in the von Neumann quantum model}\label{sec:environment}

In order to assert that a physical theory includes \em non-isolated \em (i.e~\em open\em) systems for every system $A\in |{\bf C}|$ we considered a designated process $\top_A:A\to \II$ with:
\beq\label{eq:invironmentcoher}
 \qquad\qquad\top_A\otimes\top_B=\top_{A\otimes B}
\qquad \qquad\mbox{and}\quad\top_\II=1_\II\,,
\eeq
which are such  that for all $A\in|{\bf C}|$ the mappings:
\[
\top_A\circ- :{\bf C}(\II, A)\to {\bf C}(\II, \II)
\]
assign the weights of each of these states.  We now present a result which characterizes an additional  condition that these processes have to satisfy relative to a collection of \em isolated \em (or \em closed\em) processes in order that:
\[
{\mbox{open\ processes}\over\mbox{closed\ processes}}
\, \ {\raisebox{-4pt}{$\,\cdot\,$}\over\raisebox{4pt}{$\,\cdot\,$}} \ \,
{\mbox{mixed\ state\ quantum\ theory}\over\mbox{pure\ state\ quantum\ theory}}
\, \ {\raisebox{-4pt}{$\,\cdot\,$}\over\raisebox{4pt}{$\,\cdot\,$}} \ \,
{{\sf CP}{\bf   FHilb}\over{\sf WP}{\bf   FHilb}}\,.
\]
That is, in words, if we know that our theory of closed systems is ordinary quantum theory of closed systems, what do we have to require from the  feed-into-environment processes such that the whole theory corresponds to quantum theory of open systems?  This condition turns out to be non-trival.

Consider a symmetric monoidal category ${\bf C}$ with feed-into-environment processes, that is, for each $A\in|{\bf C}|$ a designated morphism $\top_A:A\to \II$  satisfying (\ref{eq:invironmentcoher}).  Assume that
it contains a   sub symmetric monoidal category ${\bf C}_{pure}$, and we will refer to the morphisms in it as \em pure\em.  By a \em purification \em of a morphism of $f:A\to B$  in ${\bf C}$  we mean a morphism $f_{pure}:A\to B\otimes C$  in ${\bf C}_{pure}$ such that:
\beq\label{eq:purification}
(1_B\otimes \top_C)\circ f_{pure}= f\,.
\eeq
We say that ${\bf C}_{pure}$ \em generates \em ${\bf C}$ whenever each morphism in ${\bf C}$ can be purified. 

\paragraph{Example: purification in probabilistic theories.}
The power of purification as a postulate is exploited by 
Chiribella,  D'Ariano and Perinotti in \cite{DArianoPaper}.

\bigskip 

By combining the results in \cite{Selinger} with those of \cite{SelingerAxiom} we obtain:

\paragraph{Theorem.} If ${\bf C}_{pure}\simeq {\sf WP}{\bf   FHilb}$ generates  ${\bf C}$,  and if for the usual  dagger functor on ${\sf WP}{\bf   FHilb}$ we have for all $f: C\to A$ and $g:C\to B$ in ${\sf WP}{\bf   FHilb}$:
\beq\label{eq:CPMaxcondition}
\top_A\circ f = \top_B\circ g\ \Longleftrightarrow\  f^\dagger\circ f= g^\dagger\circ g\,,
\eeq
then  ${\bf C}\simeq{\sf CP}{\bf   FHilb}$.

\bigskip

More generally, for any pair ${\bf C}$ and ${\bf C}_{pure}$ the conditions (\ref{eq:invironmentcoher}, \ref{eq:purification}) and a slight generalization of (\ref{eq:CPMaxcondition}) together allow one to construct the whole category ${\bf C}$ from morphisms in ${\bf C}_{pure}$ by only using the dagger symmetric monoidal structure, together with a canonical inclusion of in ${\bf C}_{pure}$ within ${\bf C}$, a result which is obtained by combining the results in \cite{Selinger}, \cite{SelingerAxiom} and  \cite{RR}.  
If the category ${\bf C}$ is compact closed, as it is the case for ${\sf WP}{\bf   FHilb}$,  then  (\ref{eq:CPMaxcondition}) does suffice.  In this case, following Selinger in \cite{Selinger}, the open processes in the constructed category ${\sf CP}{\bf C}$ all take the form:
\[
\epsfig{figure=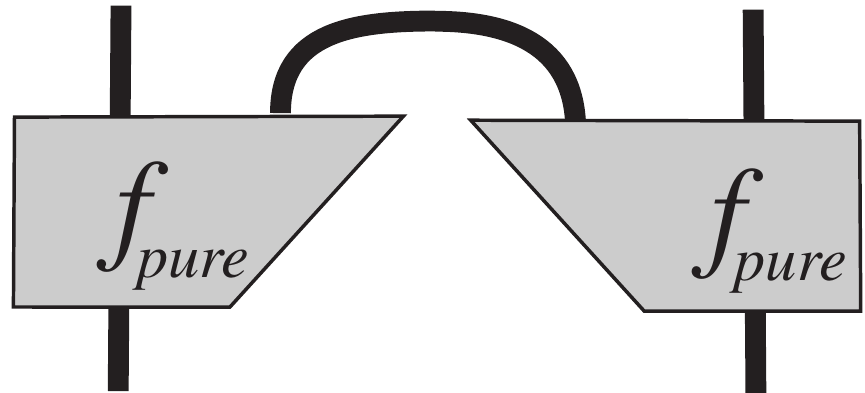,width=80pt}
\]
for some  $f_{pure}:A\to B\otimes C$ in ${\bf C}_{pure}$, where the left-right reflection represents the composite of  the dagger and \em transposition\em, explicitly:
\[
\epsfig{figure=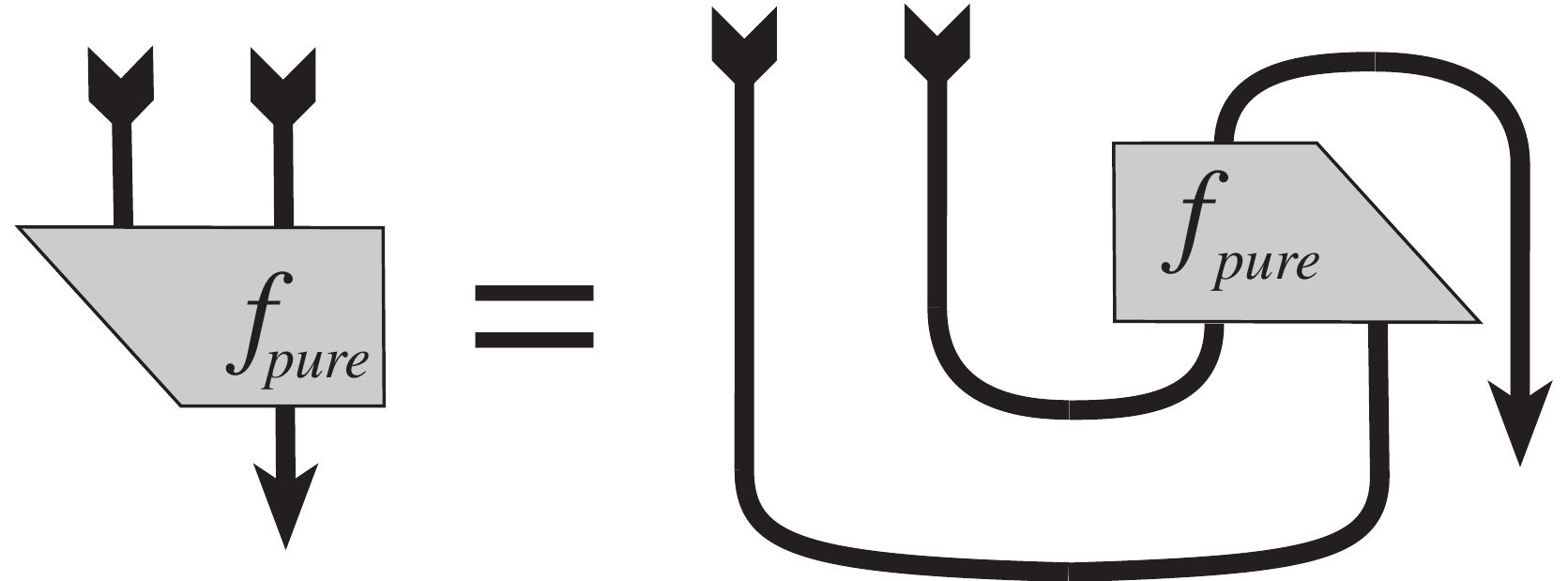,width=150pt}
\]
In ${\sf WP}{\bf   FHilb}$ this is nothing but complex conjugation (see \cite{ContPhys} for more details on this).
The subcategory of pure processes consists of those of the form:
\[
\epsfig{figure=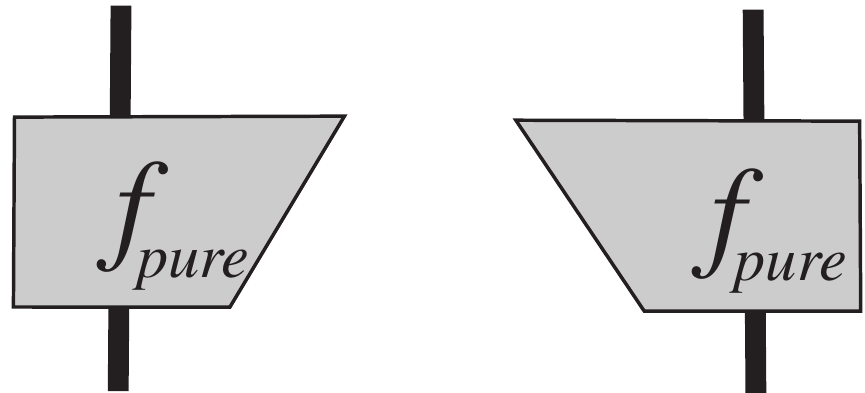,width=80pt}
\]

Graphically, condition (\ref{eq:CPMaxcondition}) can then be rewritten as:
\[
\epsfig{figure=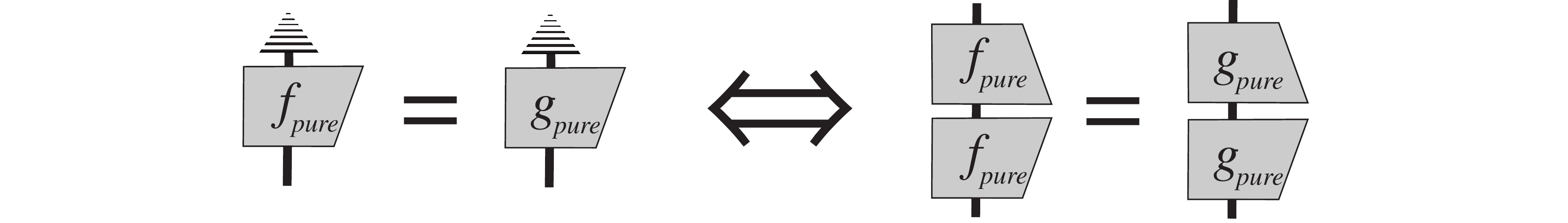,width=330pt}
\]
and from it immediately follows, setting $g:=1_A$, that:
\[
\epsfig{figure=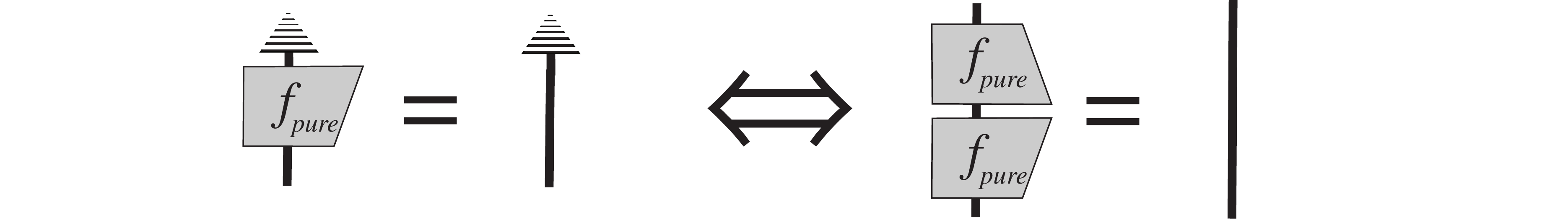,width=330pt}
\]
Calling processes $f_{pure}$ which obey 
\[
f_{pure}^\dagger\circ f_{pure}=1_A
\]
\em isometries\em, it then follows that isometries are exactly those processes which leave the feed-into-environment processes invariant.

Condition (\ref{eq:CPMaxcondition}) can in the compact case be equivalently presented as:
\[
\epsfig{figure=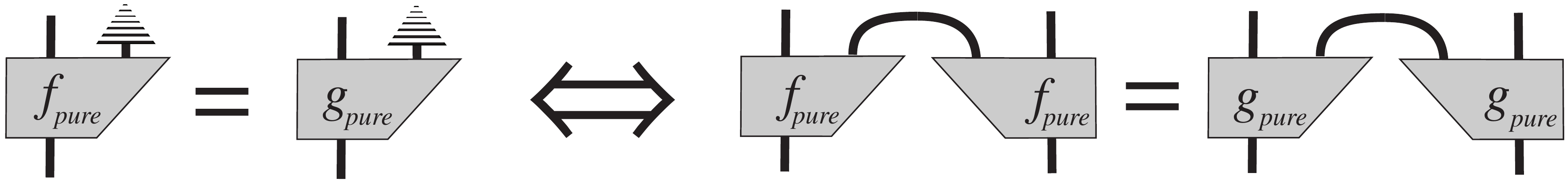,width=330pt}
\]
which provides a direct translation between Selinger's presentation of ${\sf WP}{\bf   FHilb}$ and one which relies on the feed-into-environment processes.  The more general form of the non-compact case mentioned above is now obtained by `undoing' all compact morphisms which requires introduction of symmetry morphisms:
\[
\epsfig{figure=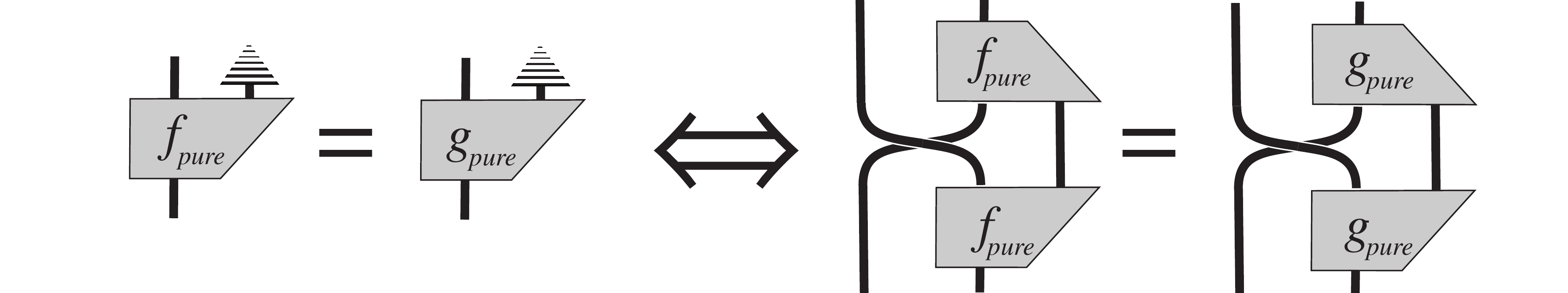,width=330pt}\,.
\]

Importantly, both $({\sf WP}){\bf   C}$ and ${\sf CP}{\bf   C}$ are symmetric monoidal (and compact) if ${\bf C}$ is,  so ${\sf CP}{\bf   C}$ admits a graphical language in its own right without reference to the underlying 
symmetric monoidal category ${\bf   C}$.

\subsection{Non-isolation and causality}

If we restrict to processes that `happen with certainty' then, as shown by Chiriballa, D'Ariano and Perinotti, uniqueness of a deterministic effect 
enforces \em causality \em in the sense that 
states of compound systems have well-defined marginals \cite{DArianoPaper}. In category-theoretic terms, this uniqueness means that $\II$ is \em terminal\em, that is, for each object $A$ there is a unique morphism of type $A\to\II$, which will then play the role of $\top_A$.   It then immediately follows that  
\[
\top_{A\otimes B}=\top_A\otimes \top_B\,,
\]
and hence, that there are no entangled effects.  The manner  in which: 
\bit
\item connectedness in graphical calculus as expressing causal connections, and 
\item this notion of causality in terms of a terminal object 
\eit
are related is currently being explored in collaboration with Ray Lal  \cite{CauCat}.

\section{Classicality and measurement}\label{sec:class}

For us, a classicality entity is one for which there are no limitations 
for to be \em shared \em among many parties, that is, using  quantum information terminology, which can be \em broadcast \em  \cite{Broadcast}.  
It is witnessed  by a collection of processes which establish this sharing/broadcasting.   
To give an example, while an unknown quantum state cannot be cloned \cite{Dieks, WZ}, this scientific fact itself is of course available to every individual of the scientific community, by means of writing a paper about it and distributing copies of the journal in which it appears. 

So our notion of classicality makes no direct reference 
to anything `material', but to the ability of a logical flow of information to admit `branching'.  In relational terms, it will be witnessed by the relation that identifies the branches as being identical.  The power of this idea for describing quantum information tasks is discussed in a paper with Simon Perdrix \cite{CPer}, where it is also discussed that \em decoherence \em
can be seen as a material embodiment of this idea. 

Our explorations have indeed made us realize that rather than starting from a classical theory which one subjects to a \em quantization \em procedure in order to produce a theory which can describe quantum systems, one obtains an elegant compositional mathematical framework when, starting from a `quantum' universe of processes,  one identifies classicality  in this manner.
Put in slightly more mathematical terms, citing John Baez in TWF 268 \cite{TWF268} on our work:
\begin{quote}
``Mathematicians in particular are used to thinking of the quantum world as a mathematical structure resting on foundations of classical logic: first comes set theory, then Hilbert spaces on top of that. But what if it's really the other way around? What if classical mathematics is somehow sitting inside quantum theory? The world is quantum, after all.''
\end{quote}
This idea of ``classical objects living in a quantum world governed by quantum rules'' was introduced by Pavlovic and myself in \cite{CPav} and further elaborated on in \cite{CPV, CPaq, CPaqPav}. In terms of symmetric monoidal categories, we are speaking the language of certain kinds of so-called \em internal Frobenius algebras \em \cite{CarboniWalters, CPav, CPV}.  
 Interestingly, these Frobenius algebras appeared first in the literature in Carboni and Walter's axiomatization of the category ${\bf Rel}$ \cite{CarboniWalters}.

\subsection{Classicality}

Below, when we (slightly abusively) denote several systems by the same symbol, we think of them as distinct identical  systems and not as the same one.

The processes which establishes the shareability that is characteristic for classicality implements an `equality' between the distinct instances of that entity, and we depict them in a `spider-like'  manner:
\[
\Xi_{n,m}\ \equiv\  \raisebox{-0.55cm}{$\underbrace{\overbrace{\epsfig{figure=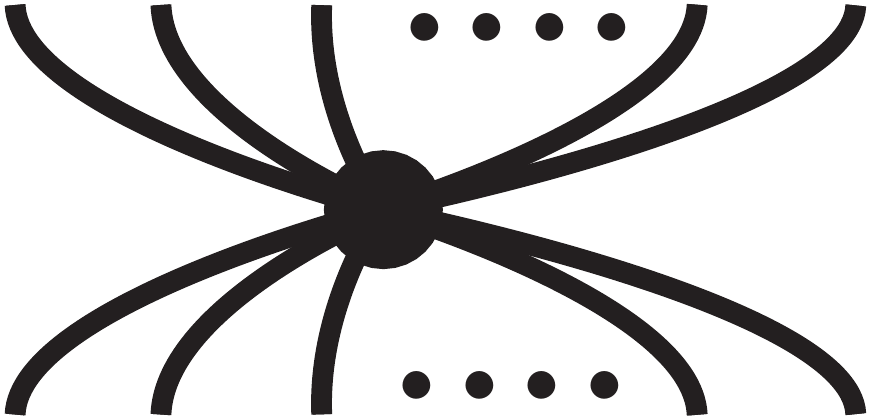,width=75pt}}^{\scriptstyle{m}}}_{\scriptstyle{n}}$}\ :\
\underbrace{X\otimes \ldots\otimes X}_{\scriptstyle{n}}\to \underbrace{X\otimes \ldots\otimes X}_{\scriptstyle{m}}
\]
By transitivity of equality it immediately follows that:
\[
\epsfig{figure=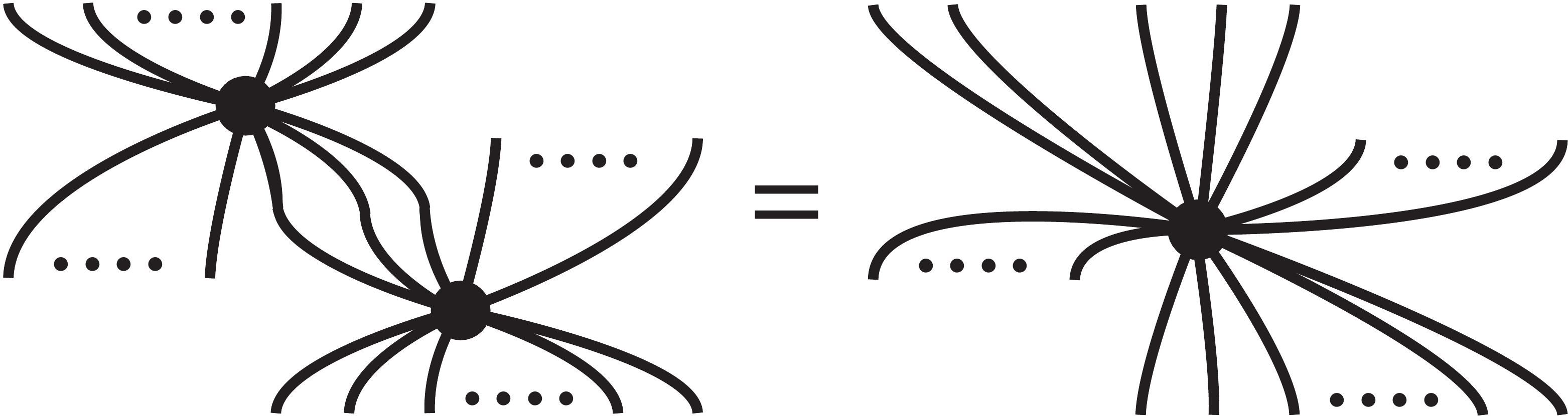,width=304pt}\,.
\]

One may distinguish two kinds of classicality. The more restrictive first kind, called \em controlled \em (or \em closed \em or \em pure\em) requires sharing to be within the domain of consideration.  This is explicitly realized by:
\[
\Xi_{1,1} = 1_X  \qquad\qquad\mbox{i.e.}\qquad\qquad
\raisebox{-5pt}{\epsfig{figure=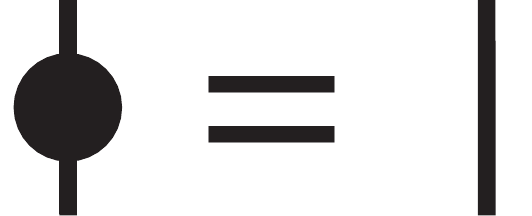,width=56pt}}\,,
\]

The second kind of classicality, called \em uncontrolled \em (or \em open \em or \em mixed\em) allows  sharing to be outside our domain of consideration, 
which in the light of the above composition rule is realized by:
\[
\Xi_{1,0}^o = \top_X  \qquad\qquad\mbox{i.e.}\qquad\qquad
\raisebox{-5pt}{\epsfig{figure=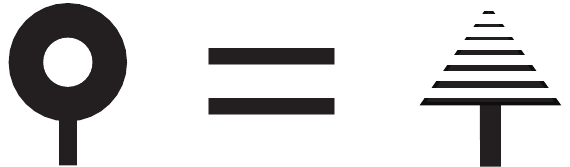,width=60pt}}\,, 
\]
since then we obain: 
\[
(\top_X\otimes 1_{X\otimes \ldots\otimes X})\circ\Xi_{n,m}^o=\Xi_{n,m-1}^o\ \ \ \ \mbox{i.e.} \ \ \raisebox{-20pt}{\epsfig{figure=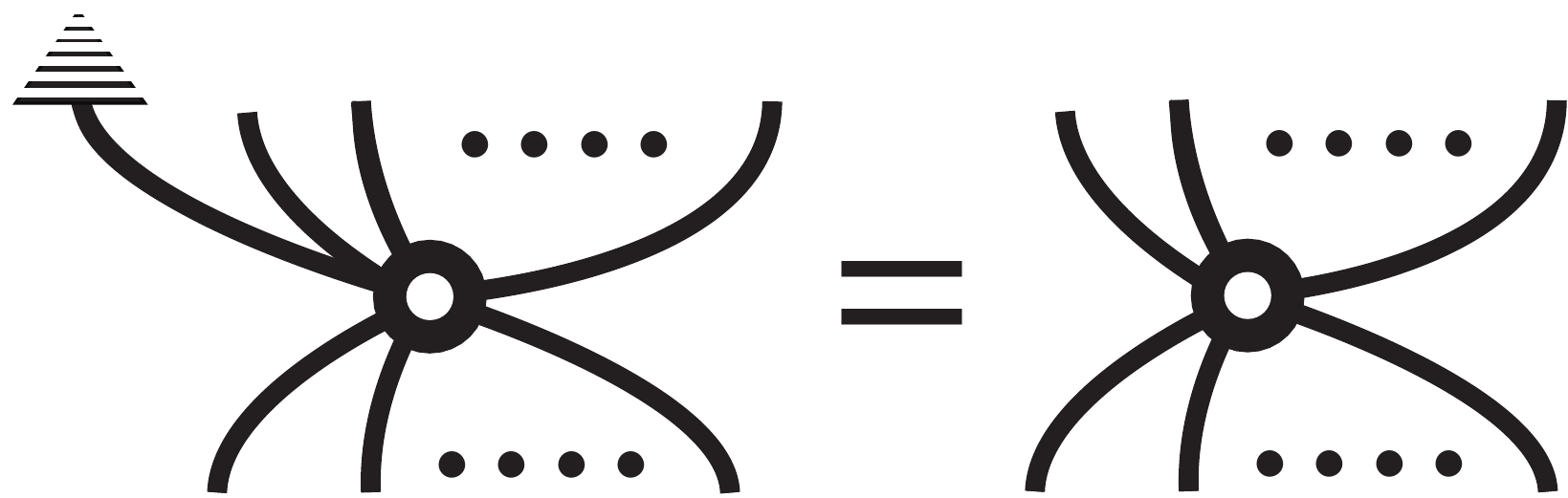,width=160pt}},
\]
i.e.~uncontrolled sharing is invariant under feed-into-environment processes.  

These two forms of classicality may be naturally related to each other by introducing feed-into-environment processes within the closed spiders, or dually put, by considering closed spiders as purifications of the open ones:
\[
(\top_X\otimes 1_{X\otimes \ldots\otimes X})\circ\Xi_{n,m}=\Xi_{n,m-1}^o\ \ \ \ \mbox{i.e.} \ \ \raisebox{-20pt}{\epsfig{figure=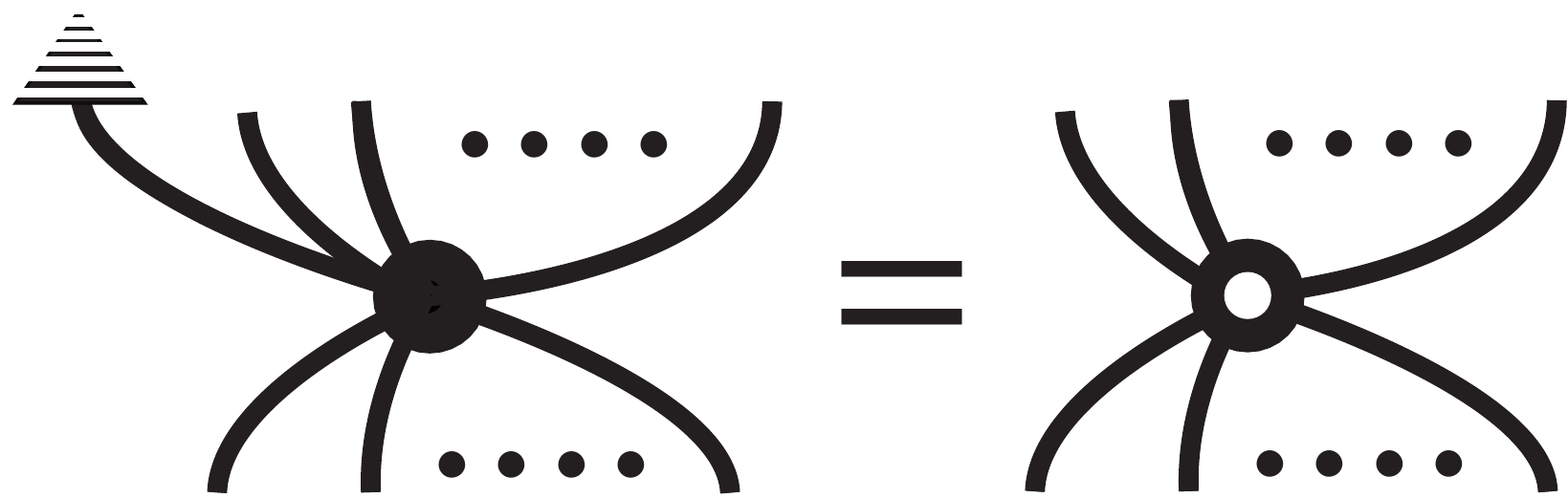,width=160pt}}.
\]
which can in fact be summarized as the following two equations:
\[
\epsfig{figure=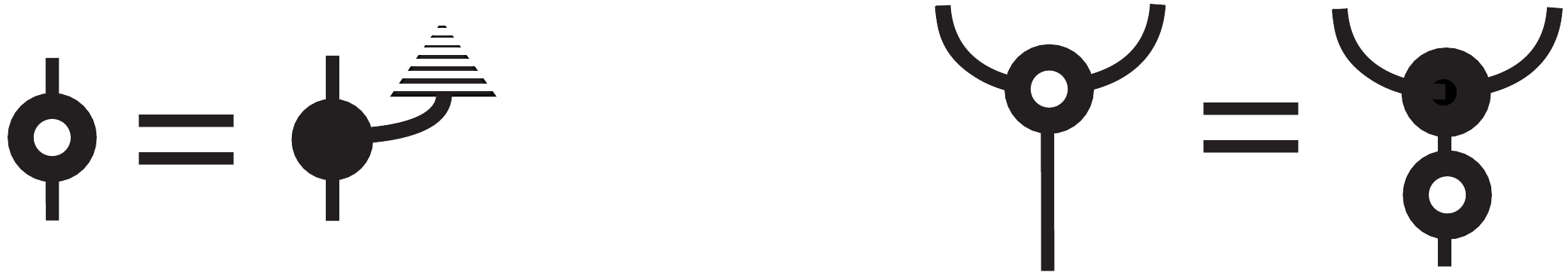,width=220pt}
\]

We can identify some special examples:
\bit
\item erasing := $\Xi_{1,0}\ \equiv\ \raisebox{-5pt}{\epsfig{figure=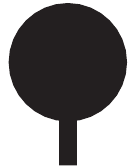,width=13pt}}$
\item cloning := $\Xi_{1,2}\ \equiv\ \raisebox{-10pt}{\epsfig{figure=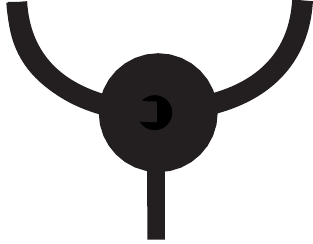,width=31pt}}$
\item correlating := $\Xi_{0,2}\ \equiv\ \raisebox{-6pt}{\epsfig{figure=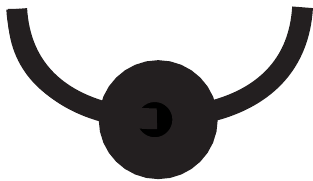,width=31pt}}$
\item comparing := $\Xi_{2,0}\ \equiv\ \raisebox{-6pt}{\epsfig{figure=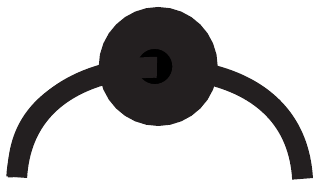,width=31pt}}$
\item matching := $\Xi_{2,1}\ \equiv\ \raisebox{-10pt}{\epsfig{figure=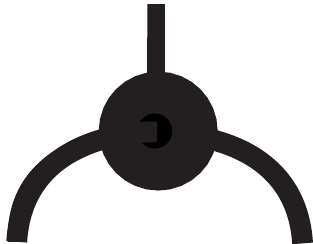,width=31pt}}$
\item either := $\Xi_{0,1}\ \equiv\ \raisebox{-5pt}{\epsfig{figure=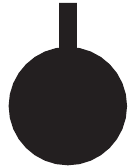,width=13pt}}$
\eit

Conceptually, it is more than fair to cast doubt on physical meaningfulness of closed spiders. To see this it suffices to consider the erasing operation in the light of Landauer's principle \cite{Landauer, Maroney}.
Also, when thinking of a cloning operation then one usually would assume some ancillary state onto which one clones, and this ancillary state by the very definition of state is an open process:  
\[
\epsfig{figure=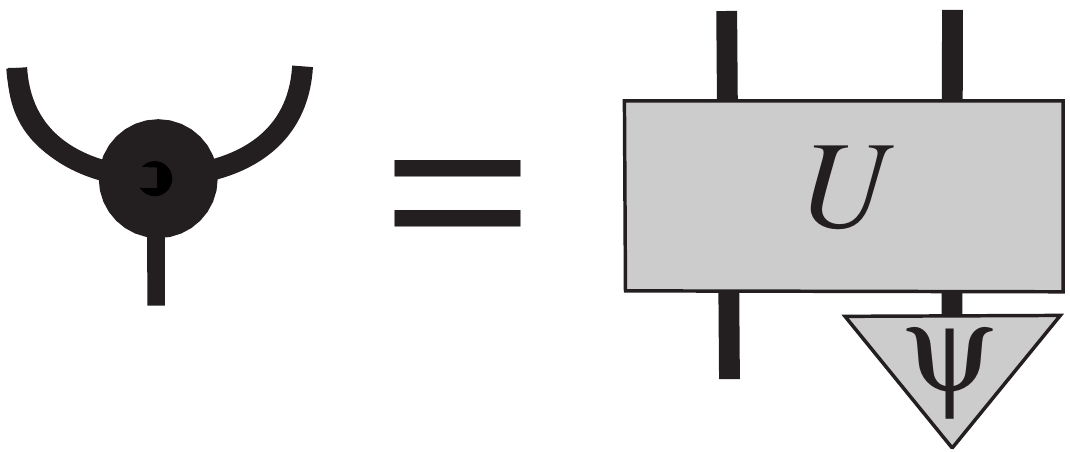,width=108pt}
\]
Still it is useful to retain the closed spiders as an idealized concept given that their behavioral specifications,  
exactly matches the well-understood mathematical gadget of commutative  Frobenius algebras 
(see Section \ref{sec:frobalg}).

A particularly relevant open spider is the purification of copying:
\bit
\item broadcasting := $\Xi_{1,1}^o\ \equiv\ \raisebox{-9pt}{\epsfig{figure=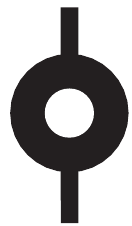,width=13pt}}$
\eit
While it has the type of an identity it is genuinely non-trivial. 

\paragraph{Example: cloning and broadcasting in quantum information.} A hint why the above may indeed characterize classicality on-the-nose comes from the \em no-cloning theorem \em \cite{Dieks, WZ} which states that the only quantum states which can be \em copied \em by a single operation have to be orthogonal. Maximal sets of these jointly copy-able states make up an orthonormal basis i.e.~a (pure) classical  `slice' of quantum theory. Similarly, the \em no-broadcasting theorem \em \cite{Broadcast} states that the only quantum states which can be \em broadcast \em by a single operation correspond to a collection of density matrices that are diagonal in the same orthonormal  basis.  This table summarizes cloneability/broadcastability:
\begin{center}
{\footnotesize\begin{tabular}{c|c|c|c|c|}
 & pure classical  & mixed classical &  pure quantum  & mixed quantum \\
\hline
broadcastable: &   {\tt yes}   &   \underline{\tt YES} &   {\tt no}   &   {\tt no}  \\
\hline
cloneable: &   {\tt yes}   &   \underline{\tt NO} &   {\tt no}   &   {\tt no}  \\
\hline
\end{tabular} }
\end{center}
Conversely, for an orthonormal basis
$\{|i\rangle\}_i$ of ${\cal H}$ the corresponding broadcast operation is the following completely positive map:
\[
|i\rangle\langle j| \mapsto \delta_{ij} |i\rangle\langle i|\,.
\]
Clearly, this completely positive map totally destroys coherence, hence broadcasting is physically embodied by \em decoherence\em.  Decoherence can indeed be seen as `sharing with  (cf.~coupling to) the environment'.

\bigskip

Given that in quantum information both copying and broadcasting enable to characterize an orthonormal  basis, the question then remains to define candidate cloning/broadcasting operations in a manner that there is a one-to-one correspondence between such operations and orthonormal bases.  That is exactly what we did above, as the theorem stated below confirms.

\paragraph{Theorem.} In $({\sf WP}){\bf FHilb}$ the above defined families 
\[
{\cal X}=\{\Xi_{n,m} \mid n,m\in\mathbb{N}\}
\]
of closed spiders are in bijective correspondence with orthogonal bases. 
If we moreover have that $\Xi_{n,m}= \Xi_{m,n}^\dagger$ for all $n, m$ then this basis is orthonormal.

\bigskip

To show this we need to combine Steve Lack's (highly abstract) account on spiders \cite{Lack} (of which a more accessible direct presentation is in  \cite{CPaq}) with a result obtained by Pavlovic, Vicary and myself \cite{CPV} (see Section \ref{sec:frobalg}).

This result states that all non-degenerate observables can indeed be bijectively represented by these sharing processes.  Now we establish that spiders are also expressive enough to associate a corresponding `classical slice' of the universe of all processes  to each family ${\cal X}$.  We \em assert classicality of a process \em by imposing \em invariance under broadcasting/decoherences \em \cite{CPaqPav}.\footnote{This  is akin to Blume-Kohout and Zurek's \em quantum Darwinism \em \cite{BlumeZurek, Zurek}.}  The copyability of pure classical data can be used to assert deterministic processes. We conveniently set $o_{\cal X}=\Xi_{1,1}^o\in{\cal X}$ and $\delta_{\cal X}=\Xi_{1,2}\in{\cal X}$. 

\paragraph{Definition.} A  \em classical process \em is a process of the form:
\[
o_{\cal Y}    \circ f\circ o_{\cal X}\ \equiv \ \raisebox{-28.5pt}{\epsfig{figure=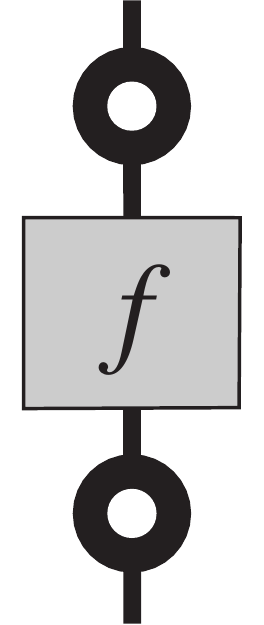,width=26pt}}
\]
where $f: X\to Y$ can be  an arbitrary process. Evidently, classical processes can be equivalently defined as processes
$f:X\to Y$ which satisfy:  
\[
o_{\cal Y}    \circ f\circ o_{\cal X}=f\qquad\mbox{that is}\qquad
\raisebox{-28.5pt}{\epsfig{figure=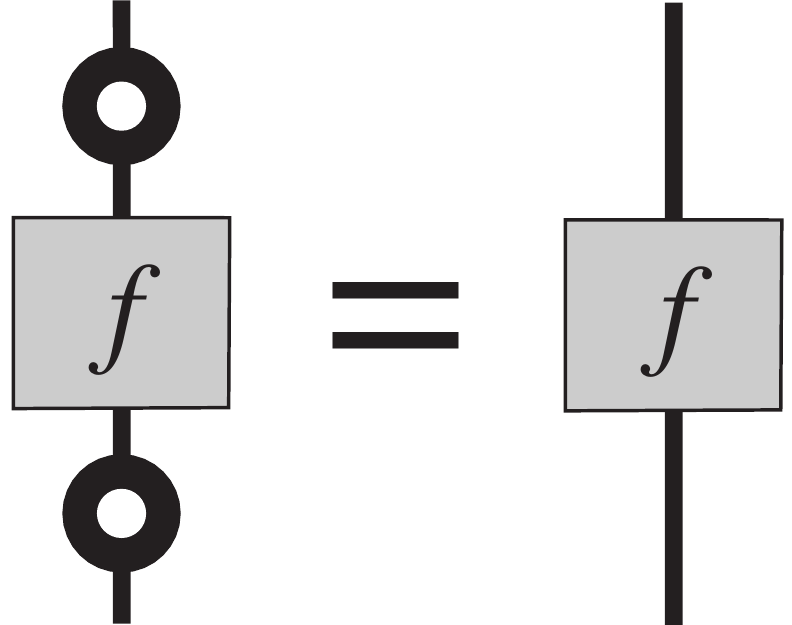,width=78pt}}\,.
\]
Such a classical process is \em normalized \em or \em stochastic \em if we have:
\[
\top_Y    \circ f =\top_X\qquad\mbox{that is}\qquad
\raisebox{-16.5pt}{\epsfig{figure=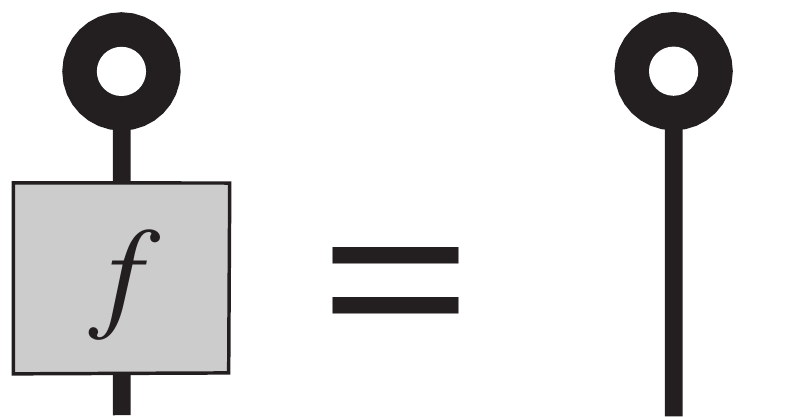,width=78pt}}\!\!\!\!\,.
\]
and it is deterministic if we moreover have that:
\[
\delta_{\cal Y}\circ f= (f\otimes f)\circ\delta_{\cal X}\qquad\mbox{that is}\qquad
\raisebox{-16.5pt}{\epsfig{figure=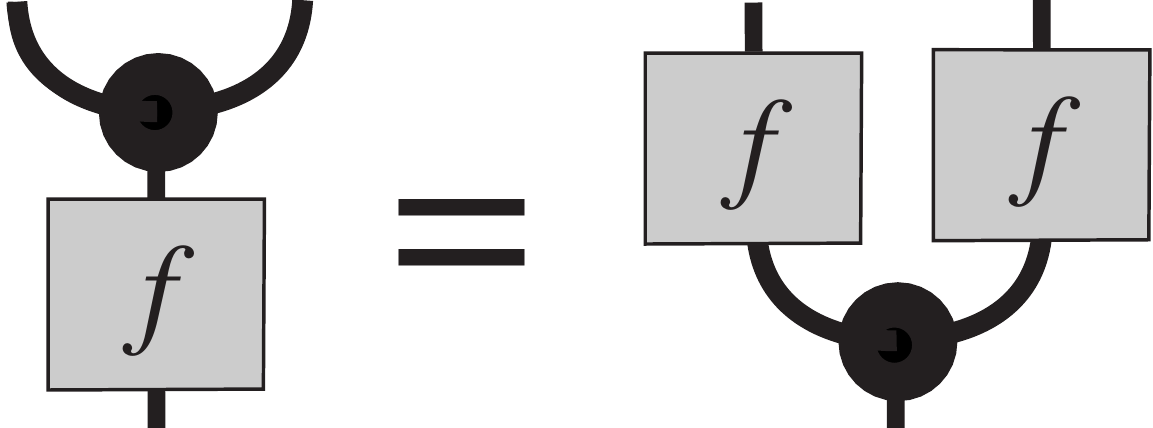,width=114pt}}\,.
\]
\medskip

\paragraph{Theorem.} In ${\sf CP}{\bf FHilb}$ normalized classical processes exactly correspond to the usual notion of stochastic maps, that is, matrices with positive real entries such that all columns add up to one, and deterministic processes correspond to functions, that is, matrices with exactly one 1-entry in each column.

\bigskip
This result was shown by Paquette, Pavlovic and myself in \cite{CPaqPav} where many other species of classical processes (doubly stochastic, partial processes, relations, ...) are defined in a similar manner.\footnote{We adopted Carboni and Walters' axiomatization of the category of relations \cite{CarboniWalters}, which also involved introducing the \em Frobenius law\em, to the probabilistic and the quantum case.}

\paragraph{Challenge.} Develop the above without any reference to closed spiders.

\subsection{Measurement}

Once we have identified these classical entities (provided they exist  at all), we 
may wish to represent general processes relative to this entity.  This is obviously what \em observables \em (or \em measurements\em) in quantum theory aim to do.  One thing we know for a fact is that it is not possible to represent the whole universe of processes by means of such an entity.  So what is the best we could aim for?

By a \em probe \em we mean a process 
\[
(1_B\otimes o_{\cal X})\circ m:A\to B\otimes X\qquad\mbox{that is}\qquad
\raisebox{-16.5pt}{\epsfig{figure=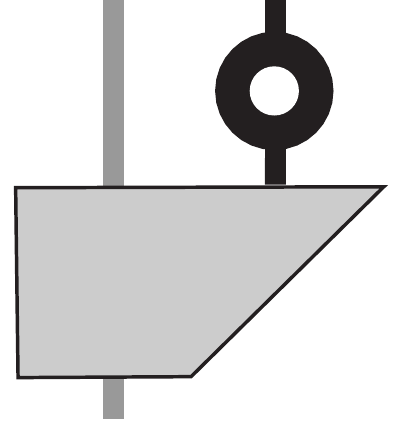,width=40pt}}\,.
\]
We call it a \em non-demolition \em probe if $A$ and $B$ are identical systems, which we denote by setting $A=B$, and we call it a \em demolition \em probe if $B$ is $\II$.  
By a \em von Neumann probe \em we mean a non-demolition one which is such that:
\beq\label{eq:meas1}
(m\otimes 1_X)\circ m = (1_A\otimes \delta_{\cal X})\circ m\ \ \mbox{that is}\ \
\raisebox{-22pt}{\epsfig{figure=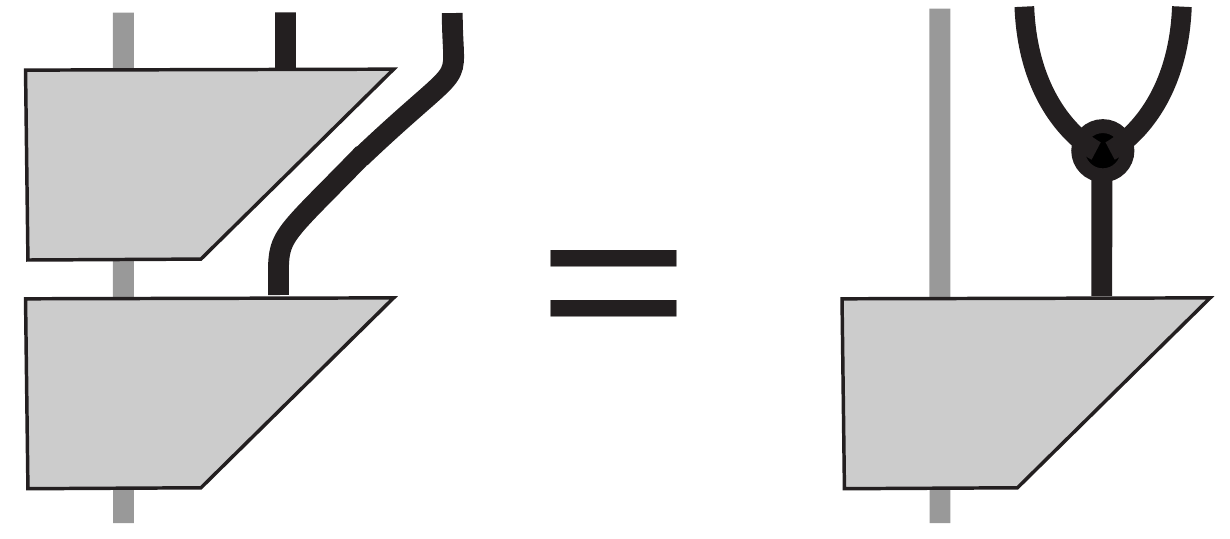,width=125pt}}
\eeq
where we used colors to distinguish the systems $A$ and $X$ for the reader's convenience.  There is a very straightforward interpretation to (\ref{eq:meas1}):
\bit
\item Applying the same probe twice, is equal to applying it  once and then copying the output.  More intuitively put,  the $A$-output after the first application 
is such that the probe produces the same $X$-output (and also the same $A$-output) after a second application.  This means that there is strict relationship between the $A$-output and the $X$-output for that probe.
\eit
Set  $\smallfrown_{\cal X}=\Xi_{2,0}\in{\cal X}$ and $e_{\cal X}=\Xi_{1,0}\in{\cal X}$.

\paragraph{Theorem.} In $({\sf WP}){\bf FHilb}$  von Neumann probes exactly correspond with spectra of \em mutually orthogonal idempotents \em $\{{\rm P}_i\}$, that is:
\[
{\rm P}_i\circ {\rm P}_j=\delta_{ij}\cdot{\rm P}_i\,.
\] 
If we moreover have that this probe is \em self-adjoint\,\em: 
\beq\label{eq:meas2}
m^\dagger = (1_A\otimes \delta_{\cal X})\circ (1_A\otimes \smallfrown_{\cal X})\quad\mbox{that is}\quad
\raisebox{-12pt}{\epsfig{figure=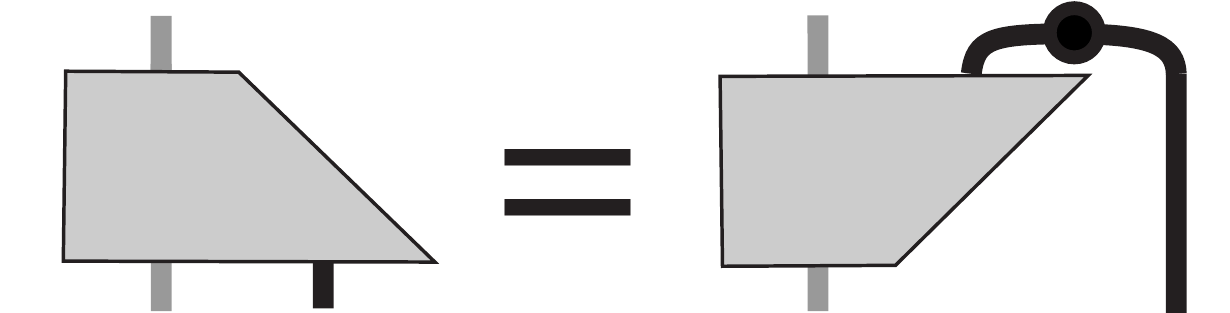,width=125pt}}
\eeq
then these idempotents are \em orthogonal \em projectors, that is:
\[
{\rm P}_i^\dagger={\rm P}_i\,, 
\]
and if: 
\beq\label{eq:meas3}
(1_A\otimes e_{\cal X})\circ m = 1_A\qquad\mbox{that is}\qquad
\raisebox{-12pt}{\epsfig{figure=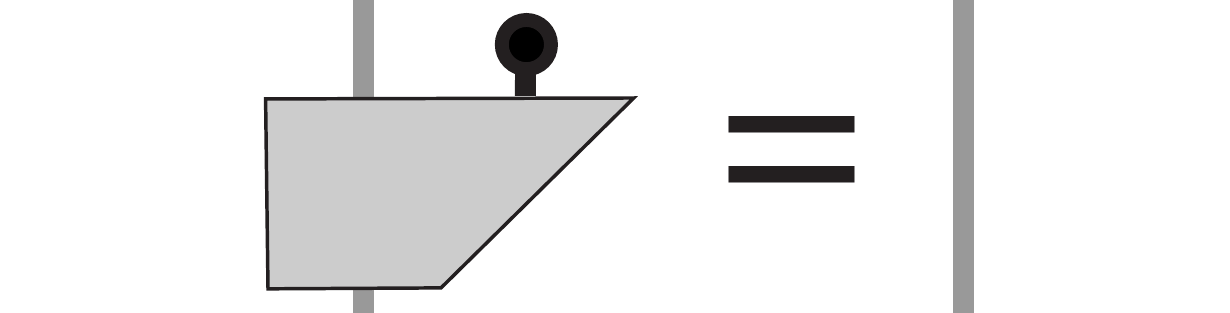,width=125pt}}\hspace{-0.8cm}
\eeq
then this spectrum is exhaustive i.e.~$\sum_i{\rm P}_i=1_{\cal H}$.

\bigskip

This result was shown by Pavlovic and myself in \cite{CPav}.  

\paragraph{Challenge.} Develop the above without any reference to closed spiders.

\paragraph{Challenge.} Express the Geneva School's properties within this framework.

\subsection{Classicality in the von Neumann quantum model}

We now discuss how classicality fits within the model of the environment (i.e.~open systems) of Section \ref{sec:environment}. First note that classicality in the above sense automatically yields \em self-dual \em (i.e.~$X=X^*$) compactness:
\[
\epsfig{figure=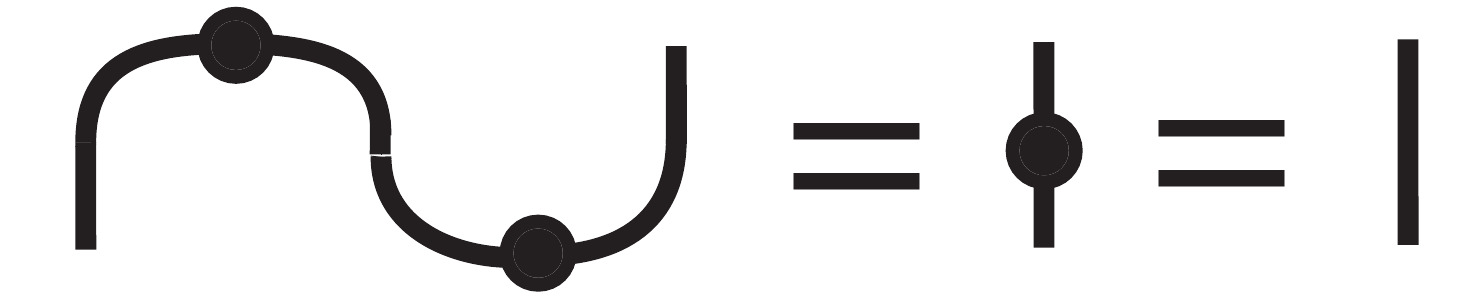,width=140pt}
\]
In ${\sf CP}{\bf C}$ decoherences $o_{\cal X}$ take the following shape:
\[
\epsfig{figure=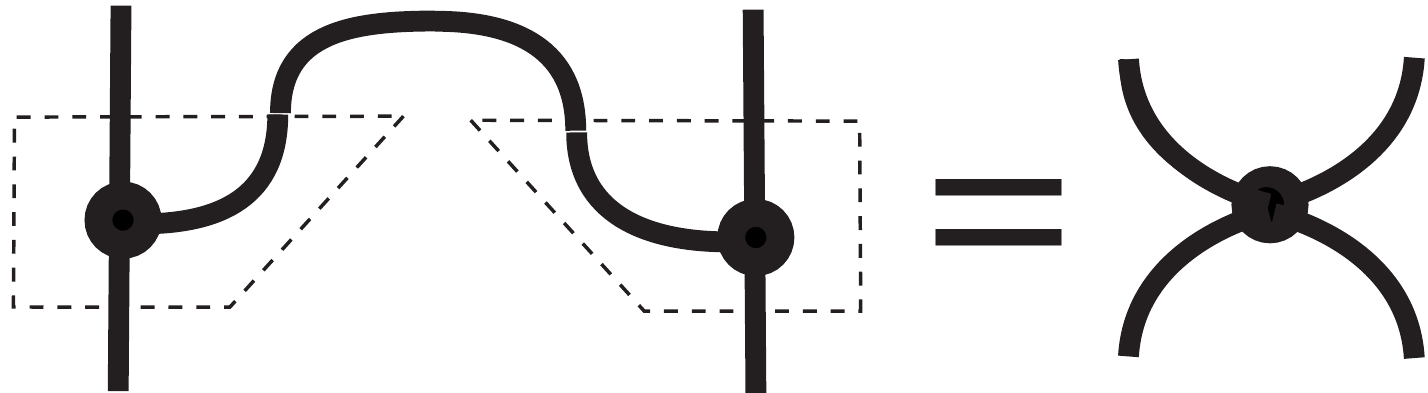,width=135pt}
\]
and consequently classical operations take the shape:
\[
\epsfig{figure=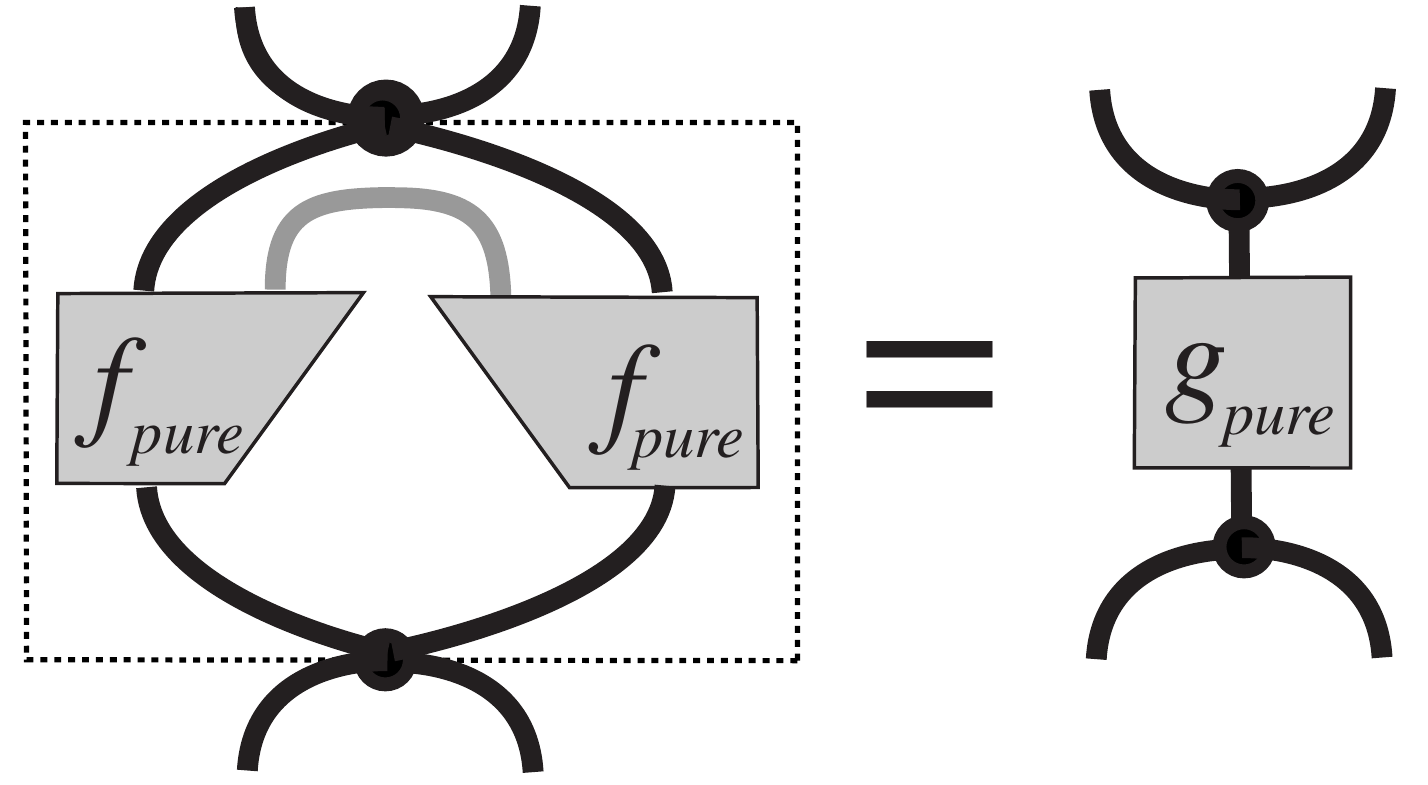,width=135pt}
\]
 for some process $g:X\to Y$.  Hence we obtain:
 \[
{\mbox{classical}\over\mbox{\bR non-classical\e}}
\, \ {\raisebox{-4pt}{$\,\cdot\,$}\over\raisebox{4pt}{$\,\cdot\,$}} \ \,
{\mbox{one\ wire}\over\mbox{\bR two\ wires\e}}\,.
\]
This fact seems to be closely related to Hardy's axiom $K=N^2$ \cite{Hardyaxioms}, which in turn is closely related to Barrett's \em local tomography \em assumption  \cite{Barrett}.

The interpretation of (\ref{eq:meas3}) in the light of (\ref{eq:meas1}) and (\ref{eq:meas2}) is intriguing:
\[
\epsfig{figure=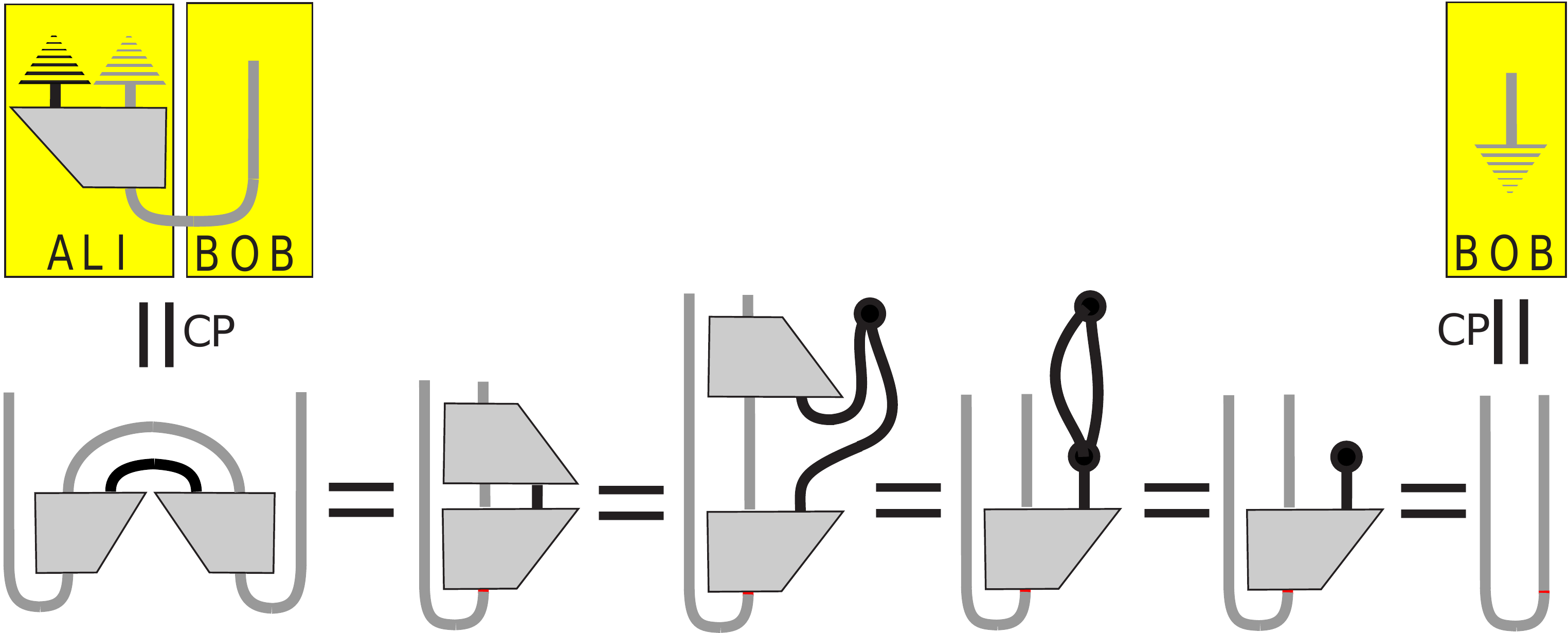,width=340pt}
\]
The left upper picture articulates a protocol where Ali and Bob share a Bell-state and Ali performs a measurement on it.  We are interested in the resulting state at Bob's end, and therefore we feed Ali's outputs into the environment.  Using (\ref{eq:meas1}),  (\ref{eq:meas2}) and in particular also (\ref{eq:meas3}) within the ${\sf CP}$-representation it then follows that what Bob sees is the dagger of a feed-into-environment process, that is, a maximally mixed state.  Hence this protocol provides Bob with no knowledge whatsoever, hence not violating no-faster-than-light-signaling.

\paragraph{Challenge.} Develop the above without any reference to closed spiders.  Then relate this to Hardy's and D'Arianio's research programs.

\subsection{The algebra of classical behaviors}\label{sec:frobalg}

The above seems to have very little to do with the structures we usually encounter in mathematics, and quantum theory in particular.  We will now relate it to `semi-familiar' mathematical structures, which are the ones that enable to establish the relation with orthonormal bases mentioned above.

\subsubsection{Commutative monoids and commutative comonoids}

A \em commutative monoid \em is a set $A$ with a binary map
\[
-\bullet-: A\times A\to A
\]
which is commutative, associative and unital i.e
\[
(a\bullet b)\bullet c=a\bullet(b\bullet c)\qquad
a\bullet b=b\bullet a\qquad a\bullet 1=a\,.
\]
In other words, which may appeal more to the physicist, it is a group without inverses.  Note  that we could also define a monoid as a one object category.  Slightly changing the $\bullet$-notation to
\[
\mu: A\times A\to A
\]
for which we now have
\[
\mu(\mu(a,b),c)=\mu(a,\mu(b,c))\qquad
\mu(a,b)=\mu(b,a)\qquad\mu(a, 1)=a\,,
\]
enables us to write these conditions in a manner that makes no reference anymore to the elements $a, b, c\in A$, namely:
\[
\mu\circ(\mu\times 1_A)=\mu\circ(1_A\times \mu)\qquad \mu=\mu\circ\sigma \qquad \mu\circ(1_A\times u)=1_A
\]
with:
\[
\sigma:A\times A\to A\times A:: (a,b)\mapsto (b,a)
\qquad\quad
u:\{*\}\to A:: *\mapsto 1
\]
where $\{*\}$ is any singleton.  This perspective emphasizes how the map $\mu$ `interacts' with itself, as opposed to how it `acts' on elements, which clearly brings us closer to the process view advocated in this paper.  

This change of perspective also allows us to define a new concept merely by reversing the order of all compositions and types. Concretely, a \em cocommutative comonoid \em  is a set $A$ with two  maps:
\[
\delta: A\to A \times A \qquad\quad\mbox{and}\qquad\quad e:A\to \{*\}
\]
which is cocommutative, coassociative and counital i.e.:
\[
(\delta\times 1_A)\circ\delta=(1_A\times\delta)\circ\delta\qquad \delta=\sigma\circ\delta \qquad (1_A\times e)\circ\delta=1_A\,.
\]

Obviously there are many well-known examples of monoids, typically monoids with additional structure e.g.~groups.
Another one is the two-element set $\{0,1\}$ equipped with the `and'-monoid:
\[
\wedge:\{0,1\}\times \{0,1\}\to\{0,1\}::
\left\{\begin{array}{l}
(0,0)\mapsto 0\\
(0,1)\mapsto 0\\
(1,0)\mapsto 0\\
(1,1)\mapsto 1
\end{array}
\right.\qquad
u_\wedge:\{*\}\to\{0,1\}::*\mapsto 1\,.
\]
Given that from the above perspective monoids and comonoids are very similar things, why do we never encounter comonoids in a standard algebra textbook?  Let us first look at an example of such a comonoid, just to show that such things do exists.  Let $X$ be a any set and 
\[
\delta:X\to X\times X::x\mapsto (x,x)\qquad\quad e: X\to\{*\}::x\mapsto *\,.
\]
The map $\delta$ \em copies \em the elements of $X$, while $e$ \em erases \em them.  Here coassociativity means that if we wish obtain three copies of something, then after first making two copies it doesn't matter which of these two we copy again.  Cocommutativity tells us that after copying we exchange the two copies we still have the same.
Counitality tells us that if we first copy and then erase one of the copies, this is the same as doing nothing.

Now, the reason why you won't encounter any comonoids in a standard algebra textbook is simply because this example is the only example of a commutative comonoid, and hence it carries no real content, i.e.~it freely arises from the underlying set.  But the reason for the trivial nature of commutative comonoids is the fact of  the following being functions:
\[
\mu: A\times A\to A\qquad \delta: A\to A \times A\qquad u: \{*\}\to A \qquad e:A\to \{*\}\,.
\]
In other words, $\mu$ and $\delta$ are morphisms in the category ${\bf FSet}$.  
While the concept of a commutative monoid is interesting in ${\bf FSet}$, that of a  cocommutative comonoid isn't in ${\bf FSet}$.  However, if we put the above definition in the language of monoidal categories and pass to other categories than ${\bf FSet}$, then the situation changes.  In fact, if this category has a $\dagger$-functor, then to each commutative monoid corresponds a cocommutative comonoid. This already happens when we relax the condition that $\mu$ and $\delta$ are functions to $\mu$ and $\delta$ being relations.  

Let ${\bf C}$ be any symmetric monoidal category.  A \em commutative ${\bf C}$-monoid \em is an object $A\in|{\bf C}|$ with morphisms:
\[
\mu: A\otimes A\to A \qquad\quad u:\II\to A
\]
which is commutative, associative and unital i.e.:
\[
\mu\circ(\mu\otimes 1_A)=\mu\circ(1_A\otimes \mu)\qquad \mu=\mu\circ\sigma \qquad \mu\circ(1_A\otimes u)=1_A\,.
\]
Similarly, a \em cocommutative ${\bf C}$-comonoid \em is an object $A$ with morphisms
\[
\delta: A\to A\otimes A \qquad\quad e:A\to \II
\]
which is cocommutative, coassociative and counital i.e.:
\[
(\delta\otimes 1_A)\circ\delta=(1_A\otimes\delta)\circ\delta\qquad \delta=\sigma\circ\delta \qquad (1_A\otimes e)\circ\delta=1_A
\]

Now putting all of this diagrammatically, a commutative ${\bf C}$-monoid is  a pair:
\[
\raisebox{-0.38cm}{\epsfig{figure=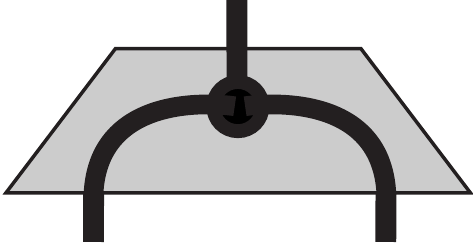,width=50pt}}:A\otimes A\to A\qquad\quad
\raisebox{-0.40cm}{\epsfig{figure=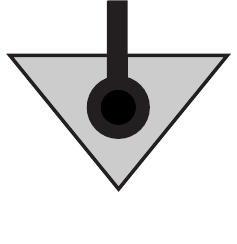,width=25pt}}\,: {\rm I}{\to} A
  \]
satisfying:
\begin{center}
\epsfig{figure=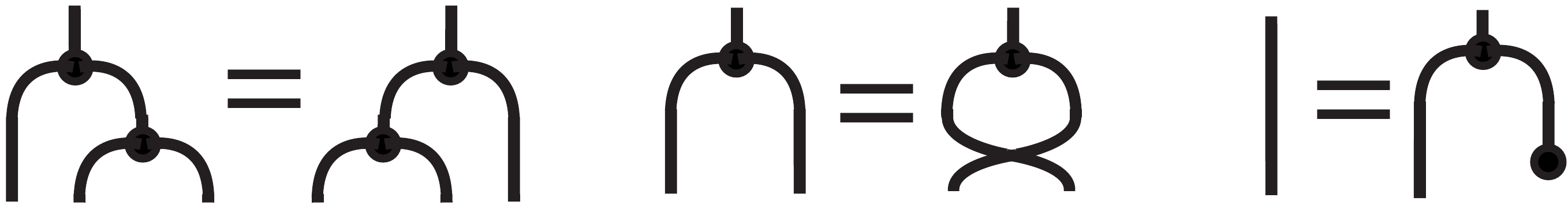,width=300pt}  
\end{center}
and a cocommutative ${\bf C}$-comonoid is a pair:
\[
\raisebox{-0.38cm}{\epsfig{figure=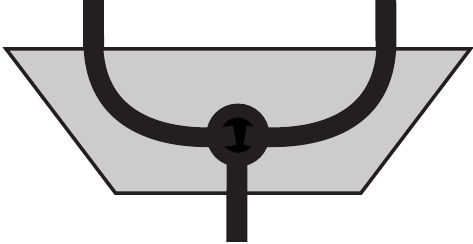,width=50pt}}:A\to A\otimes A\qquad\quad
\raisebox{-0.54cm}{\epsfig{figure=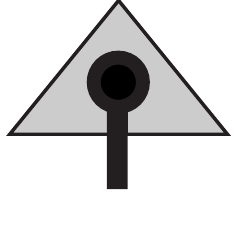,width=25pt}}:  A{\to} {\rm I}
  \]
satisfying:
\begin{center}
\epsfig{figure=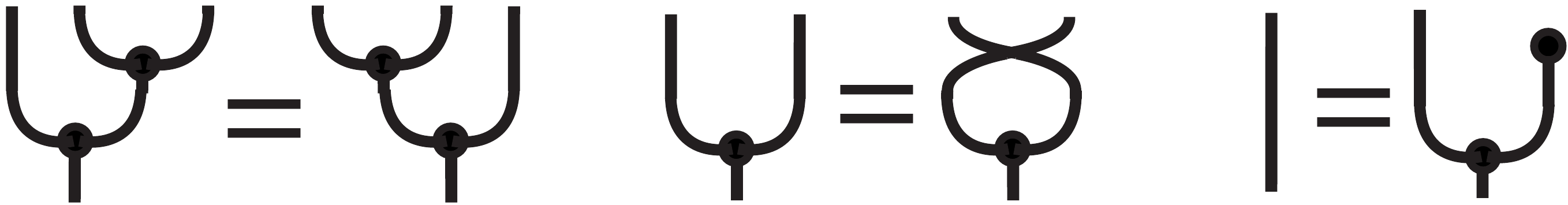,width=300pt}  
\end{center}

Recall here also that it is a general fact in algebra that if a binary operation $-\bullet-$ both has a left unit $1_l$ and right unit $1_r$, then these must be equal:
\[
1_l = 1_l\bullet 1_r = 1_r\,.
\]
It then also follows that a commutative multiplication can only have one unit, i.e.~if it has a unit then it is completely determined by the multiplication.  This fact straightforwardly lifts to the more general kinds of monoids and comonoids that we discussed above, and therefore, we will at several occasions omit specification of the (co)unit.

Here is an example of commutative  monoids and corresponding cocommutative comonoids in ${\bf   FHilb}$ on a two-dimensional Hilbert space:
\begin{center}
\begin{tabular}{l|l}
$\raisebox{-0.50cm}{\epsfig{figure=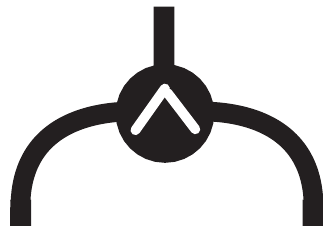,width=38pt}}::
\left\{\begin{array}{l}
|00\rangle,|01\rangle,|10\rangle\mapsto|0\rangle\\
|11\rangle\mapsto|1\rangle\\
\end{array}\right.$
&
$\raisebox{-0.50cm}{\epsfig{figure=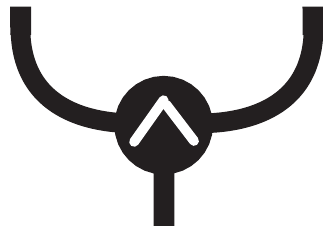,width=38pt}}::
\left\{\begin{array}{l}
|0\rangle\mapsto|00\rangle+|01\rangle+|10\rangle\\
|1\rangle\mapsto|11\rangle\\
\end{array}\right.$
\end{tabular}
\end{center}
The first monoid has the `and'-operation applied to the $\{|0\rangle, |1\rangle\}$-basis as its multiplication.  The comultiplication is the corresponding adjoint.  

\subsubsection{Commutative dagger Frobenius algebras}

Now consider the following three comultiplications: 
\begin{center}
\begin{tabular}{l|l}
$\mu_Z= \raisebox{-0.50cm}{\epsfig{figure=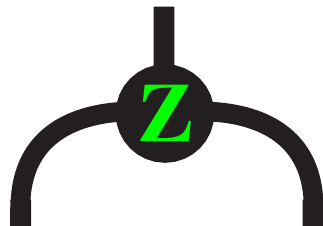,width=38pt}}::
\left\{\begin{array}{l}
|00\rangle\mapsto|0\rangle\\
|11\rangle\mapsto|1\rangle\\
\end{array}\right.$
 \ & \
$\delta_Z= \raisebox{-0.50cm}{\epsfig{figure=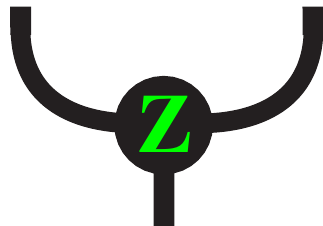,width=38pt}}::
\left\{\begin{array}{l}
|0\rangle\mapsto|00\rangle\\
|1\rangle\mapsto|11\rangle\\
\end{array}\right.$
\\ \\
$\mu_X= \raisebox{-0.50cm}{\epsfig{figure=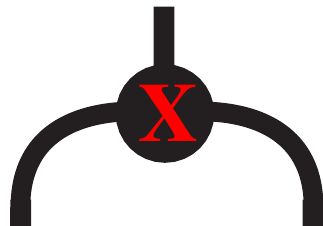,width=38pt}}::
\left\{\begin{array}{l}
|++\rangle\mapsto|+\rangle\\
|--\rangle\mapsto|-\rangle\\
\end{array}\right.$
 \ & \
$\delta_X= \raisebox{-0.50cm}{\epsfig{figure=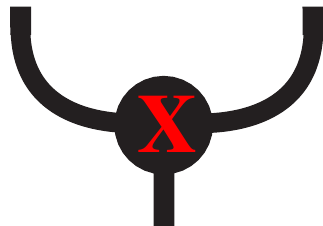,width=38pt}}::
\left\{\begin{array}{l}
|+\rangle\mapsto|++\rangle\\
|-\rangle\mapsto|--\rangle\\
\end{array}\right.$
\\ \\
$\mu_Y= \raisebox{-0.50cm}{\epsfig{figure=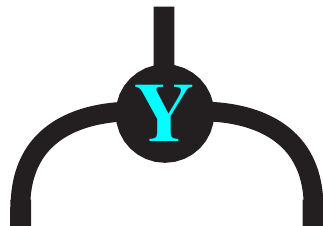,width=38pt}}::
\left\{\begin{array}{l}
|\ \sharp\ \,\sharp\,\rangle\mapsto|\,\sharp\,\rangle\\
|\!=\,=\rangle\mapsto|\!=\rangle\\
\end{array}\right.$
 \ & \
$\delta_Y= \raisebox{-0.50cm}{\epsfig{figure=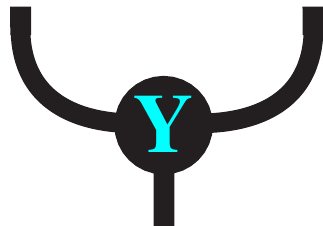,width=38pt}}::
\left\{\begin{array}{l}
|\,\sharp\,\rangle\mapsto|\ \sharp\ \,\sharp\,\rangle\\
|\!=\rangle\mapsto|\!=\,=\rangle\\
\end{array}\right.$
\end{tabular}
\end{center}
Each of these is defined as a copying operation of some basis, respectively 
\[
{\cal Z}=\{|0\rangle, |1\rangle\}\qquad {\cal X}=\{|+\rangle=|0\rangle+ |1\rangle, |-\rangle=|0\rangle-|1\rangle\}
\]
\[
{\cal Y}=\{|\sharp\rangle=|0\rangle+i |1\rangle, |\!=\rangle=|0\rangle-i|1\rangle\}\,,
\]
that is, the eigenstates for the usual Pauli operators.  Each of these encodes a basis in the sense that we can recover the basis from the comultiplication as those vectors that satisfy:
\[
\delta(|\psi\rangle)=|\psi\rangle\otimes|\psi\rangle\,.
\]
The fact that no other vector besides those that we by definition copy are in fact copied is a consequence of the above mentioned no-cloning theorem  \cite{Dieks,WZ}. 

The corresponding multiplications are again their adjoints.   These last examples embody the reason why we are interested in commutative comonoids.  What is already remarkable at this stage is that each of these encodes an orthonormal basis in a language only involving composition and tensor.  There is no reference whatsoever to either sums or scalar multiples in contrast to the usual definition of an orthonormal basis $\{|i\rangle\}$ on a Hilbert space: 
\[
\forall |\psi\rangle\in{\cal H}, \exists (c_i)_i\in\mathbb{C}^n: |\psi\rangle=\sum_i c_i |i\rangle
\qquad \forall i,j: \langle i|j\rangle=\delta_{ij}\,.
\]

But there is more.  In fact, one can endow these monoids and comonoids with some additional properties, expressible in a language only involving composition, tensor and now also adjoint, such that they are in bijective correspondence with orthonormal bases.

\paragraph{Definition.} A \em commutative algebra \em in a symmetric monoidal category is a pair consisting of a commutative monoid and a cocommutative  comonoid on the same object. A \em special commutative Frobenius algebra \em is a commutative algebra
which is moreover \em special \em and satisfies the \em Frobenius law\em, respectively: 
\begin{center}
\epsfig{figure=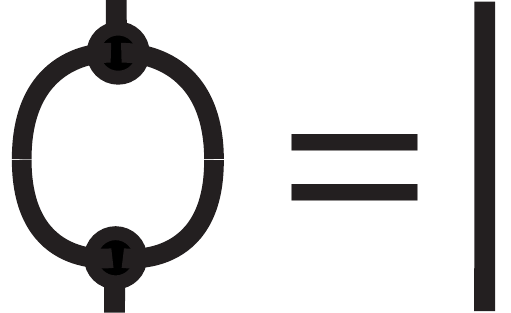,width=57pt}    \qquad\qquad \epsfig{figure=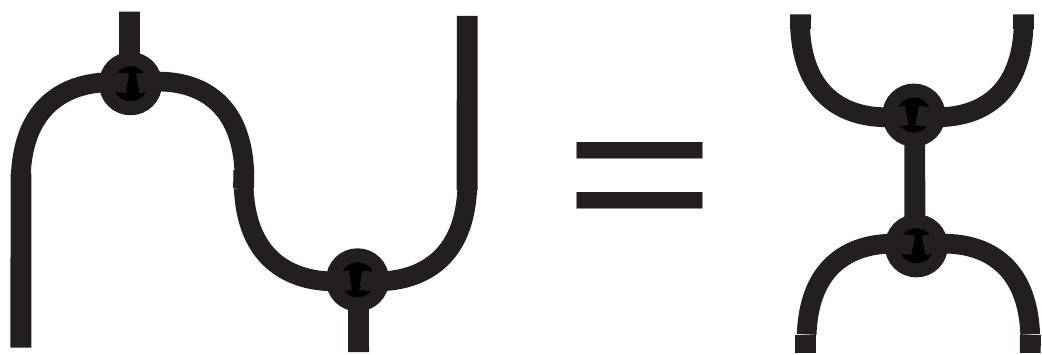,width=114pt}\,.
\end{center}
A \em special commutative dagger Frobenius algebra \em  or \em classical structure \em or \em basis structure \em in a dagger symmetric monoidal category  is a \em special commutative Frobenius algebra \em for which the monoid is the dagger of the comonoid.

\medskip

Think of an orthonormal basis ${\cal B}=\{|b_i\rangle\}_i$ for a Hilbert space ${\cal H}$ as the pair $({\cal H}, {\cal B})$ consisting of the Hilbert space which carries this basis as additional structure.   Since the multiplication and the comultiplication are related by the dagger, and since having a unit is rather a property than a structure, we denote a special commutative dagger Frobenius algebra on an object $A$ as $(A,\delta)$.

\paragraph{Theorem.} 
There is a bijective correspondence between orthogonal bases for finite dimensional Hilbert spaces and special commutative Frobenius algebras in ${\bf   FHilb}$.  This correspondence is realized by the mutually inverse mappings:
\bit
\item Each special commutative Frobenius algebra $({\cal H},\delta)$ is mapped on $({\cal H},{\cal B}_\delta)$ where ${\cal B}_\delta$ consists of the set of vectors that are copied by $\delta$.
\item Each orthonormal basis $({\cal H}, {\cal B})$ is mapped on $({\cal H}, \delta_{\cal B}:{\cal H}\to {\cal H}\otimes {\cal H})$ where $\delta_{\cal B}$ is the linear map which copies the vectors of ${\cal B}$.
\eit
Restricting to orthonormal bases corresponds to restricting to  special commutative dagger Frobenius algebras.

\bigskip

This result was shown by Pavlovic, Vicary and myself in \cite{CPV}. 

That classicality boils down to families of spiders is a consequence of the fact that special commutative dagger Frobenius algebras are in bijective correspondence with spiders, as our notation already indicated. This was shown by Lack \cite{Lack}, but in a manner which is so abstract that it may not be accessible to the reader.  A more direct presentation of the proof is in \cite{CPaq}.

\subsubsection{Varying the coordinate system}

But what if we change the category?  It turns out that this mathematical concept, when we look through coordinate systems other than ${\bf   FHilb}$, allows us to discover important quantum mechanical concepts in places where one doesn't expect it, most notably `complementarity' or `unbiasedness' \cite{Schwinger}. 

In \cite{CD} the author and Duncan defined \em complementarity \em in terms of special commutative dagger Frobenius algebras, i.e.~still in terms of a language only involving composition, tensor and adjoint, in a manner which yields the usual notion in  ${\bf   FHilb}$.  Concretely, it was shown that classical structures:
\[
\Bigl({\cal H}\,,\delta_G=\raisebox{-0.30cm}{\epsfig{figure=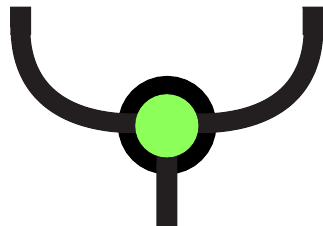,width=28pt}}\Bigr)\qquad \mbox{and} \qquad \Bigl({\cal H}\,,\delta_R=\raisebox{-0.30cm}{\epsfig{figure=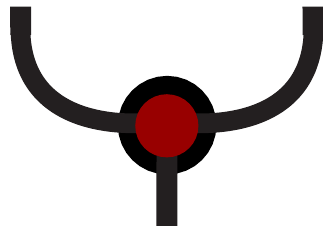,width=28pt}}\Bigr)
\]
in ${\bf   FHilb}$ are complementary if and only if we have:
\[
\epsfig{figure=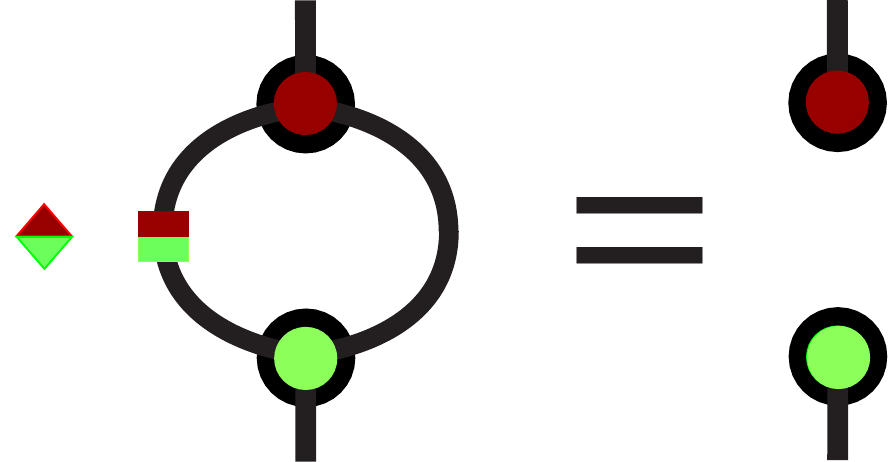,width=110pt} 
\]
where $\raisebox{-0.07cm}{\epsfig{figure=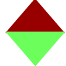,width=8pt}}$ is a normalizing scalar and $\raisebox{-0.14cm}{\epsfig{figure=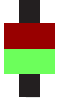,width=8pt}}$ is a so-called \em dualizer \em \cite{CPaqPer, CD}, which both are obtained by composing $\delta_g^\dagger$, $\delta_r$, $u_g$ and  $u_r^\dagger$ in a certain manner.\footnote{Their explicit definition is not of importance here; it suffices to know that formally it witnesses the role played by complex conjugation in adjoints, in the sense that it becomes trivial (i.e.~identity) when only real coefficients are involved, which is for example the case for the $Z$- and $X$-classical structures for which we have: 
\[
\epsfig{figure=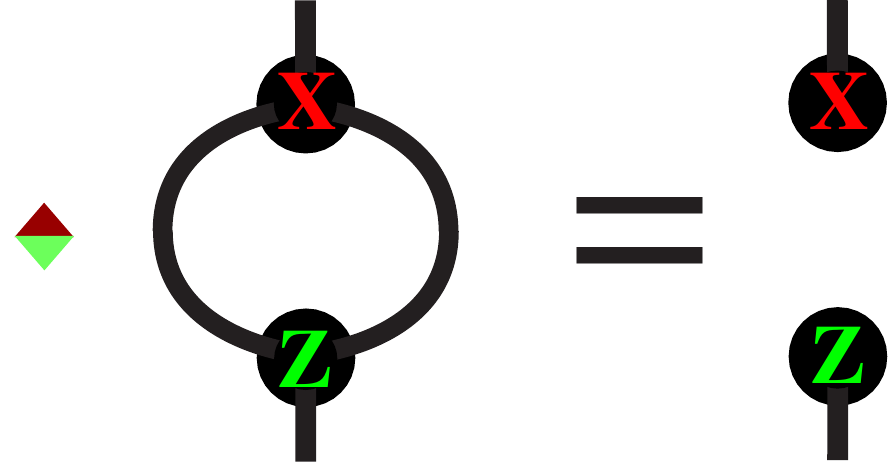,width=100pt}
\]
Those acquainted with the field of quantum algebra \cite{StreetBook} might recognize here the defining equation of a Hopf-algebra, with the dualizer playing the role of the antipode.  The apparent non-symmetrical left-hand-side picture becomes symmetric if we represent the bases in terms of the unitary operations which transform a chosen standard basis into them \cite{CWWWZ}.} 

The quite astonishing fact discovered by Edwards and the author in \cite{CE} was that even in ${\bf FRel}$ one encounters such complementary classical structures, even already on the two-element set $\{0, 1\}$:
\begin{center}
\begin{tabular}{l|l}
$\raisebox{-0.50cm}{\epsfig{figure=Mon2a.pdf,width=38pt}}::
\left\{\begin{array}{l}
(0,0)\mapsto 0\\
(1,1)\mapsto 1\\
\end{array}\right.$
\ \ \ &
$\raisebox{-0.50cm}{\epsfig{figure=Mon2b.pdf,width=38pt}}::
\left\{\begin{array}{l}
0\mapsto(0,0)\\
1\mapsto(1,1)\\
\end{array}\right.$
\\ \\
$\raisebox{-0.50cm}{\epsfig{figure=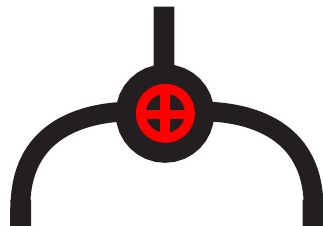,width=38pt}}::
\left\{\begin{array}{l}
(0,0),(1,1)\mapsto 0\\
(0,1),(1,0)\mapsto 1\\
\end{array}\right.$
\ \ \ &
$\raisebox{-0.50cm}{\epsfig{figure=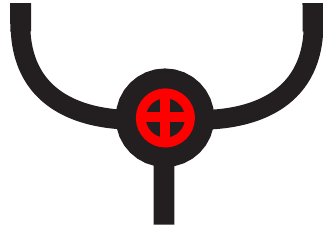,width=38pt}}::
\left\{\begin{array}{l}
0\mapsto(0,0),(1,1)\\
1\mapsto(0,1),(1,0)\\
\end{array}\right.$
\end{tabular}
\end{center}
Meanwhile, Pavlovic, Duncan and Edwards, and Evans et al., have classified all classical structures and complementarity situations in ${\bf FRel}$ \cite{Dusko, DE, Evans}.  

\paragraph{Example: Spekkens' toy qubit theory.}
It is a particular case of these complementarity situations in ${\bf FRel}$ which gives rise to  Spekkens' toy qubit theory discussed above and hence its striking resemblance to quantum theory.  
This exploration of  ${\bf FRel}$ is still an unfinished story.   
While, for example Spekkens' toy theory is a local theory, we strongly suspect that we can discover non-locality  (in the sense of \cite{CES}) somewhere within ${\bf FRel}$.\footnote{For completeness let us mention that in  \cite{Baez} Baez emphasizes structural similarities between ${\bf   FHilb}$ and the category ${\bf 2Cob}$ of 1-dimensional closed manifolds and cobordisms between these, which play an important role in  topological quantum field theory \cite{Atiyah}.  While in ${\bf 2Cob}$ each object comes with a classical structure, there is never more than one, so in this coordinate system there are no complementarity situations.}









\section{Acknowledgements}

This paper benefited from discussions with Samson Abramsky, John Baez, Rick Blute, Giulio Chiribella, Mauro D'Ariano, Andreas D\"oring, Chris Fuchs, Chris Isham, Keye Martin, David Moore, Constantin Piron, Phil Scott, Rob Spekkens, Prakash Panangaden, Simon Perdrix, Jamie Vicary, Frank Valckenborgh and Alex Wilce at some point in the past, from discussions with my students Bill Edwards, Benjamin Jackson and Raymond Lal, and in particular from duels with Lucien Hardy, be it either on whatever they call beer in England, his obsession with the shape of tea bags, or his `appreciation' of mathematics, but in particular for  one which can be seen at \cite{Hardy1}.  Credits go to Howard Barnum for recalling the Radiohead song lyric of \em Karma Police \em in an email to Chris Fuchs, Marcus Appleby and myself, just when I was about to finish this chapter. We also in particular like to thank Hans Halvorson and my students Ray Lal and and Johan Paulsson for proofreading the manuscript .  We acknowledge the Perimeter Institute for Theoretical Physics for a Long Term Visiting Scientist position.  It was during my stay there that some of the ideas in this paper were developed.  This work is supported by the authors'  EPSRC Advanced Research Fellowship, by the EU FP6 STREP QICS, by a Foundational Questions Institute Large Grant (FQXi) and by US Office of Naval Research (ONR).


%
\end{document}